\documentclass[onecolumn]{article} 
\pdfoutput=1

\usepackage{blindtext}
\usepackage[utf8]{inputenc} 

\usepackage[top=.75in,bottom=.75in,left=.5in,right=.8in]{geometry}

\usepackage{calc}
\usepackage{booktabs} 
\usepackage{array} 
\usepackage{paralist} 
\usepackage{verbatim} 
\usepackage{subfig} 
\usepackage{amsmath,bbm} 
\usepackage{amsfonts} 
\usepackage{graphicx} 
\usepackage{amssymb}
\usepackage{hyperref}
\usepackage{authblk, units}
\usepackage{lineno}
\usepackage{braket}
\usepackage[numbers,sort&compress]{natbib}

\usepackage{fancyhdr} 
\usepackage{floatrow}
\newfloatcommand{capbtabbox}{table}[][\FBwidth]

\usepackage{sectsty}
\usepackage{titlesec}

\usepackage[nottoc,notlof,notlot]{tocbibind} 
\usepackage[titles,subfigure]{tocloft} 

\newcommand{\nbar}{\overline{n}}

\begin{document}


\title{Demonstrating Quantum Error Correction that Extends the Lifetime of Quantum Information}

\author[$*$,$\dag$,1]{Nissim Ofek}
\author[$*$,$\dag$,1]{Andrei Petrenko}
\author[1]{Reinier Heeres}
\author[1]{Philip Reinhold}
\author[1]{Zaki Leghtas}
\author[1]{Brian Vlastakis}
\author[1]{Yehan Liu}
\author[1]{~~~~~~~~~Luigi Frunzio}
\author[1]{S.\ M.\ Girvin}
\author[1]{L.\ Jiang}
\author[1,2]{Mazyar\ Mirrahimi}
\author[1]{M.\ H.\ Devoret}
\author[1,$\dag$]{R.\ J.\ Schoelkopf}
\affil[$*$]{These authors contributed equally to this work.}
\affil[1]{Departments of Physics and Applied Physics, Yale University, New Haven, CT 06510, USA}
\affil[2]{QUANTIC team, INRIA de Paris, 2 Rue Simone Iff, 75012 Paris, France}
\affil[$\dag$]{email: andrei.petrenko@yale.edu, nissim.ofek@yale.edu, robert.schoelkopf@yale.edu}
\date{}
\maketitle


\noindent
The remarkable discovery of Quantum Error Correction (QEC), which can overcome the errors experienced by a bit of quantum information (qubit) ~\cite{Shor:1995vn}, was a critical advance that gives hope for eventually realizing practical quantum computers.  In principle, a system that implements QEC can actually pass a``break-even" point and preserve quantum information for longer than the lifetime of its constituent parts.  To implement QEC, one redundantly encodes a qubit in a higher dimensional space using quantum states with carefully tailored symmetry properties. By performing a projective measurement of this symmetry, or error syndrome, errors can be detected and then corrected via simple operations~\cite{NielsenQI}. Reaching the break-even point and demonstrating the extension of a qubit's lifetime, however, has thus far remained an outstanding and challenging goal~\cite{Devoret:2013jz}.  Several previous works have demonstrated elements of QEC in NMR~\cite{Cory:1998kw,Knill:2001tr,Moussa:2011jj,Leung:1999vz}, ions~\cite{Chiaverini:2004hy,Schindler:2011ch,Nigg:2014eb}, nitrogen vacancy (NV) centers~\cite{Waldherr:2014kt,Taminiau:2014up,Cramer:2015uk}, photons~\cite{Pittman:2005du, Aoki:2009ew}, and superconducting transmons~\cite{Reed:2012hu,Kelly:2015tg,Corcoles:2015bg,Riste:2015uh}.  However, these works primarily illustrate the signatures or scaling properties of QEC codes rather than test the capacity of the system to extend the lifetime of quantum information over time.  Here we demonstrate a QEC system that reaches the break-even point by suppressing the natural errors due to energy loss for a qubit logically encoded in superpositions of coherent states, or cat states~\cite{book:haroche06} of a superconducting resonator~\cite{Leghtas:2013ff,Mirrahimi:2014js,Vlastakis:2013ju,Sun:2013ha}. Moreover, the experiment implements a full QEC protocol by using real-time feedback to encode, monitor naturally occurring errors, decode, and correct.   As measured by full process tomography, the enhanced lifetime of the encoded information is $320\mathrm{\mu s}$ without any post-selection. This is longer than the lifetime of any of the system's parts: $20$ times greater than that of the system's transmon, over twice as long as an uncorrected logical encoding, and $10\%$ longer than the highest quality element of the system (the resonator's $\ket{0}_f, \ket{1}_f$ Fock states).  Our results illustrate the power of novel, hardware efficient qubit encodings over traditional QEC schemes. Furthermore, they advance the field of experimental error correction from confirming the basic concepts to exploring the metrics that drive system performance and the challenges in implementing a fault-tolerant system.


The promise of QEC lies in its ability to suppress errors in a quantum computer, thus empowering it to perform larger and more complex algorithms.  Moreover, if one implements an error correction system within the framework of a fault-tolerant architecture, in which the performance of all the components also exceeds the architecture's threshold, the size of the computation can be scaled exponentially with only a polynomial overhead in hardware.  Challenges abound, however, given the typically large overhead requirements, the stipulation that errors do not propagate, and the demanding error thresholds~\cite{Steane:1996wk,Fowler:2012fi}.  Experimentally, with any scheme one must first demonstrate a logical encoding that corrects the dominant sources of natural errors while operating beyond the break-even point.  As logical qubits typically contain many physical qubits, each with their own errors, surpassing the break-even point requires high levels of performance in every aspect of the system, from the encoding and decoding operations into and out of the logical space to the correction of logical errors with fast, repeated error syndrome measurements in time.  As no source of decoherence can ever be completely suppressed within a single round of correction, one still always witnesses an exponential decay in the process fidelity of any QEC system over many such rounds.  The system operates beyond the break-even point, however, when the performance of all its aspects is sufficient to overcome the inherent overhead of encoding. Only then can the characteristic decay time of the information exceed that of the system's best component, so that employing QEC is  advantageous. Despite the progress in improved coherence~\cite{Gambetta:2015ue} and high fidelity operations~\cite{Chow:2012ug,Barends:2014fu}, realizing such a scenario with traditional QEC architectures has thus far remained an outstanding task.

Previous works have demonstrated parts of a full QEC solution by either correcting artificially induced errors; focusing on one out of several dominant error processes; assessing the performance of known protected or particularly vulnerable states; employing post-selection as a means to study specific decoherence mechanisms; or some combination thereof~\cite{Cory:1998kw,Knill:2001tr,Moussa:2011jj,Leung:1999vz,Chiaverini:2004hy,Schindler:2011ch,Nigg:2014eb,Waldherr:2014kt,Taminiau:2014up,Cramer:2015uk,Pittman:2005du, Aoki:2009ew,Reed:2012hu,Kelly:2015tg,Corcoles:2015bg,Riste:2015uh}.  By isolating subsets of a general error process to study the viability of a QEC scheme, however, these works indicate a correction of specific errors under restricted circumstances.  Indeed, they do not quantify the exponential decay in time one would see in the process fidelity of a quantum bit subjected to repeated rounds of error correction.  In this work we instead implement a full QEC system to protect a qubit exposed to its natural environment.  We study how all sources of decoherence, including those arising from an imperfect ancillary system (ancilla), together contribute to the probability that a single round of correction fails.  By minimizing this probability, we suppress the process fidelity decay rate to the break-even point.  Furthermore, we pinpoint the dominant limitation on performance to be that of forward propagation of errors from the ancilla, thus motivating the future steps necessary to realize a fault-tolerant QEC system.

\begin{figure*}[!ht]
\centering
\includegraphics[width=5.5in]{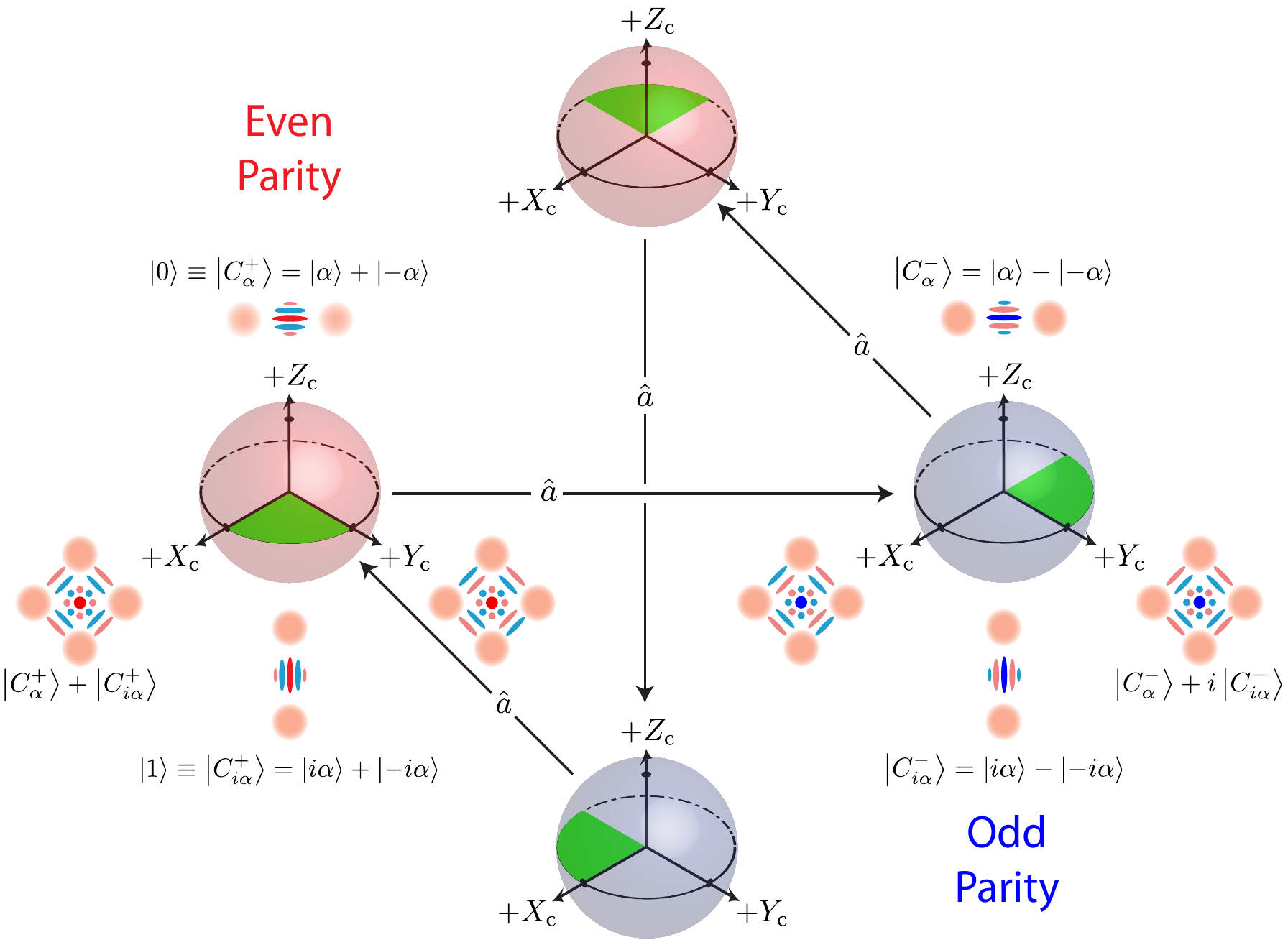}
\caption{\footnotesize
    \textbf{The cat code cycle.}
        In the logical encoding of $\ket{0}\equiv\ket{C^+_{\alpha}}=\ket{\alpha}+\ket{-\alpha}$ and $\ket{1}\equiv\ket{C^+_{i\alpha}}=\ket{i\alpha}+\ket{-i\alpha}$ (normalizations omitted), the two ``2-cats" $\ket{C^+_{\alpha}}$ and $\ket{C^+_{i\alpha}}$ are both eigenstates of even photon number parity (an ``$n$-cat" is a superposition of $n$ coherent states).  For large enough $|\alpha|$ they are also effectively orthogonal to one another.  In this basis, the states along $+X_c$ and $+Y_c$ are both ``4-cats" of even parity as well.  The different patterns in the fringes of their cartoon Wigner functions signify the different phase relationship between the basis states.  These features allow one to store a qubit in a superposition of 2-cats and at the same time monitor the parity as the error syndrome without projecting the state out of this encoding.  The loss of a single photon changes not just the parity of the basis states, but the phase relationship between them by a factor of $i$ ($\ket{C^+_{\alpha}}+\ket{C^+_{i\alpha}}\rightarrow\ket{C^-_{\alpha}}+i\ket{C^-_{i\alpha}}$).  Decoding after one jump, one finds the initial qubit rotated by $\pi/2$ about the $Z_c$ axis (indicated by green shading).  Thus, with each application of $\hat{a}$, the encoded state cycles between the even and odd parity subspaces (shaded in red and blue), while due to each consequent factor of $i$, the encoded information rotates about the $Z_c$ axis by $\pi/2$, returning to the original state after four photon jumps.  Between the stochastic applications of $\hat{a}$, the cat states deterministically decay toward vacuum: $\alpha\rightarrow\alpha e^{-\kappa t/2}$ (not depicted here).  As long as the coherent states do not overlap appreciably, this effectively constitutes the identity operation on the encoded state. }
\label{fig1}
\end{figure*}

The cat code we implement here is a hardware-efficient scheme that requires fewer physical resources and introduces fewer error mechanisms~\cite{Leghtas:2013ff,Mirrahimi:2014js} than traditional QEC proposals.  Designed to operate within a continuous-variable framework~\cite{Lloyd:2003kv,Gottesman:2001jb,Menicucci:2006ir,LundRalph:2008}, the cat code exploits the fact that a coherent state $\ket{\alpha}$ is an eigenstate of the resonator lowering operator $\hat{a}$: $\hat{a}\ket{\alpha}=\alpha\ket{\alpha}$.  Using a logical basis comprised of superpositions of cat states, which are eigenstates of photon number parity, the cat code requires just a single ancilla to monitor the dominant error due to single photon loss induced by resonator energy damping (Fig.~\ref{fig1}).  This error channel gives rise to two effects: deterministic energy decay of the resonator field to vacuum and the accompanying stochastic application of $\hat{a}$, which results in a change of photon number parity of any state within the cat code.  The former becomes a limiting factor only at small resonator field amplitudes when coherent state overlap can no longer be neglected (Methods); it can be addressed through either dissipative pumping approaches~\cite{Leghtas:2015uf} or unitary gates, imposing no time limit on the duration of the cat code.  Photon loss is accompanied by phase shifts of $\pi/2$ about the $Z_c$ axis within the codeword, indicating that by monitoring photon parity as the error syndrome, we adhere to the prescriptions for error correction by translating single photon loss into a unitary operation on the encoded state:
\begin{align}
\hat{a}(c_0\ket{C^+_{\alpha}}+c_1\ket{C^+_{i\alpha}})\label{cat_code}=&\frac{c_0}{\sqrt{2}}(\ket{\alpha}-\ket{-\alpha})+i\frac{c_1}{\sqrt{2}}(\ket{i\alpha}-\ket{-i\alpha})\nonumber \\
=&c_0\ket{C^-_{\alpha}}+ic_1\ket{C^-_{i\alpha}}\nonumber,
\end{align}
By detecting photon jumps in real-time with Quantum Non-Demolition parity measurements~\cite{Sun:2013ha}, we learn how the phase relationship between the basis states changes, thereby protecting the encoded qubit from the system's dominant error channel.

\begin{figure}[!ht]
\centering
\includegraphics[width=3in]{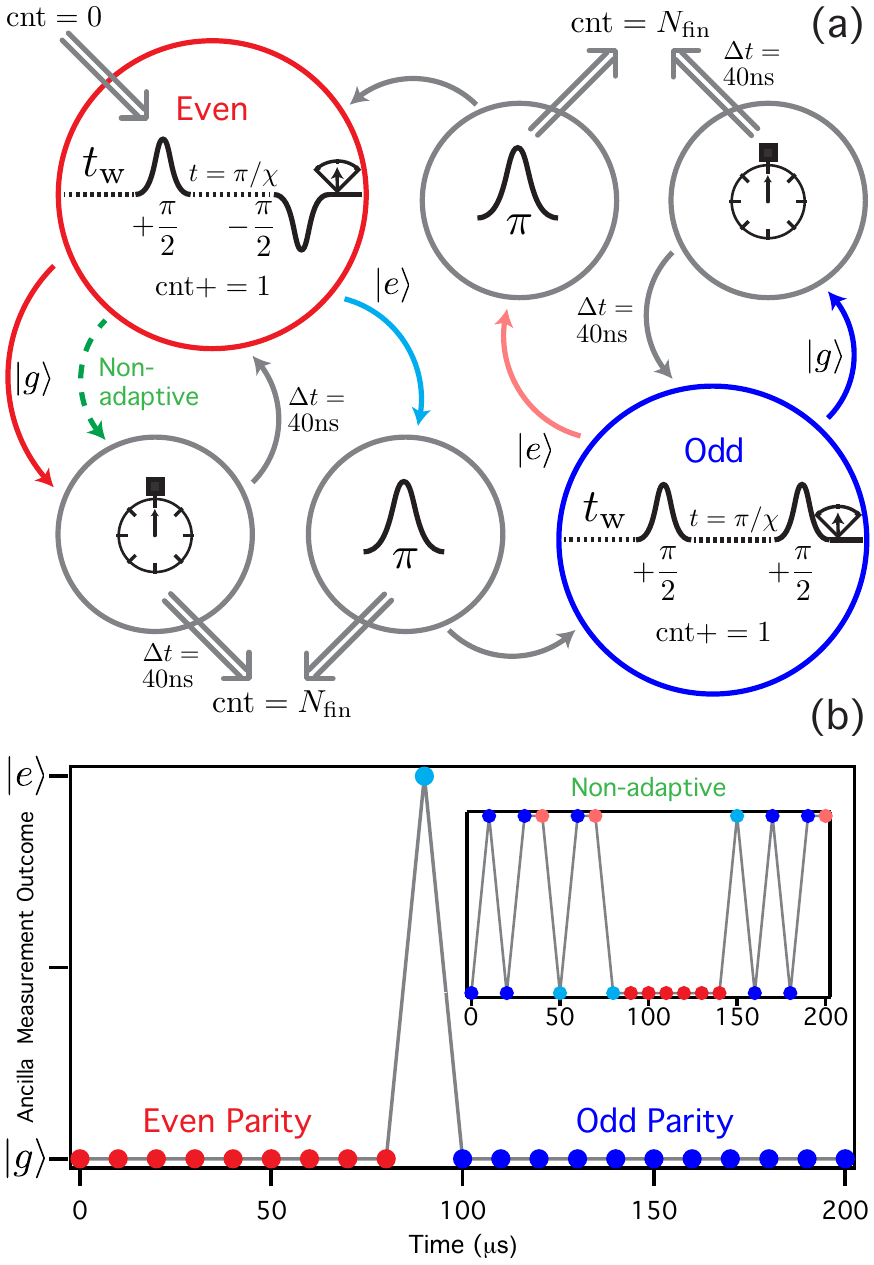}
\caption{\footnotesize
    \textbf{Quantum state machine for adaptive error monitoring.}
        \textbf{(a)} Schematic of the logical flow.  This quantum state machine implements an adaptive parity monitoring scheme in which the parity mapping protocol is updated in real-time to maximize the probability to measure $\ket{g}$.  This reduces the time the ancilla spends in $\ket{e}$ by up to a factor of $50$ per error (Methods), enhancing parity measurement fidelities and making it an essential component in the experimental workflow.  Entering the state machine (double-arrow pointing inward) consists of initializing a counter ($\mathrm{cnt}=0$) and using the protocol that maps even parity to $\ket{g}$ with a simple Ramsey-style pulse sequence~\cite{Bertet:2002df,Haroche:2007uc} (red circle) followed by a projective measurement of the ancilla; prior to the pulse sequence there is an idling time $t_\mathrm{w}$.  In addition, the counter is incremented.  If the measurement result is $\ket{g}$, the system idles for $40\mathrm{ns}$ (denoted by the stopwatch) and then returns to the previous state.  A measurement of $\ket{e}$ implies a change in parity, or the occurrence of an error.  In this case, a $\pi$ pulse is applied (Gaussian envelope; $\sigma=2\mathrm{ns}$; total duration $40\mathrm{ns}$) and the system moves to using a pulse sequence that maps odd parity to $\ket{g}$ (blue circle), and again increments the counter.  The controller returns to the initial state after another error, completing the state machine cycle.  When the counter reaches the pre-loaded number $\mathrm{cnt}=N_\mathrm{fin}$, the system exits (double-arrows pointing out). Throughout a single measurement trajectory, counting the errors amounts to counting the number of times $\ket{e}$ occurs.  Measurement infidelities are emphasized by lighter shades of red and blue between parity transitions, corresponding to a lower confidence when the meter measures $\ket{e}$.  The non-adaptive protocol simply cycles between using a fixed parity mapping sequence and the short idling time (green dotted arrow).  \textbf{(b)} Results.  An example single-shot record of parity measurement results, with $t_\mathrm{w}=9\mu$s between each measurement, demonstrates the difference between the adaptive and non-adaptive protocols.  In the former, the ancilla is found to be in $\ket{e}$ just once out of 20 measurements regardless of the parity.  With the latter, $\ket{e}$ is measured 8 times; given $t_\mathrm{w}/T_1\approx0.3$, the odds of ancilla decay in this trace are so high that it is unclear how many errors occurred.}
\label{fig2}
\end{figure}

Comparing traditional QEC schemes to the cat code, the former typically protect a qubit from decoherence by projecting components of the redundant encoding into spaces defined by four unitary operators: identity $\hat{I}$, and the Pauli operators $\hat{\sigma}_x$, $\hat{\sigma}_y$ and $\hat{\sigma}_z$.  In the latter, however, within the logical encoding of the cat code there are only two such operators: $\hat{I}$ (no photon loss and amplitude decay $\ket{\alpha}\rightarrow\ket{\alpha e^{-\kappa\delta t/2}}$) and $(\hat{I}+i\hat{\sigma}_z)/\sqrt{2}$ (application of $\hat{a}$).  Furthermore, given that the overlap between coherent states $|\left\langle\alpha|i\alpha\right\rangle|$ falls off exponentially with increasing $|\alpha|$~\cite{book:haroche06}, we can use basis states of average photon number $\bar{n}\approx 2$, which increases the error rate within the codeword by only a factor of $\sim 2$~\cite{book:haroche06} rather than by orders of magnitude as in traditional schemes~\cite{Steane:1996wk,Fowler:2012fi}.  Finally, we use real-time feedback to not only drastically enhance the fidelity of the parity measurements with an ``adaptive" monitoring scheme (Fig~\ref{fig2}), but moreover monitor and record the occurrence of errors as they happen.  This enables us to simply log the changes in parity without having to insert any photons back into the resonator, leaving ample time to perform the correction once we decode the state.


We employ a 3D circuit quantum electrodynamics (QED) architecture~\cite{Wallraff:2004dy} comprised of a single transmon qubit coupled to two waveguide resonators~\cite{Paik:2011hd}.  The transmon is used as an ancilla to both interrogate the error syndrome and to encode and decode the logical states.  One of the resonators stores the logical states while the other is used for ancilla readout.  The dominant interaction terms are described by the following Hamiltonian:
\begin{equation}
    \label{eq:hamiltonian}
\hat{H}/\hbar = \omega_s \hat{a}^\dagger\hat{a} + (\omega_{a} - \chi_{sa}\hat{a}^\dagger\hat{a} ) \ket{e}\bra{e} - \frac{K}{2}\hat{a}^{\dag 2}\hat{a}^2,
\end{equation}
\noindent
with $\ket{e}\bra{e}$ being the ancilla excited state projector; $\omega_s,~\omega_a$ the storage resonator (henceforth just the resonator) and ancilla transition frequencies; $\chi_{sa}/2\pi\sim1.95\mathrm{MHz}$ the dispersive interaction strength between the two; and $K/2\pi\sim4.5\mathrm{kHz}$ the resonator anharmonicity, or Kerr.  The readout resonator is excluded as it is only used to measure the ancilla at the end of each parity-check.  The ancilla has coherence times $T_1\approx35\mathrm{\mu s}$ and $T_2\approx13~\mathrm{\mu s}$, while the resonator has a single-photon Fock state relaxation time of $\tau_c\approx250\mu$s, and $T^c_2\approx 330\mathrm{\mu s}$ (Methods).  To perform high-fidelity single-shot measurements of the ancilla~\cite{Vijay:2011gl,Hatridge:2013ke}, we set the readout resonator to have a $1~\mathrm{MHz}$ bandwidth and use a nearly quantum-limited phase-preserving amplifier, the Josephson Parametric Converter (JPC)~\cite{Bergeal:2010iu}, as the first stage of amplification, which allows for a readout fidelity of $99.3\%$ and an error syndrome measurement fidelity of $98.5\%$ (Methods).  The duration of ancilla state readout totals $\sim 700\mathrm{ns}$, which includes integration times, cable latencies and feedback delays.  The total duration of each error syndrome measurement is thus just $1\mathrm{\mu s}$, or $\sim0.8\%$ of the average time between photon jumps for cat states of $\bar{n}= 2$.

\begin{figure*}[!ht]
\centering
\includegraphics[width=7.2in]{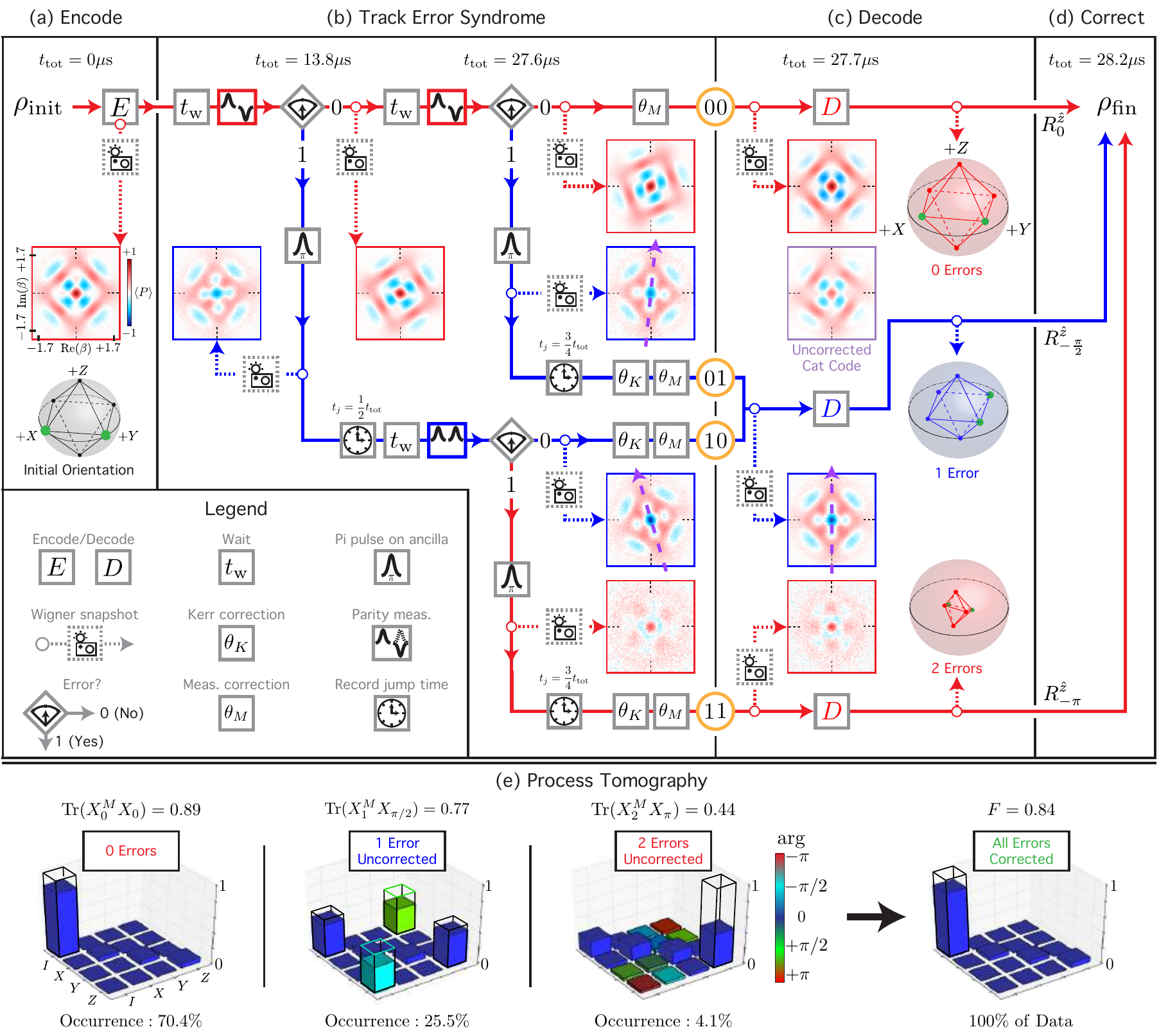}
\caption{\footnotesize
    \textbf{Example of a two-step quantum trajectory executed by the QEC state machine.}
        Two steps shown for a total monitoring time of $\sim 28\mathrm{\mu s}$.  \textbf{(a)} Six cardinal points on the Bloch sphere $\rho_\mathrm{init}$, initially encoded in the ancilla, are encoded onto a resonator state; green markers indicate the initial coordinate system orientation.  A ``Wigner [tomography] snapshot'' is shown for an initial state $\frac{1}{\sqrt{2}}(\ket{0}-\ket{1})$ mapped onto $\frac{1}{\sqrt{2}}(\ket{C^+_{\alpha}}-\ket{C^+_{i\alpha}})$; $\bar{n}_0=|\alpha|^2=3$.  \textbf{(b)}  The controller employs the quantum state machine for adaptive parity monitoring protocol with delays of $t_\mathrm{w}\sim13\mu$s between each measurement.  Parity measurement rectangles: pulse patterns that map even (odd: dotted line) parity onto ancilla $\ket{g}$ ($\ket{e}$); diamonds: the controller branches on the ancilla measurement result ($0\rightarrow$ no error, $\ket{g}$; $1\rightarrow$ error, $\ket{e}$); $\pi$ pulse rectangle: ancilla reset ($\ket{e}\rightarrow\ket{g}$).  The controller records the time at which an error occurs $t_j$ (clock icon); purple arrows emphasize the phase rotation due to the non-commutativity of $\hat{a}$ and the Kerr Hamiltonian: $\theta_K=Kt_j$, which leads to a phase difference between trajectories $10$ and $01$.  Deterministic rotations $\theta_M$ are due to cross-Kerr interactions between the readout and storage resonators during projective measurements of the ancilla.  The program returns one of the four possible records $\{00,01,10,11\}$ with probabilities $\{70.4\%,13.7\%,11.8\%,4.1\%\}$, in agreement with expected statistics.  The parity (origin of the Wigner tomogram) matches the controller's best estimate at any time (border color), and each tomogram matches the expected resonator state as seen in simulations.  \textbf{(c)} The feedback rotates all states back to the initial reference frame, where $01$ and $10$ are averaged together after the $\theta_K$ correction.  Ancilla state tomography after decoding (D) returns octahedrons similar to the one in (a), which exhibit the expected rotations of $\pi/2$ about $Z$ per error (green markers).  The controller decides in real-time whether to apply decoding pulses for even (red D), or odd (blue D) parity pulses.  \textbf{(d)}  The correction to obtain final state $\rho_\mathrm{fin}$ is made in software through coordinate system rotations by $0$ (0 errors), $-\pi/2$ (1 error), and $-\pi$ (2 errors), emphasizing that correcting errors, whether active or passive, amounts to just knowing how many errors occurred.  \textbf{(e)}  Process tomography results for $j=0$, $1$, and $2$ errors prior to correction.  Ideal $X_{j\pi/2}$ process matrices are shown in wire-outlined bars.  Experimental data for $X^M_j$ is shown in solid bars; the values are complex numbers with amplitude on the vertical axis and an argument specified by the bar color.  Amplitudes less than $0.01$ are not depicted.  Process tomography after correction is shown to the right of the arrow.}
\label{fig3}
\end{figure*}

Real-time feedback has been previously demonstrated to be a powerful tool for realizing control and performance enhancements of a quantum system~\cite{Sayrin:2011jx,Vijay:2012bv,Riste:2013if,Shulman:2014ua,Liu:2015}, but has typically been limited by latencies dominated by long inter-equipment communication times.  In our solution, the speed of real-time decision-making is possible because all of the feedback controller's components, including a DAC to generates pulses, an ADC to sample readout signals, and an FPGA loaded with in-house logic that orchestrates the timings and drives the experiment, are on a single piece of hardware.  It is thus a new, multi-purpose computer designed to execute programs for quantum control (Methods).  In this experiment, we translate the cat code into a program the controller understands, load it onto the FPGA, and press ``play."  A carefully choreographed stream of pulses on the ancilla, resonator, and readout resonator ensues to take a qubit through a series of error syndrome measurements and return a corrected qubit (Fig.~\ref{fig3}a-d), thus realizing a full QEC system.


Every repetition of the program (Fig.~\ref{fig3}) begins with an encoding (Methods) of one of the six cardinal points on a Bloch sphere in the logical basis states, enough to perform full process tomography of this QEC system~\cite{NielsenQI}.  The waiting time $t_\mathrm{w}$ between syndrome measurements is set to an optimal value that depends on $\bar{n}$; for cat states of $\bar{n}\approx 3$, $t_\mathrm{w}\approx 13\mathrm{\mu s}$ (more detailed discussion to follow; also, see Methods).  The program employs the quantum state machine for adaptive parity monitoring with the addition of a second application of real-time feedback: an instruction to record not just the occurrence, but the time at which an error occurs.  The necessity of this feature stems from the non-commutativity of the Kerr Hamiltonian $\frac{K}{2}\hat{a}^{\dag 2}\hat{a}^2$ and the dissipation operator $\hat{a}$, which leads to an extra undesired effect of random photon jumps: a phase shift of the resonator state in phase space that is proportional to $K$ and the time at which the jump occurs~\cite{Leghtas:2013ff} (Methods).

The program stores in memory a single measurement record of 0s (no error) and 1s (error) that specifies the monitoring history of the encoded state; Fig.~\ref{fig3} shows the four possibilities for two steps: $\{00$, $01$, $10$, $11\}$ with probabilities $\{70.4\%,13.7\%,11.8\%,4.1\%\}$.  The asymmetry in the occurrence of $01$ and $10$ is due to parity measurement infidelity and is well-modeled by a Bayesian analysis (Methods).  Wigner tomography is obtained by direct measurements of the resonator Wigner function in the continuous variables basis~\cite{Lutterbach:1997cn} and is equivalent to the density matrix of the encoded state~\cite{Cahill:1969vq}.  Conditioned on obtaining one of these four records, each tomogram provides a striking visual demonstration of the cat code in action.  Interference fringes, signatures of quantum coherence~\cite{book:haroche06}, continue to be sharp and extremal as the program proceeds in time, as compared with the case of performing no parity monitoring at all.  Indeed, at each point in the program, the tomograms agree well in parity contrast, phase, and amplitude as seen in simulations.  These levels of predictability highlight the advantages of this hardware-efficient scheme: knowing the Hamiltonian parameters together with a measurement fidelity of a single error syndrome is sufficient to encapsulate the evolution of an error-corrected logical qubit.

Prior to decoding, the feedback takes just $100$ns to align all frames of reference to the orientation defined at the outset, consolidating trajectories of equal error number yet different error timestamp into a single effective resonator state; e.g. $01$ and $10$ become a single ``1 error" state.  As even and odd parity basis states are mutually orthogonal, no one set of decoding pulses can map both back onto the ancilla (Methods), motivating another crucial application of feedback.  Based on the parity of the final state, the controller decides in real-time which set of decoding pulses to apply.  Fig.~\ref{fig3} shows qubit state tomography of the ancilla after decoding but before correction, conditioned on the number of errors.  The rotation of the six cardinal points by $\pi/2$ for one error and $\pi$ for two errors indicates that the cat code successfully maps photon loss errors in the resonator onto a unitary operation on the encoded qubit.

Upon completion of the program's execution, the knowledge of how many errors occurred is equivalent to having corrected the state.  Although aligning the Bloch spheres of all error trajectories to the same orientation requires a simple phase adjustment on the ancilla drive in the decoding sequence, here we instead choose to explicitly emphasize how the cat code maps errors to logical rotations.  The program thus returns the corrected qubit, now stored again in the ancilla, completing the full QEC cycle.


We benchmark the performance of our QEC system by performing process tomography of the QEC system, enforcing no prior knowledge of the initial quantum state and making no assumptions as to the prevalent decoherence mechanisms.  We use the chi matrix representation for a single qubit~\cite{NielsenQI}, in which state tomography of the output density matrix $\rho_\mathrm{fin}$ is used to calculate the measured chi matrix $X^M$.  This complex $4\times4$ matrix completely describes the action of our QEC system on the arbitrary input state $\rho_\mathrm{init}$.  The fidelity $F=\mathrm{Tr}(X^MX_0)$ is defined as the overlap of $X^M$ with $X_0$, the chi matrix for the identity operator $\hat{I}$, the ideal case in which a QEC system corrects a state perfectly.  For large enough cat state amplitudes, we can study each error case individually prior to correction while still maintaining the linearity of the process.  Shown in Fig.~\ref{fig3}e are the process matrices for the QEC program demonstrated in Fig.~\ref{fig3}a-d.  The form of $X^M_j$ ($X^M$ for $j=0$, $1$, and $2$ errors) matches the process matrix for ideal rotations about the $Z$ axis by $j\pi/2$, $X_{j\pi/2}$.  Signatures of developing incoherent mixture are evident from the non-zero values in all diagonal elements.  Trajectories with two errors have a lower process fidelity due to the low confidence that a record $11$ reflects the true error history of the encoded state.  They occur $4.1\%$ of the time after $28\mathrm{\mu s}$, however, and thus have little bearing on the final output.  With more syndrome measurements, the fidelity of the two-error case in fact increases toward a maximal value attained at a later time, which can be understood with an application of Bayes' rule to measurement statistics (Methods).  Upon correction, the resemblance of $X^M$ to $X_0$ reflects the simplicity of the cat code; by accounting for single photon jumps we witness no dominant processes besides $\hat{I}$ and emerging depolarization~\cite{NielsenQI} within the logical space.

\begin{figure*}[!ht]
\centering
\includegraphics[width=7.2in]{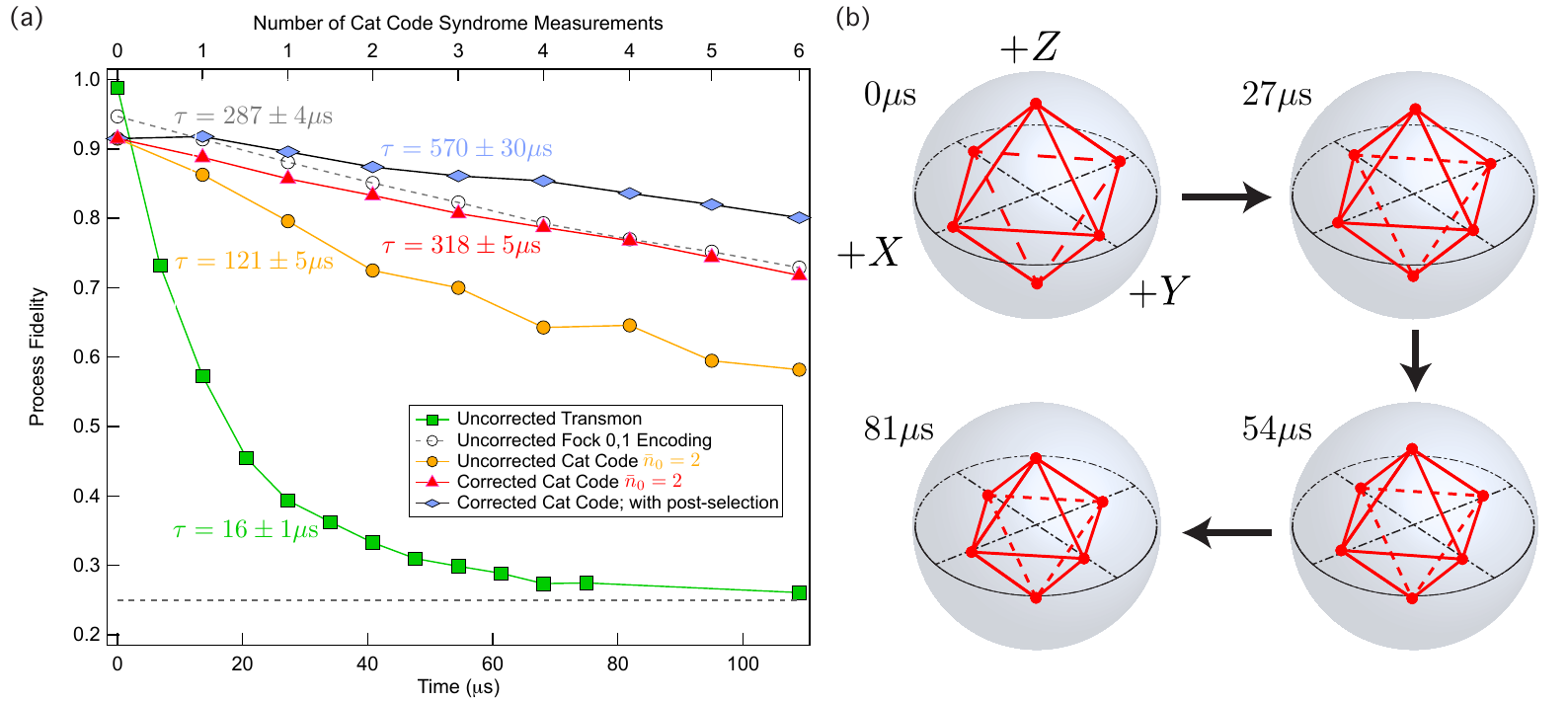}
\caption{\footnotesize
    \textbf{QEC program process tomography.}
        \textbf{(a)} Process fidelity decay of four possible qubit storage schemes in our system: just the ancilla transmon (green squares), an uncorrected superposition of cat states with $\bar{n}_0=2$ (orange circles), the cat code QEC system with $\bar{n}_0=2$ (red triangles), and a superposition of resonator Fock states $\ket{0}_f,\ket{1}_f$ (gray empty circles).  Each point in the cat code data was averaged 100,000 times, and each point in the qubit and Fock state encodings was averaged 50,000 times; error bars are smaller than marker sizes.  All curves are fit to single exponentials, $F(t)=0.25+Ae^{-t/\tau}$, except for the uncorrected cat code, which is fit to $F(t)=0.25+Ae^{-\bar{n}_0(1-e^{-t/\tau_c})}$; for short times this decay is well-approximated by a single exponential with $\tau\approx\tau_c/\bar{n}_0$.  Fluctuations beyond the error bounds can be explained by the effects of Kerr and are reproduced in simulation.  Uncertainties are given by errors in the fit.  When trajectories of low measurement confidence are excluded (purple diamonds), the qubit decay rate decreases by nearly a factor of two.  The top axis shows the number of syndrome measurements used to obtain each point in the cat code data.  \textbf{(b)}  Qubit state tomography after the correction step of the QEC system (corresponding to the cat code data in red triangles), shown for four different times.  The uniform shrinking of the Bloch sphere in time demonstrates that the residual loss of fidelity is well-represented by a depolarization channel.}
\label{fig4}
\end{figure*}

Our feedback controller can in fact perform an unlimited number of consecutive syndrome measurements while still maintaining all aforementioned feedback capabilities.  Moving to an initial encoding size of $\bar{n}_0=2$ to reduce the probability of errors in the codeword, we implement the cat code with up to six repetitions of the QEC program's monitoring step over $\sim 110\mathrm{\mu s}$ (Fig.~\ref{fig4}a).  Each repetition is separated by an optimized waiting time ranging from $t_\mathrm{w}\approx 15\mathrm{\mu s}$ to $t_\mathrm{w}\approx 25\mathrm{\mu s}$.  Employing the QEC system pays substantial dividends immediately.  Without any post-selection or renormalization, the cat code outperforms the uncorrected transmon with a time constant of exponential decay that is a factor of $\sim 20$ higher.  It also surpasses the decay of the uncorrected cat code by over a factor of $2$, in line with predictions (Methods).  Furthermore, it matches the performance of the simple encoding in the $\ket{0}_f$, $\ket{1}_f$ Fock states after $\sim80\mathrm{\mu s}$, and by decaying with a time constant that exceeds that of the Fock state by $10\%$, it reaches the break-even point.  The history of errors, however, also provides us with a valuable measure of confidence that the result of an error syndrome measurement faithfully reflects the actual trajectory of the encoded state.  Using this information to accept only high-confidence trajectories, which occur nearly $80\%$ of the time even after $100\mathrm{\mu s}$ (Methods), yields a decay constant of over half a millisecond.  Thus, if the measurement strategy allows for post-selection, our QEC system can lose as little as $\sim 10\%$ in process fidelity over more than $100\mathrm{\mu s}$ of correction.

We find the performance of the cat code to be most limited by the coherence properties of the ancilla, rather than other forms of resonator dephasing possible with continuous variables systems~\cite{Reagor:2015}.  As a result, the loss in fidelity stems primarily from a depolarization channel, shown in Fig.~\ref{fig4}b.  In contrast to the amplitude damping of the transmon or Fock states $\ket{0}_f,\ket{1}_f$, all cardinal points on the Bloch sphere shrink uniformly towards a completely mixed state at the origin.  Table 1 details an infidelity budget for the cat code, which lists the contributions to depolarization arising from the dominant avenues of code failure, applicable in fact to any QEC system.  Although the contribution of each source is small, the sources are many and add up quickly, bluntly encapsulating the challenges one faces in realizing fault-tolerant QEC.  Many of these effects can be mitigated by measuring more quickly and perhaps employing a quantum filter to retrieve a best estimate of the parity at any given time~\cite{Sun:2013ha}.  For example, infidelities due to missing the occurrence of an error; ancilla preparation and readout; and excursions out of the logical space due to Kerr, all approach zero as the measurement rate increases.  These thus impose no intrinsic limitation on cat code performance.

Crucially, however, errors due to the ancilla $T_1$ still persist regardless of measurement rate.  Due to its dispersive coupling to the resonator, a change in the energy of the ancilla at an unknown time imparts an unknown rotation to the resonator state in phase space (Methods); this is the forward propagation of an error.  Any such rotation commutes with the parity measurement, indicating that we learn nothing about these errors from our one syndrome measurement and that the cat code in its current implementation is not fault-tolerant~\cite{Gottesman:1998gt}.  Measuring the syndrome more frequently only increases the likelihood of ancilla-induced dephasing, necessitating an optimized measurement cadence to balance the risk of missing errors during $t_\mathrm{w}$ with the probability of error propagation during each syndrome interrogation.  Given that the cat code still performs at the break-even point even in the presence of all of these sources of loss, however, we are optimistic about the prospect of realizing a fault-tolerant QEC system.  Indeed, supplementing the cat code with a scheme that abates just a single source of loss, ancilla back-action, promises to allow increased error syndrome measurement rates and thus greater gains in lifetime.  This is the next task in our ongoing research.


Our results show that QEC can actually protect an unknown bit of quantum information, and extend its lifetime by active means. By employing the cat code as the foundation of a novel QEC system, we demonstrate the advantages of the hardware efficiency: the capacity of a single resonator to store a logical qubit; its intrinsically high coherence times; and the need to monitor just one error syndrome with just one ancilla.  Furthermore, we demonstrate the crucial role of real-time feedback, an addition to the experimental setup that vastly improves error correction performance and allows us to realize the cat code at the break-even point of QEC.  Future goals include combining the cat code with mechanisms to re-inflate cat state amplitudes and to equip the parity monitoring protocol to handle changes in ancilla energy, thereby addressing issues of non-fault-tolerance head-on.  We believe our results, however, already motivate the adaptation of QEC schemes to exploit the efficiencies of hardware platforms beyond the purview of traditional architectures, and the promise of cat states as components integral to future quantum computing applications.


\subsection*{Acknowledgements}
We thank K. Sliwa, A. Narla, and L. Sun for helpful discussions. This research was supported by the U.S. Army Research Office (W911NF-14-1-0011).  A.P. was supported by the National Science Foundation (NSF) (PHY-1309996).  S.M.G. acknowledges additional support from NSF DMR-1301798.  Facilities use was supported by the Yale Institute for Nanoscience and Quantum Engineering (YINQE), the Yale SEAS cleanroom, and the NSF (MRSECDMR 1119826).


\subsection*{Author Contributions}
A.P. and N.O. performed the experiment and analyzed the data.  N.O. designed and built the feedback architecture with help from Y.L. under the supervision of R.J.S. and M.H.D.  R.H. and P.R. developed the optimal control pulses.  M.M., Z.L., L.J., S.G., and B.V. provided theoretical support. R.H. and L.F. fabricated the transmon qubit.  R.J.S. supervised the project.  A.P., N.O., L.F., and R.J.S. wrote the manuscript with feedback from all authors.

\clearpage

\begin{figure}[!ht]
\centering
\includegraphics[width=3.3in]{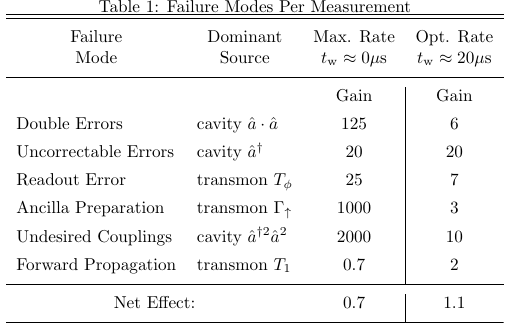}
\captionsetup{labelformat=empty}
\caption{\footnotesize
    \textbf{The Optimal Measurement Strategy.}
        The dominant modes of failure in the cat code are double errors ($\hat{a}$ followed by $\hat{a}$) between consecutive syndrome measurements separated by a time $t_\mathrm{w}$; possible errors that the cat code does not address, such as additions of a single photon ($\hat{a}^\dag$); a failed parity mapping resulting from ancilla dephasing ($T_\phi$); incorrect ancilla initialization prior to syndrome measurement resulting from unknown excitations ($\Gamma_\uparrow$) of the ancilla during $t_\mathrm{w}$; undesired couplings that result in dephasing due to Kerr ($\hat{a}^{\dag2}\hat{a}^{2}$); and finally ancilla decoherence that directly propagates to unrecoverable errors in the codeword, a result of ancilla decay or excitation ($T_1$).  This table shows the individual and total ramifications of these failure modes, where expected multiplicative gains in lifetime over a Fock state encoding ($290\mathrm{\mu s}$) are shown for two different measurement strategies: as quickly as possible (predicted), and the optimal monitoring time for an initial $\bar{n}=2$ (observed).  As detailed in the Methods, these channels of loss are not independent.  The performance of our QEC system is most limited by the non-fault tolerance of the error syndrome measurements.}
\label{tab1}
\end{figure}


\makeatletter
\renewcommand{\thefigure}{S\@arabic\c@figure}
\makeatother

\setcounter{figure}{0}
\setcounter{equation}{0}
\clearpage


\section{Methods}\label{setup}
\subsection{Setup}\label{Hamiltonian}
\begin{figure*}[!ht]
\centering
\includegraphics[width=4in]{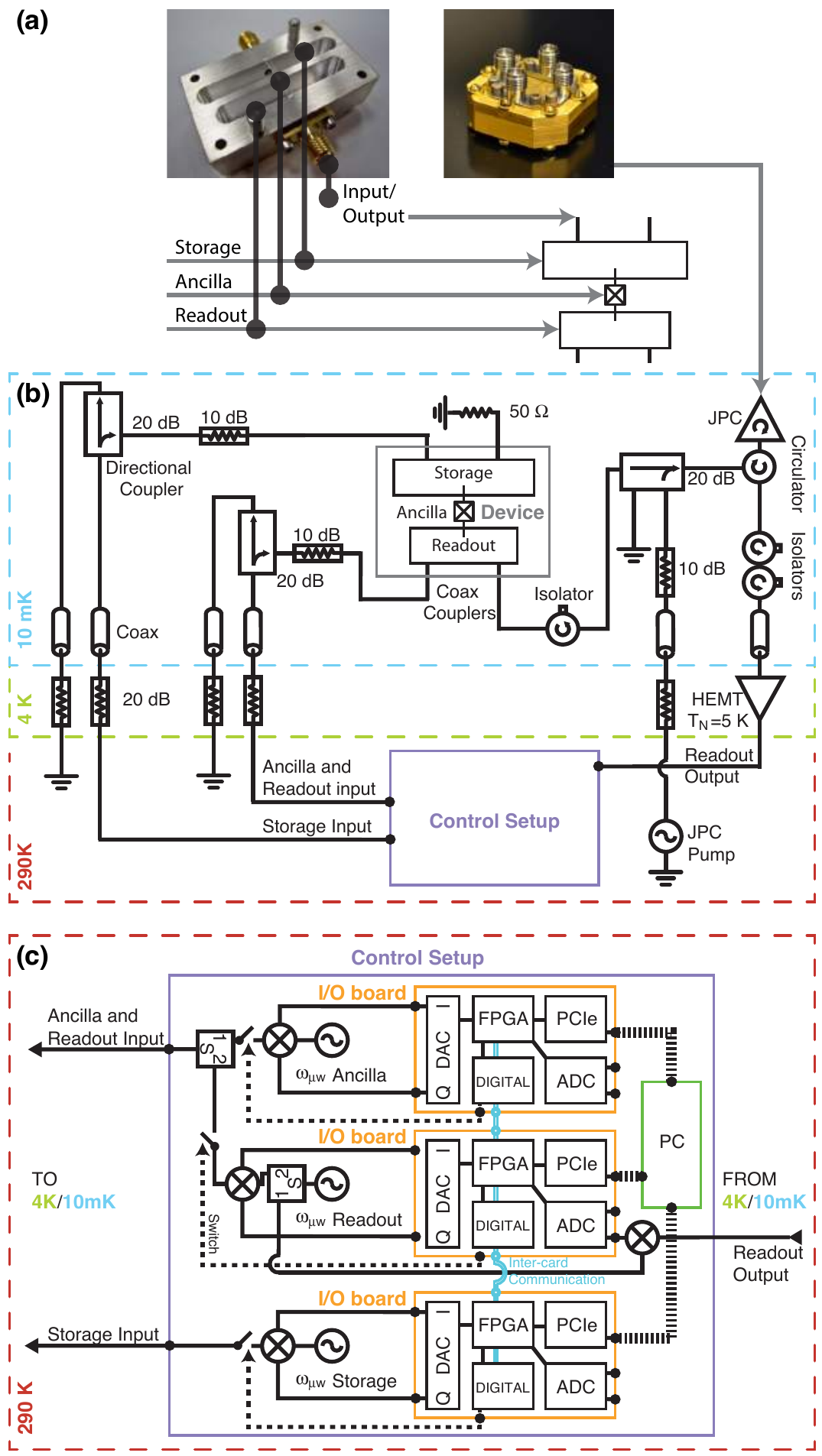}
\caption{\footnotesize
    \textbf{Setup.  (a)}  Shown on the left is a picture of one half of the experimental device, consisting of two resonators machined out of high purity 4N aluminum and a transmon coupled to both of them.  The resonators, one for storage and one for readout, are separated by a 2mm wall.  The transmon is patterned on a thin piece of sapphire, and is comprised of a Josephson junction with and two antennas of different lengths that extend from the junction into either cavity.  All system modes (storage, ancilla, readout) are realized and coupled to one another through these antennas.  Input and output couplers are used to excite each mode with microwave tones.  The image on the right depicts the JPC amplifier, used as the first stage of amplification in our measurement chain.  We operate it at a gain of $\sim 20dB$ and bandwidth of $\sim 5MHz$.  \textbf{(b)}  Schematic of the experimental setup, from room temperature at $290\mathrm{K}$ to the base temperature of the dilution unit ($\sim 10\mathrm{mK})$.  All microwave drives are produced and collected using our feedback setup, outline in purple.  \textbf{(c)} A detailed schematic of our feedback architecture, which is comprised of three Input/Output (I/O) boards.  The primary components of each board are: DACs to generate pulse envelopes, modulated at $50\mathrm{MHz}$ for the purposes of single-sideband modulation of local oscillators corresponding to each mode frequency; ADCs that digitize the incoming analog signal of the returning readout pulse, which is down-converted back to 50MHz after exiting the fridge; digital markers that operate microwave switches and are used for inter-board communication; an FPGA that synchronizes all components to realize the full experimental flow of pulse generation, readout signal integration, and real-time feedback and calculations; and finally the PCIe that allows the FPGA to send data to the PC. }
\label{fig:setup}
\end{figure*}

We perform our experiments in a cryogen-free dilution refrigerator that operates in a temperature range of $10-20$mK.  Our setup is identical to that described in~\cite{Vlastakis:2015tw}, aside from a $2\mathrm{mm}$ rather than $4\mathrm{mm}$ wall separating the storage and readout resonators (Fig.~S\ref{fig:setup}a), a Josephson Parametric Converter (JPC)~\cite{Bergeal:2010iu} replacing a Josephson Bifurcation Amplifier (JBA)~\cite{Vijay:2009ko} as the first stage of amplification (Fig.~S\ref{fig:setup}b), and a quantum control architecture replacing the hybrid FPGA-Tektronix AWG configuration (Fig.~S\ref{fig:setup}c; see sec.~\ref{quantum_control}).

\subsection{Hamiltonian Parameters}\label{hamiltonian}
Employing the rotating wave approximation (RWA), the Hamiltonian of our system in the strong dispersive regime~\cite{Schuster:2007ki} is approximated by:
\begin{align}
    \label{eq:hamiltonian}
\hat{H}/\hbar =  (\omega_{a} - \chi_{sa}\hat{a}_s^\dagger\hat{a}_s) \hat{b}^\dag\hat{b}+ (\omega_r -\chi_{ra}\hat{b}^\dag\hat{b})\hat{a}_r^\dag\hat{a}_r + (\omega_s-\chi_{sr}\hat{a}_r^\dag\hat{a}_r)\hat{a}_s^\dagger\hat{a}_s - \frac{K_s}{2}\hat{a}_s^{\dag 2}\hat{a}_s^2 - \frac{K_r}{2}\hat{a}_r^{\dag 2}\hat{a}_r^2 - \frac{K_a}{2}\hat{b}^{\dag 2}\hat{b}^2,
\end{align}
where the annihilation operator for the storage resonator from the main text is relabeled, $\hat{a}\rightarrow\hat{a}_s$, that of the readout resonator $\hat{a}_r$ is introduced, and that of the ancilla $\hat{b}$ replaces the projector $\ket{e}\bra{e}$.  In a similar manner, anharmonicities for all three components are defined as $K_s$, $K_r$, and $K_a$.  Higher-order corrections to dispersive shifts and anharmonicities were found to be negligible in this experiment.

The Hamiltonian is organized to highlight some of the dominant mechanisms in our system.  As described in the main text, the frequency of the ancilla is dispersively shifted by a frequency $\chi_{sa}$ for every photon in the storage resonator~\cite{Schuster:2007ki}; with this mechanism we realize an effective controlled-NOT (cNOT) gate on the ancilla depending on the photon number parity of the storage (see secs.~\ref{cohstate_basis},~\ref{smart_tracking}). Similarly, the frequency of the readout is shifted by $\chi_{ra}$ depending on the ancilla state; this is the typical dispersive readout mechanism~\cite{Blais:2004tl} that is now ubiquitous in superconducting cQED systems.  Finally, the frequency of the storage is shifted by $\chi_{sr}$ per photon in the readout, indicating that every time we measure the state of the ancilla we shift the phase of the storage by an amount that depends on the strength of the readout pulse.

The mode frequencies, dispersive shifts, and the ancilla anharmonicity are measured using the experimental methods described in~\cite{Vlastakis:2015tw}.  The Kerr interaction of the storage resonator $K_s$ is measured by monitoring the errors of a cat code logical state (see sec.~\ref{cat_code}) and finding the difference in phase between trajectories where errors are measured to occur at different times: $\Delta\theta=K_s\Delta t_j$ (see secs.~\ref{undesired_couplings},~\ref{record_error_time}).  In Fig.~3 of the main text, we show two Wigner functions~\cite{book:haroche06} for the case of a single parity jump: $01$ and $10$, where $0\equiv \mathrm{``no~error"}$ and $1\equiv \mathrm{``error."}$  We Fourier transform circular cuts at a fixed radius of these Wigner functions that show pronounced interference fringes to compare the phase of the oscillations for $01$ vs. $10$ and in so doing find $\Delta\theta$.  On average, photon jumps for $01$ versus $10$ are separated in time by $t_{M}$, where $t_M$ is the total time between syndrome measurements (see sec.~\ref{exptflow}); we thus find the average difference between jump time $\overline{\Delta t_j}=t_\mathrm{M}$ to find $K_s$.  Table~\ref{table_params} summarizes the measured (predicted~\cite{Nigg:2012jja}) parameters.

\subsection{Coherence Times and Measurement Fidelities}\label{fidelities}
Coherence times and thermal populations of all modes, obtained using the methods outlined in~\cite{Vlastakis:2015tw}, are summarized in Table~\ref{coherence}.  The single-shot ancilla measurement fidelity is $99.3\%$.  The parity measurement fidelity is $98.5\%$ for no photons in the storage resonator, $98.1\%$ for an average photon number $\bar{n}=2$, and $97.7\%$ for $\bar{n}=3$; these fidelities are also obtained using the methods in~\cite{Vlastakis:2015tw}.  The average rate of thermal excitation of the ancilla from its ground state $\ket{g}$ to excited state $\ket{e}$, $\Gamma_\uparrow$, is given by $\Gamma_\uparrow=1/(T_1n^a_{th})$, where $n^a_{th}=0.04$ is the steady-state ancilla excited state occupation.  A lower bound on the dephasing rate of the storage resonator, $\Gamma^s_{\phi}$, is given by $\Gamma^s_{\phi}=\Gamma_\uparrow$~\cite{Reagor:2015}, akin to the dephasing one expects of a qubit (e.g. transmon) coupled to a low-Q readout resonator with a finite thermal population~\cite{Sears:2012cm}.  The storage resonator coherence time $T^s_2$ is thus given by $(T^s_2)^{-1}=(2\tau_s)^{-1}+\Gamma^s_{\phi}$, where $\tau_s$ is the average lifetime of the single photon Fock state $\ket{1}$.  The coherence time $T^s_2$ is consistent with the observed time constant in the decay of the process fidelity of a qubit stored in Fock states $\ket{0}_f$, $\ket{1}_f$ (see sec.~\ref{depolarization}).  Henceforth we refer to the storage resonator as just ``the resonator."


\section{Building the Cat Code}
\subsection{Encoding in a Continuous-Variable System}

Universal quantum computation with continuous-variable encoding schemes is possible~\cite{Lloyd:2003kv,Gottesman:2001jb,Menicucci:2006ir} and can in fact offer advantages over those employing collections of discrete physical qubits.  Although a continuous-variable quantum computer formally has the same power as a discrete-variable system, there are regimes in which it could be more efficient.  For example, a single oscillator can in principle accommodate an unlimited amount of information, owing to the infinite size of its Hilbert space.  The hardware requirements can be more favorable as well, calling for linear elements and photon detectors in the optical platforms~\cite{Knill:2001is} or simple microwave resonators with long coherence times that are easy to assemble in superconducting cQED systems~\cite{Reagor:2015}.  Furthermore, the natural relation between continuous variables and communication can in principle simplify the transmission of quantum information for the purposes of teleportation~\cite{Lloyd:1998ub,Braunstein:1998uo}, cryptography~\cite{Ralph:1999ex}, and dense coding~\cite{Braunstein:2000cw}, to name a few.

There are trade-offs as well, however, which include challenges resulting from possible non-orthogonality of basis states in experimental realizations, the possibility of continuous excursions from a logical sub-space, and manipulating encoded states with high fidelity.  Nonetheless, several promising continuous variable QEC protocols exist~\cite{Braunstein:1998cd,LundRalph:2008,Ralph:2011ct,Leghtas:2013ff,Mirrahimi:2014js,Michael:2016}, and substantial progress has been made in demonstrating that continuous-variable encodings can be a powerful resource for the storage, control, and measurement of quantum information~\cite{Pittman:2005du,Deleglise:2008gt, Aoki:2009ew,Hofheinz:2009ba,Jensen:2011ba,Vlastakis:2013ju,Sun:2013ha,Leghtas:2015uf,Heeres:2015,Vlastakis:2015tw}.  Moreover, with recent progress demonstrating cQED architectures that offer a natural path toward scalability~\cite{Brecht:2015_1,Brecht:2015_2,Minev:2015}, we see continuous variables systems as a promising platform for realizing a practical quantum computer. 

\subsection{Single Photon Loss as the Dominant Error}
We consider the system introduced in section~\ref{setup}.  The time evolution of the density matrix $\rho_s$ of a resonator field, which has some equilibrium thermal photon population $n^s_{th}$, a rate $\kappa_s$ associated with single photon creation and annihilation operators $\hat{a}_s^\dag$ and $\hat{a}_s$, is well-modeled by the following Lindblad operators in the master equation formalism~\cite{book:haroche06}:

\begin{align}
L_-&=\sqrt{\kappa_s (1+n^s_{th})}\hat{a}_s\\
L_+&=\sqrt{\kappa_s n^s_{th}}\hat{a}_s^\dag,
\end{align}
where the master equation reads:

\begin{align}
\frac{d\rho_s}{dt}=-i\omega_s[\hat{a}_s^\dag\hat{a}_s,\rho_s]-\frac{\kappa_s(1+n^s_{th})}{2}(\hat{a}_s^\dag\hat{a}_s\rho_s+\rho_s \hat{a}_s^\dag\hat{a}_s-2\hat{a}_s\rho_s\hat{a}_s^\dag)-\frac{\kappa_s n^s_{th}}{2}(\hat{a}_s\hat{a}_s^\dag\rho_s+\rho_s \hat{a}_s\hat{a}_s^\dag-2\hat{a}_s^\dag\rho_s a)
\end{align}

One can show that such a formulation returns the expected prediction that on average the occupation of the resonator mode $\bar{n}=\mathrm{Tr}[\rho_s\hat{a}_s^\dag\hat{a}_s]$ simply decays exponentially in time to thermal equilibrium with a characteristic time constant $\tau_s=1/\kappa_s$:

\begin{align}
\bar{n}(t)=n_0e^{-t/\tau_s}+n^s_{th}(1-e^{-t/\tau_s})\label{energy_decay}
\end{align}

Treating $n^s_{th}$ as negligible for the remainder of this discussion, one may conclude that the evolution of the density matrix $\rho_s$ can be simply described by the stochastic application of the lowering operator $\hat{a}_s$ on the resonator field.  This assertion, that there are essentially just two dominant processes within the resonator, the application of $e^{-\frac{\kappa_s}{2}\hat{a}^\dag_s\hat{a}_s\Delta t}$ (for time steps $\Delta t$) and the stochastic application of $\hat{a}_s$, is a powerful incentive to consider storing quantum information in a superconducting resonator rather than a two level system, such as a transmon, which is susceptible to both amplitude ($\sigma_-$) and phase damping ($\sigma_z$).  Here $\sigma_x,\sigma_y,\sigma_z$ are the standard Pauli operators and $\sigma_\pm=\sigma_x\mp i\sigma_y$.  Indeed, it has been experimentally demonstrated~\cite{Reagor:2015} that 3D superconducting resonators have no currently measurable source of inherent dephasing arising from a Lindblad operator of the form $L_\phi=\sqrt{\kappa_\phi}\hat{a}_s^\dag\hat{a}_s$, which corresponds to a frequency jitter, or higher order photon loss mechanisms such as $L_{2ph}=\sqrt{\kappa_{2ph}}\hat{a}_s^2$~\cite{Sun:2013ha}.  In practice, some resonator dephasing is induced by its dispersive coupling to occupation fluctuations of other modes in the system, particularly to that of the transmon qubit used as the ancilla in the error correction; see sec.~\ref{losses}.  The governing goal is to thus construct a code that can track the occurrence of single photon jumps, as this would correct for the dominant error channel in the system.  In implementing a QEC system to realize this goal, we look to translate the discretized energy dissipation of the resonator field into a unitary operation on an encoded state, the occurrence of which can be deduced from an appropriate error syndrome measurement.  

\subsection{A Basis of Coherent States}\label{cohstate_basis}
Coherent states $\ket{\alpha}$ are an attractive option for a logical encoding scheme as they are eigenstates of $\hat{a}_s$, where $\ket{\alpha}=e^{-|\alpha|^2/2}\sum_{n=0}^{\infty}\frac{\alpha^n}{\sqrt{n!}}\ket{n}$ for a complex amplitude $\alpha$, and $\hat{a}_s\ket{\alpha}=\alpha\ket{\alpha}$.  This feature suggests that one could try encoding a qubit in a superposition of coherent states: $\ket{\psi}=c_0\ket{0}+c_1\ket{1}\rightarrow c_0\ket{\alpha}+c_1\ket{-\alpha}$.  As the overlap between two coherent states falls off exponentially in the difference of their amplitudes~\cite{book:haroche06}, choosing an $|\alpha|^2=\bar{n}\gtrsim 1.5$ would be sufficient for basis states $\ket{\alpha}$ and $\ket{-\alpha}$ to be almost completely orthogonal, with an overlap of $\sim0.2\%$.  The penalty we pay is that the rate of photon jumps $\gamma$ scales with the mean photon number $\bar{n}=|\alpha|^2$.  When $c_0=c_1=\pm1/\sqrt{2}$, $\ket{\psi}\rightarrow1/\sqrt{2}(\ket{\alpha}\pm\ket{-\alpha})$, the logical encoding is an equal superposition of coherent states that we refer to in this work as ``2-cat" states, which are eigenstates of even ($+$) or odd ($-$) photon number parity $\hat{P}=e^{i\pi\hat{a}_s^\dag\hat{a}_s}$:
\begin{align}
\ket{C^{+}_{\alpha}}=&\frac{1}{\sqrt{2}}(\ket{\alpha}+\ket{-\alpha})=e^{-|\alpha|^2/2}\sum_{n=0}^{\infty}\frac{\alpha^{2n}}{\sqrt{2}\sqrt{(2n)}!}\ket{2n}\\
\ket{C^{-}_{\alpha}}=&\frac{1}{\sqrt{2}}(\ket{\alpha}-\ket{-\alpha})=e^{-|\alpha|^2/2}\sum_{n=0}^{\infty}\frac{\alpha^{2n+1}}{\sqrt{2}\sqrt{(2n+1)}!}\ket{2n+1},
\end{align}
where $\bra{C^{\pm}_{\alpha}}\hat{P}\ket{C^{\pm}_{\alpha}}=\pm1$.  In fact, this parity is a quantity that can be measured in our system with a simple Ramsey-style pulse sequence~\cite{Bertet:2002df,Haroche:2007uc} (see sec.~\ref{smart_tracking}).  Measurements of photon parity have already been used to track the loss of single photons in real-time with high fidelity, demonstrating their efficiency in extracting an error syndrome from a resonator field~\cite{Sun:2013ha}.  The problem with this encoding, however, is that aside from the special case of a 2-cat, for arbitrary $c_0$ and $c_1$ there is no parity symmetry and no other measurable symmetry property that would indicate the loss of a photon.

\subsection{The Cat Code: A Basis of Cat States}\label{cat_code}
\begin{figure*}[!ht]
\centering
\includegraphics[width=7in]{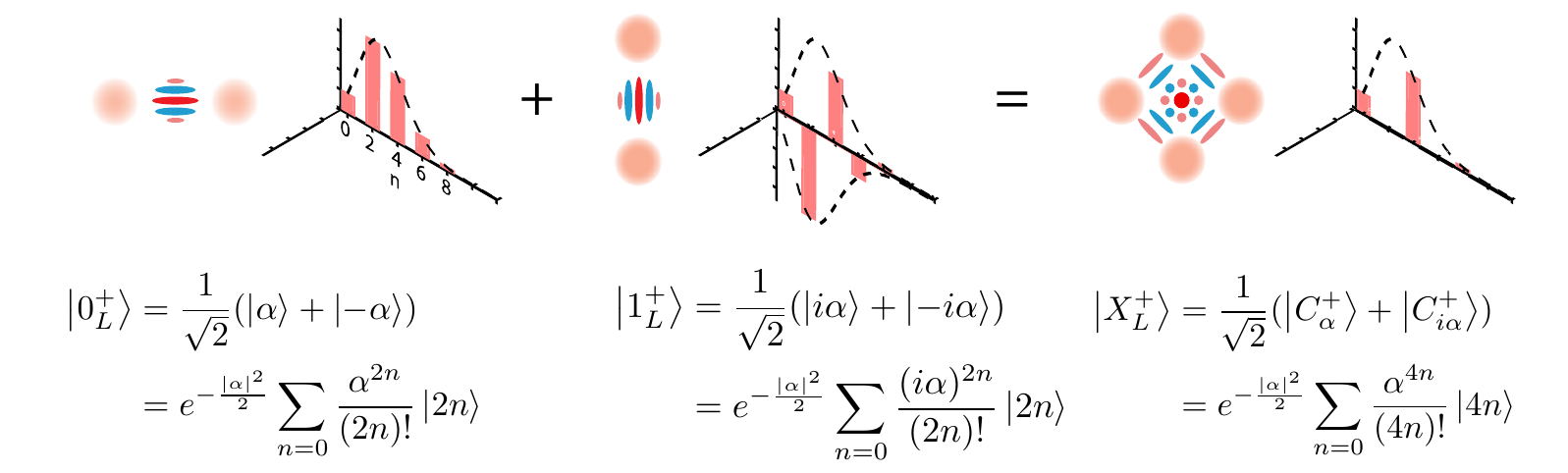}
\caption{\footnotesize
    \textbf{Cat states in the Fock basis.}  The logical basis states $\ket{C_\alpha^+}, \ket{C_{i\alpha}^+}$ are termed ``2-cats" because they are superpositions of two coherent states.  They can be expanded in the Fock basis to elucidate certain cat code features.  In particular, although $\ket{C_\alpha^+}$ and $\ket{C_{i\alpha}^+}$ are both eigenstates of even parity with non-zero amplitudes only for even Fock state components $\ket{2n}$ ($n$ an integer), the phase of $\ket{1_L^+}$ results in an extra minus sign on $\ket{2+4n}$ in its Fock state expansion.  When one has the equal superposition of $\ket{X_L^+}=\ket{C_\alpha^+}+\ket{C_{i\alpha}^+}$ (normalization omitted), consequent destructive interference results in a ``4-cat" state that has non-zero components every fourth Fock state, $\ket{4n}$.  This is still an eigenstate of even parity, but unlike the basis states individually, which are eigenstates of $\hat{a}_s^2$, $\ket{X_L^+}$ and indeed any arbitrary superposition of $\ket{C_\alpha^+}$ and $\ket{C_{i\alpha}^+}$ are eigenstates of $\hat{a}_s^4$.  In the simple example here, one can see that by applying $\hat{a}_s$ four time on $\ket{X_L^+}$ in the Fock basis, one returns to a $\ket{4n}$ expansion.  Hence the modulo-four behavior of the cat code.}
\label{fig:Circuit_a}
\end{figure*}

This necessitates us to move on to the cat code~\cite{Leghtas:2013ff,Mirrahimi:2014js}, wherein we access a larger part of the resonator's Hilbert space in order to accommodate an encoding scheme where the individual basis states are themselves 2-cats along the real and imaginary axes in phase space: $\ket{C^\pm_{\alpha}}\equiv\mathcal{N^\pm_\alpha}(\ket{\alpha}\pm\ket{-\alpha})$ and $\ket{C^\pm_{i\alpha}}\equiv\mathcal{N^\pm_\alpha}(\ket{i\alpha}\pm\ket{-i\alpha})$, where $\mathcal{N^\pm_\alpha}\rightarrow1/\sqrt{2}$ for large $\alpha$.  To prevent appreciable basis overlap (see sec.~\ref{orthogonality}), one must now have $|\alpha|^2=\bar{n}\gtrsim 2$ and thus $\gamma\gtrsim2\kappa_s$.  Such a modification allows us to encode a quantum state with arbitrary $c_0$ and $c_1$ in an eigenstate of parity.  Figure S\ref{fig:Circuit_a} shows an explicit example where $c_0=c_1=1/\sqrt{2}$.  This in turn allows changes in parity to serve as the error syndrome for the loss of a photon in a logical qubit:
\begin{align}
\ket{\psi_{\mathrm{init}}}=&c_0\ket{0}+c_1\ket{1}\rightarrow c_0\ket{C^+_{\alpha}}+c_1\ket{C^+_{i\alpha}}\label{cat_code}\\
\hat{a}_s(c_0\ket{C^+_{\alpha}}+c_1\ket{C^+_{i\alpha}})=&\mathcal{N^-_\alpha}[c_0(\ket{\alpha}-\ket{-\alpha})+ic_1(\ket{i\alpha}-\ket{-i\alpha})]\nonumber \\
=&c_0\ket{C^-_{\alpha}}+ic_1\ket{C^-_{i\alpha}}\nonumber\\
\hat{a}_s(c_0\ket{C^-_{\alpha}}+ic_1\ket{C^-_{i\alpha}})=&\mathcal{N^+_\alpha}[c_0(\ket{\alpha}+\ket{-\alpha})-c_1(\ket{i\alpha}+\ket{-i\alpha})]\nonumber \\
=&c_0\ket{C^+_{\alpha}}-c_1\ket{C^+_{i\alpha}}\nonumber
\end{align}

Equation~\ref{cat_code} shows that the cat code maps a photon loss error in the resonator field into a rotation by $\pi/2$ about the logical $Z$ axis, as seen from the factor of $i$ that comes out in front of $c_1$.  With each successive error the parity of the basis states cycles between even and odd, while the encoded information continues to rotate about $Z$ in increments of $\pi/2$, returning to the initial state after four errors (Fig.~1, main text).  Left uncorrected, an encoded state devolves into a mixture of cat states, and equivalently, a classical mixture of coherent states.  By performing single-shot parity measurements, however, we repeatedly update our knowledge as to the parity of the state and infer the occurrence of an error when the parity changes~\cite{Sun:2013ha}.  We thereby follow the stochastic evolution of the resonator state through the cat code loop, maintaining the coherence of the qubit despite errors.

In addition to the stochastic loss of single photons, as shown in eq.~\ref{energy_decay}, the energy of the resonator field decays deterministically to vacuum at a rate $\kappa_s$.  Therefore, in the experimental implementation of the cat code, whenever we decode a resonator state back onto the ancilla transmon (see sec.~\ref{oc_pulses}) we must always take into account the decay of the cat state amplitude after a finite time of monitoring $t$:
\begin{align}
\ket{C^{\pm}_{\alpha}}\rightarrow&\ket{C^{\pm}_{\alpha e^{-\kappa_s t/2}}}=\mathcal{N^\pm_\alpha}_{(t)}(\ket{\alpha(t)}\pm\ket{-\alpha(t)})\\
\mathcal{N^\pm_\alpha}_{(t)}&=\frac{1}{\sqrt{2(1\pm e^{-2|\alpha(t)|^2})}}\nonumber\\
\alpha(t)&=\alpha e^{-\kappa_s t/2}\nonumber,
\end{align}
where again $\mathcal{N^\pm_\alpha}_{(t)}\rightarrow1/\sqrt{2}$ for large $\alpha(t)$.  Of course, without any intervention, any state stored in the resonator eventually decays to vacuum, thereby erasing any stored information.  This effect is not irreversible, however, as energy can be periodically repumped into the resonator using unitary gates.  Moreover, it has recently been demonstrated that cat state amplitudes can be preserved indefinitely through the application of off-resonant pumps at carefully chosen frequencies~\cite{Leghtas:2015uf}.  We have not yet implemented such a system in this work.

Perhaps surprisingly, the loss of a photon has no effect on the amplitude of a coherent state, as can be seen from a simple argument in~\cite{book:haroche06}, section 4.4.4.  The authors explain that losing single photons simply updates one's knowledge that there must have been more photons in the resonator to begin with.  By virtue of this curious property of coherent states, the amplitude of our logical states is independent of the number of photon jumps we detect.

\section{Encoding and Decoding Pulses}\label{grape}
\subsection{Optimal Control Pulses}\label{oc_pulses}
We employ optimal control pulses to encode and decode logical states in the resonator based on the Gradient Ascent Pulse Engineering (GRAPE) algorithm originally developed for pulse sequences in NMR spectroscopy~\cite{Khaneja:2005vd,deFouquieres:2011wm}.  This algorithm is designed to numerically find a set of pulses that most accurately realizes a unitary operation or state transfer, taking an initial state $\ket{\psi(t=0)}$ to a final state $\ket{\psi(T)}$.  We define the fidelity of the simulated state transfer $F_{oc}$ to be:
\begin{align}
F_{oc}=\frac{1}{K^2}|\sum_k^K\left<\psi_k(T)|\psi_k^\mathrm{tar}\right>|^2,
\end{align}
for a target state $\psi^\mathrm{tar}$, where $K$ is the total number of state transfers we wish to realize.  In order to model the physical limits in output amplitude imposed by our electronics hardware, we add an amplitude constraint of the form $\lambda\sum_{t=1}^T e^{(a_t/h)^2}$, where $\lambda$ is an overall scaling; $a_t$ is the amplitude at each point in time of the pulse (discretized into $1\mathrm{ns}$ steps); and $h$ is an amplitude threshold, which we choose to be slightly below the maximum output amplitude our waveform generators can produce.  This penalty term turns on sharply when $a_t$ reaches $h$.  The scaling factor $\lambda$ is a proportionality constant that makes the total penalty much smaller than $1$ for pulses that have all amplitudes below $h$.  We also include a derivative penalty to give preference to smoother pulses, similarly defined as $\lambda_d\sum_{t=1}^T e^{(a_t-a_{t-1})^2/h_d^2}$.  With such a term included, the simulation favors pulses with changes smaller than $h_d$ between neighboring control points.  The criterion we enforce is that $F_{oc}$ must exceed a value typically set to be $98\%$, although this constraint is relaxed when the overlap of basis states becomes non-negligible (see sec.~\ref{orthogonality}).  

In our implementation, we use the Hamiltonian defined in sec.~\ref{Hamiltonian} (excluding any terms involving the readout resonator) along with driving terms on the ancilla and resonator of the form $\varepsilon_a(t)\hat{b}^\dag+\varepsilon^*_a(t)\hat{b}$ and $\varepsilon_s(t)\hat{a}^\dag_s+\varepsilon^*_s(t)\hat{a}_s$.  Temporal envelopes $\varepsilon_a(t)$ and $\varepsilon_s(t)$, which specify $\hat{U}(t)$, are discretized into $1\mathrm{ns}$ pieces.  It is the shape and amplitude of these envelopes that we wish to numerically optimize in order to realize the desired state transfer.  More explicitly, for the encoding pulses we wish to find a single unitary $\hat{U}_\mathrm{tar}$ such that for all $c_0$ and $c_1$ we have:
\begin{align}
\hat{U}_\mathrm{tar}(c_0\ket{g}+c_1\ket{e})\ket{0}\rightarrow&\ket{g}(c_0\ket{C_\alpha^+}+c_1\ket{C_{i\alpha}^+})
\end{align}
This unitary takes a quantum bit initially stored in the ancilla (with resonator in the vacuum) to a superposition of cat states in the resonator with the same amplitudes $c_0$ and $c_1$ (returning the ancilla to $\ket{g}$).  Figure S\ref{fig:GRAPE}a shows a set of such encoding pulses on the ancilla and resonator.

The decoding pulses simply reverse this mapping.  A single decoding pulse, however, cannot take two resonator states of different parity back to the same state of the ancilla since a unitary operation cannot bring two orthogonal states to a single state.  For a given monitoring time we thus prepare two sets of decoding pulses, one for even states and one for odd states, and apply the correct one depending on the final parity of the state (see sec.~\ref{adaptive_decoding}).  After monitoring errors for arbitrary lengths of time the decoding pulse must also take into account the substantial deviations of the coefficients in the Fock state expansion of the basis states from their original Poisson values.  We thus use these pulses to remove any distortions in the resonator state due to the deterministic action of the Kerr Hamiltonian and deterministic amplitude damping due to energy decay.  For example, after a monitoring time $T$ the decoding pulse for even parity realizes the following state transfer:
\begin{align}
\ket{g}\{e^{-i\frac{K_s}{2}\hat{a}_s^{\dag 2}\hat{a}_s^2T}\mathcal{N^+_\alpha}_{(T)}[c_0(\ket{\alpha(T)}+\ket{-\alpha(T)})+c_1(\ket{i\alpha(T)}+\ket{-i\alpha(T)})]\}\rightarrow&(c_0\ket{g}+c_1\ket{e})\ket{0}
\end{align}
For the data presented in Fig.~4 of the main text we therefore require a different pair of decoding pulses for each of the nine data points in the plot.  The feedback controller stores these in memory and applies them at the appropriate time.

\subsection{Implementation}\label{grape_implementation}
\begin{figure*}[!ht]
\centering
\includegraphics[width=7.2in]{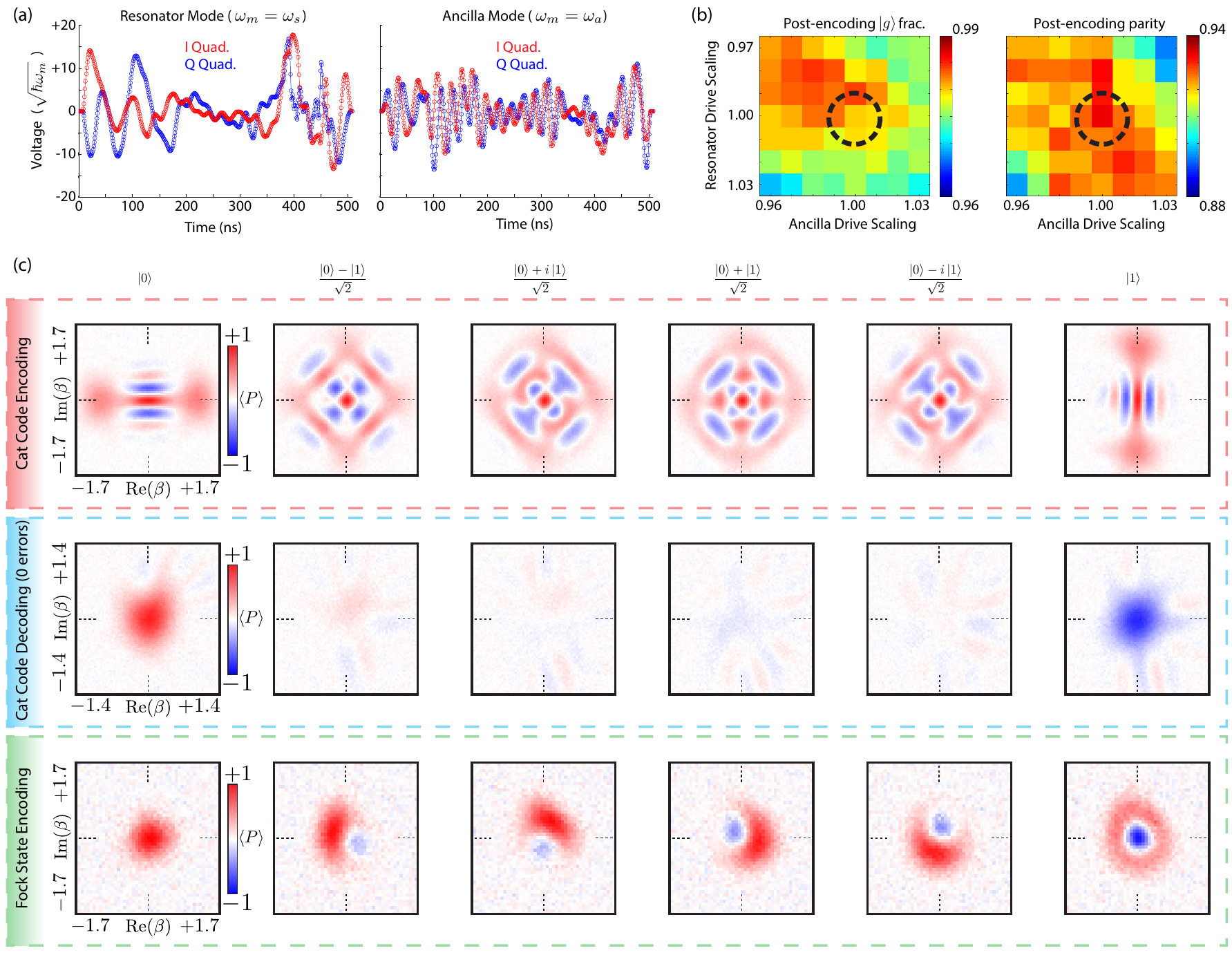}
\caption{\footnotesize
    \textbf{Optimal control pulses.  (a)}  Example encoding pulses for cat states of initial size $\bar{n}_0=3$.  The pulses are played by the controller at the same time, minus a $2\mathrm{ns}$ offset to account different lengths of line going down to the resonator versus the ancilla.  The y-axis is given in units of the voltage that needs to be applied on a particular mode (frequency $\omega_m$) for the same amount of time to insert one quantum of energy.  For the resonator, this value is determined by finding the voltage needed to create a coherent state of amplitude $|\alpha|=1$ for a square pulse of duration $508\mathrm{ns}$; likewise, with the ancilla it's the voltage needed to perform a $\pi$ pulse in $508\mathrm{ns}$.  Mixer quadratures $I$ and $Q$ are shown in red and blue, respectively.  \textbf{(b)}  The numerical optimization has no knowledge of our experimental imperfections, such as frequency-dependent reflections in our microwave lines and components.  We calibrate a scaling factor on the amplitudes for both the resonator and ancilla drives by performing a 2D voltage sweep to see at which scalings we find both the maximum percentage of ancilla in its ground state $\ket{g}$ (ideally $100\%$) and the maximum parity of the resonator state (ideally $+1$).  Shown here are images where we have already found what looks to be the optimal drive scaling, outlined in the black dotted circle.  \textbf{(c)}  Using joint Wigner tomography~\cite{Vlastakis:2015tw}, we plot $W_z(\beta)=\braket{\hat{\sigma}_z\hat{P}(\beta)}$ to we demonstrate the capability of a single pair of pulses to encode an arbitrary vector on the qubit Bloch sphere.  The first (third) row shows all six cardinal points encoded in the cat code basis (Fock states $\ket{0}_f,\ket{1}_f$).  The only difference in the pulse sequence is the initial qubit preparation pulse.  The second row shows the action of cat state decoding pulses that immediately follow the encoding shown in the first panel.  The first and sixth tomogram demonstrate that we map the 2-cat along the real axis back to vacuum with the ancilla in $\ket{g}$ ($W(\beta)=\braket{\hat{P}(\beta)}$) and a 2-cat along the imaginary axis back to vacuum with the ancilla in $\ket{e}$ ($W(\beta)=-\braket{\hat{P}(\beta)}$).  The remaining four joint tomograms should ideally have no visible features for perfect encoding and decoding since $\braket{\hat{\sigma}_z}=0$.  Experimental imperfections and primarily ancilla decoherence, however, result in residual resonator-ancilla entanglement at the end of the sequence and thus slightly visible interference features in the Wigner functions.}
\label{fig:GRAPE}
\end{figure*}

Crucial to the success of finding an optimal control pulse with high fidelity is an accurate knowledge of the dominant Hamiltonian parameters.  Furthermore, careful microwave hygiene at all points in the experimental chain is necessary to prevent undesired reflections and dispersions that can distort the pulse as it goes from room temperature to the setup inside the dilution refrigerator.  Indeed, the fluctuations in the process fidelity decay curves can be mostly attributed to the imperfect application of the decodings.  Figure S\ref{fig:GRAPE}b demonstrates a calibration sequence we use to tune the amplitudes on individual ancilla and resonator drives.  Ideally, the encoding pulse returns the ancilla to the ground state and creates a cat state with mean parity $\left<\hat{P}\right>=+1$.  In practice, both the parity and the final ground state occupation are slightly lower than their ideal values, and are sensitive to errors in pulse power.  By sweeping the relative powers for both the ancilla and resonator drives, we find the maximum ground state occupation and parity value to occur at roughly equal scalings.  The ultimate figure of merit comes down to the process fidelity of encoding and decoding with no time delay in between.  As seen in Fig.~4a of the main text, the drop in process fidelity at $t=0$ is $\sim8\%$; assuming both pulses are similarly afflicted by all sources of error, this implies that each one contributes $\sim4\%$ infidelity, in line with expectations given that the ancilla coherence time $T_2=12\mathrm{\mu s}$ and the duration of both pulses combined is $\sim 1\mathrm{\mu s}$.  This fixed cost of entrance, to realize the augmented codeword, afflicts any QEC implementation to some degree.

The full Wigner tomography shown in Fig.~S\ref{fig:GRAPE}c, defined as $W(\alpha) = \frac{2}{\pi}\braket{\hat{D}_\alpha \hat{P}\hat{D}_\alpha^\dagger}$ for a resonator displacement operator $\hat{D}_\alpha $~\cite{book:haroche06}, illustrates visually how we do indeed have the capability to encode any arbitrary state with the same pulses, where the only difference lies in the preparation of the initial qubit.  Shown in the first two rows are examples of encoding and decoding all six cardinal points on the Bloch sphere with cat states of average photon number $\bar{n}=3$.  In the third row we show the result of encoding a qubit into the Fock states $\ket{0}_f$, $\ket{1}_f$. We note that our numerical optimization can find encoding and decoding pulses for $\ket{0}_f$, $\ket{1}_f$ that are about half the length of those used for the cat code, owing to the smaller fraction of the Hilbert space that needs to be accessed.  The shorter pulse length reduces the time the ancilla is entangled with the resonator and thus improves the process fidelity, accounting for the initial offset between the Fock state and cat code encodings.  These results demonstrate that beyond offering the convenience of fast encoding and decoding that take into account distortions due to higher order Hamiltonian parameters, optimal control pulses provide a striking example of the levels of control possible with continuous-variable systems in a cQED framework.

\section{Sources of Loss}\label{losses}
In this section we expand upon the calculations that produce the predicted gains in lifetime of the cat code over the Fock state encoding listed in Table 1 of the main text.  Each dominant source of decoherence, whether arising from errors in the codeword or in the syndrome interrogation, contributes to the probability that the cat code will fail in a single round of error correction.  When isolated in a hypothetical situation as the only source of loss in the system, it can be quantified with simple estimates based on coherence times, thermal populations, and measurement fidelities.  We stress, however, that the sources of loss detailed in the sections below do not act independently when considering the experimental reality.  Indeed, simply adding all of the rates in parallel leads to an underestimate of the cat code performance.  In section~\ref{optimal_rate} we analyze the system as a whole and show that we can analytically predict the data for the cat code decay shown in Fig.~4a of the main text.

In section~\ref{forward_prop} we arrive at a result that has substantial bearing on our understanding of fault tolerance regarding this implementation of the cat code.  We find that forward propagation of errors into the codeword due to the ancilla $T_1$ constitutes the single dominant source of non-fault tolerance in the current implementation of the cat code.  Indeed, changes in ancilla energy at unknown times result in unknown excursions of our encoded states out of their logical space.  This is the central limiting feature of our QEC system that necessitates a lower measurement cadence, one that effectively balances the probability of dephasing due to ancilla decay with the probability of dephasing due to all other sources of error combined (quantified in sec.~\ref{optimal_rate}).  We show that mitigating this form of ancilla decoherence promises to offer considerable improvements in cat code performance.

\subsection{Double-Errors}
The cat code is a first-order code, which means that the error syndrome we employ cannot detect the occurrence of multiple errors between two consecutive measurements.  The probability of such events, $p_{2\varepsilon}$, can be calculated from the Poisson distribution: 
\begin{align}\label{form:double_jumps}
p_{2\varepsilon}(t_M)=\frac{(\bar{n}\kappa_s t_M)^2}{2}e^{-\bar{n}\kappa_s t_M},
\end{align}
where we take the approximation that $\bar{n}$ is constant throughout the small time interval $t_M$.  In this expression, $t_M\approx t_\mathrm{w}+1\mathrm{\mu s}$ is the total measurement time; $t_\mathrm{w}$ is the time delay between the end of one syndrome measurement and the beginning of the next; and the parity mapping together with ancilla readout totals $\sim1\mathrm{\mu s}$ (see sec.~\ref{exptflow} for exact timings).  The average time between photon jumps is given by $1/\bar{n}\kappa_s$.  A simple calculation using eq.~\ref{form:double_jumps} for measurement intervals $t_M\approx1\mathrm{\mu s}$ and $t_M\approx21\mathrm{\mu s}$ returns the predicted gains in the process fidelity lifetime over a Fock state encoding, $\tau_{f01}$, one would expect to see if missing such events were the only source of error.  Defining a gain $G_{2\varepsilon}(t_M)=t_M/(p_{2\varepsilon}\tau_{f01})$, we find:
\begin{align}
G_{2\varepsilon}(1\mathrm{\mu s})\approx 125\\
G_{2\varepsilon}(21\mathrm{\mu s})\approx 6,
\end{align}
reproducing the results presented in Table 1 of the main text.

\begin{figure*}[!ht]
\centering
\includegraphics[width=5in]{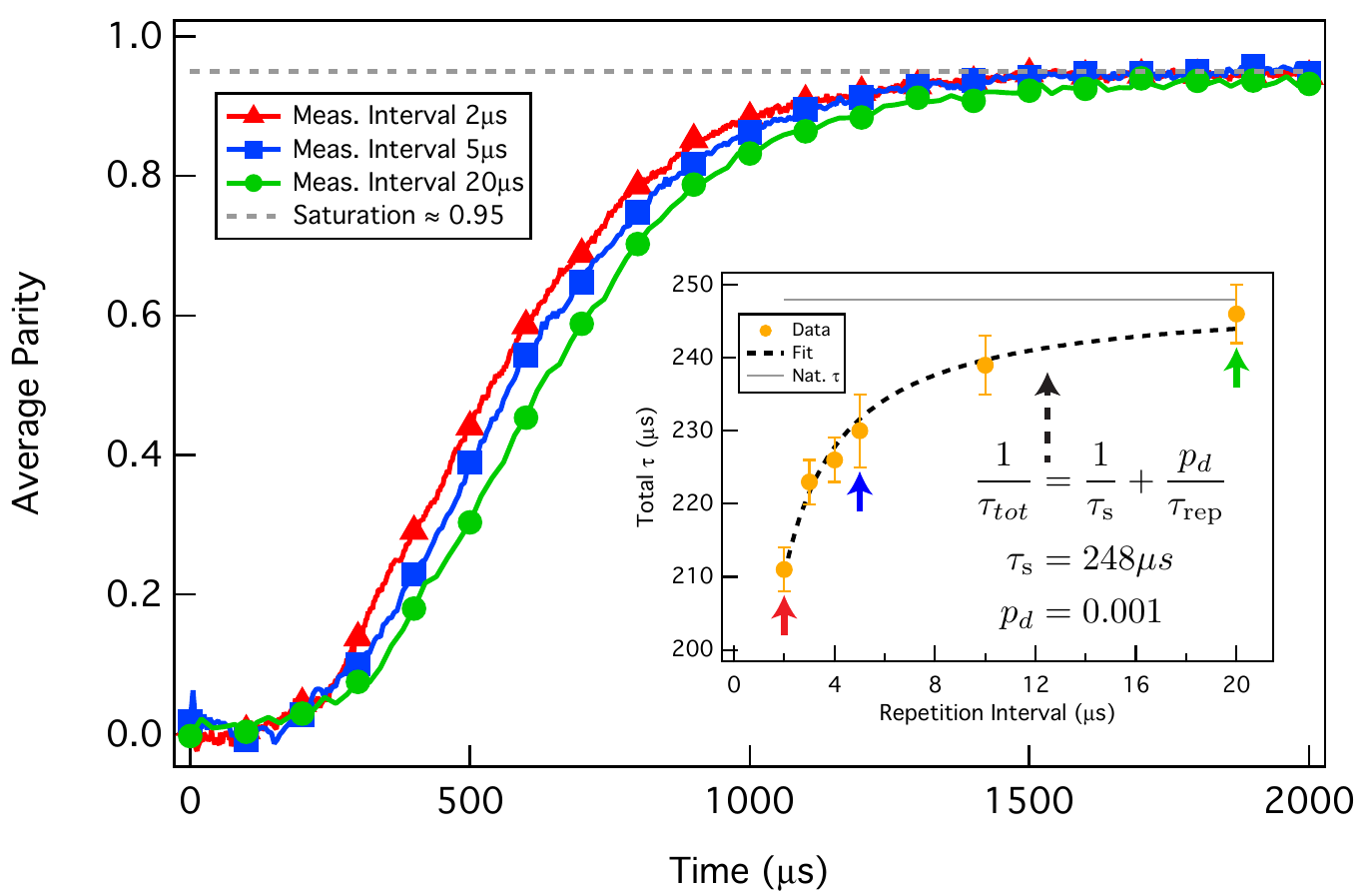}
\caption{\footnotesize
    \textbf{QND parity measurements.}  We repeat the same experiment demonstrated in~\cite{Sun:2013ha}, which quantifies the effect of measuring parity on the effective resonator decay rate $1/\tau_{tot}$.  Ideally, for perfectly QND measurements $1/\tau_{tot}$ should match the natural decay rate $1/\tau_s$, regardless of the measurement rate.  In reality, there is a small probability $p_d$ that by measuring parity we induce more photon jumps.  The main plot shows three decay curves of average parity $\braket{P}$ versus time for three different measurement repetition intervals: $2\mathrm{\mu s}$, $5\mathrm{\mu s}$, and $20\mathrm{\mu s}$; the initial displacement is $|\alpha|=2$.  The inset shows a plot of the extracted time constants for these three and other points.  The data fits well to a model in which the natural decay rate acts in parallel with an induced decay rate $p_d/\tau_{\mathrm{rep}}$, for $\tau_{\mathrm{rep}}$ the repetition interval.  In this experiment we find $p_d$ to be $0.1\%$ per measurement, in agreement with the results in~\cite{Sun:2013ha}.  Note that by using the adaptive parity monitoring protocol (see sec.~\ref{smart_tracking}), regardless of measurement rate each curve saturates at a measured $\braket{P}\approx0.95$.  This is consistent with a thermal population of the storage resonator $n^s_{th}<2\%$, which reduces the average parity from ideally $+1$ of the vacuum to $\braket{P}\approx1-2n^s_{th}$, and a parity measurement fidelity of $98.5\%$ for no photons in the resonator.  Such performance would be impossible without the crucial application of the adaptive parity monitoring protocol.}
\label{fig:QND}
\end{figure*}

We also quantify how Quantum Non-Demolition (QND) our parity measurements are, or in other words, with what probability of demolition $p_d$ do we induce a photon jump by measuring parity.  We use the methods studied extensively in~\cite{Sun:2013ha} to find that $p_d=0.1\%$ in our system (Fig.~S\ref{fig:QND}), comparable to the result in~\cite{Sun:2013ha}.  The probability of dephasing, however, is $p_d^2$, since this is the probability of inducing two jumps in a row; this effect is negligible.  The mechanism behind $p_d$ is a subject of our on-going research.
 
\subsection{Uncorrectable Errors}
The cat code cannot distinguish between photon loss ($\hat{a}_s$) and photon gain ($\hat{a}_s^\dag$).  The probability of excitation due to $\hat{a}_s^\dag$ is given by $p_{\uparrow s}(t_M)=t_M n^s_{th}\bar{n}/\tau_s$.  Given the low thermal population and high coherence of the resonator (Table~\ref{coherence}), we expect an $\hat{a}_s^\dag$ event on average every $\sim6\mathrm{ms}$ for $\bar{n}=2$, a rate of thermal excitation that is negligible compared to all other sources of loss.  If this were the only source of code failure, the gain $G_{\uparrow s}=t_M/(p_{\uparrow s}\tau_{f01})$ would be independent of $t_M$ and equal to approximately $20$, as given in Table 1 of the main text.

When these currently uncorrectable sources of error become dominant, the redundancy of the cat code can be augmented by increasing the size of the logical basis stated to superpositions of three coherent states (and higher) ~\cite{Leghtas:2013ff}.  Although coherent states are not eigenstates of $\hat{a}_s^\dag$, for large enough amplitudes the addition of a single photon results in a distortion in the Poisson coefficients that can still be corrected by the pumping scheme described in~\cite{Leghtas:2015uf}.

\subsection{Readout Error}\label{readout_errors}

During the parity mapping sequence, ancilla dephasing due to $T_2$ is the dominant contribution to the overall drop in parity measurement fidelity.  The parity of the state of course does not change upon an errant measurement, but our reaction to the result within the experimental flow does (see sec.~\ref{feedback_applications}).  Were it not for the detrimental effects of ancilla back-action (see sec.~\ref{forward_prop}), the optimal approach would be to measure as quickly as possible to build up measurement statistics (see sec.~\ref{confidence}).  Indeed this was the strategy implemented in~\cite{Sun:2013ha}, where the goal was to understand the dynamics of photon jumps and so error propagation was not considered.  In that work, a quantum filter used Bayesian statistics to best estimate the parity of the state at any given time, and it would take about three consecutive agreeing measurements for the filter to switch from one parity to another.  In this work, with improved fidelities and lifetimes we understand that with a measurement cadence of $t_M\approx1\mathrm{\mu s}$ it would take roughly $2\mathrm{\mu s}$ for an equivalent filter to converge on a parity with high probability.  If a photon jump occurs within this effective bandwidth, the filter will not detect it, resulting in a readout error.  With average photon jump times on the order of $120\mathrm{\mu s}$ for an $\bar{n}=2$ in the resonator, the probability of missing a jump is therefore $p_{mj}\approx2/120\approx1.5\%$.  The gain is therefore approximately equal to $120\mathrm{\mu s}/(p_{mj}\tau_{f01})\approx25$, as in Table 1 of the main text.

For the optimal measurement cadence $t_M\approx21\mathrm{\mu s}$ for $\bar{n}=2$, given that the probability to have photon jumps within $t_M$ is approximately $20\%$, the optimal strategy is to trust each result implicitly (see sec.~\ref{optimal_rate}).  If ancilla dephasing were the only source of error, after the syndrome mapping time $\pi/\chi_{sa}$ the purity of the ancilla state would decrease to approximately $\pi/(\chi_{sa}T_2)\approx0.98$.  The remaining $2\%$, which is mixture, would be measured to have the correct syndrome mapping result half of the time.  The probability to dephase the qubit due to an errant syndrome measurement result becomes $p_\phi\approx1\%$.  The gain $G_\phi(t_M)=t_M/(p_\phi\tau_{f01})\approx7$ returns the result presented in Table 1 of the main text.
\subsection{Ancilla Preparation}\label{anc_prep}
After every syndrome measurement, we reinitialize the ancilla to $\ket{g}$ regardless of the result (see sec.~\ref{reset}).  Given its finite rate of excitation, $\Gamma_\uparrow$ (see sec.~\ref{setup}), after $t_M$ the ancilla may no longer be in the ground state with a probability $p_{\uparrow a}=\Gamma_\uparrow t_M$, which leads to an errant subsequent syndrome measurement.  With a maximal syndrome measurement rate, ancilla preparation errors are negligible as some type of majority voting can be performed on groups of measurements to filter out this effect.  Errors only occur on the order of $p_{\uparrow a}^2$ when majority voting in groups of three.  The gain is therefore $G_{\uparrow a}(t_M)=t_M/(p_{\uparrow a}^2\tau_{f01})\approx2000$, leading to the high gain seen in Table 1 of the main text.  For the optimal rate (see sec.~\ref{optimal_rate}), however, this gain is limited by $\Gamma_\uparrow$ only, and thus $G_{\uparrow a}(t_M)=1/(\Gamma_\uparrow\tau_{f01})\approx3$.  This mechanism is of course also responsible for resonator dephasing, as is described in section~\ref{forward_prop}, and could be mitigated by stabilizing the ancilla ground state during  $t_\mathrm{w}$.

\subsection{Orthogonality of Basis States}\label{orthogonality}
The non-orthogonality of the basis states in the cat code is an important consideration in deciding the initial amplitude of the encoded state.  Larger cat states mean that the cat code can be employed for longer periods of time without applying any unitary gates or dissipative pumps~\cite{Leghtas:2015uf} to restore the amplitude.  As this is largely a technical point that has been demonstrated not to be a fundamental limitation, the more salient question is at what point does increasing the cat state amplitude begin to adversely affect the performance of the code due to the increased rate of errors.  The trade-off between non-orthogonality and average error rate is in fact very generous, however, since the overlap between two coherent states falls off exponentially with the difference between them in a resonator's phase space~\cite{book:haroche06} while the error rate increases linearly (in $\bar{n}$):
\begin{align}
\left<\alpha|\beta\right>&=e^{-|\alpha|^2/2}e^{-|\beta|^2/2}e^{-\alpha^*\beta}\label{eq:overlap}\\
|\left<\alpha|\beta\right>|^2&=e^{-|\alpha-\beta|^2}
\end{align}
Using eq.~\ref{eq:overlap}, one can perform a similar calculation for the cat code basis states to obtain the following overlaps:
\begin{align}
|\left<C_\alpha^+|C_{i\alpha}^+\right>|^2&=\left(\frac{2e^{-\alpha^2}\cos(\alpha^2)}{1+e^{-2\alpha^2}}\right)^2\\
|\left<C_\alpha^-|C_{i\alpha}^-\right>|^2&=\left(\frac{2e^{-\alpha^2}\sin(\alpha^2)}{1-e^{-2\alpha^2}}\right)^2,
\end{align}
where $\alpha$ is understood to be a real number here.  The trigonometric terms are a result of the interference changing with $\alpha$, or equivalently in this experiment, with time.  Using a Python quantum simulation software package called QuTiP~\cite{Johansson20131234,Johansson20121760}, we simulate the effect of the increasing non-orthogonality on the efficacy of the optimized decoding pulses to faithfully transfer an encoded state in the resonator back onto the ancilla.  We find that with an initial encoding size of $\bar{n}_0=2$ and after a time of $\sim100\mathrm{\mu s}$, the error in the decoding pulses for even parity states approximately equals $2\textrm{-}3\%$, while for the odd parity states it approximately equals $6\textrm{-}7\%$.  Using the Poisson distribution to calculate the percentage of even and odd parity states after $\sim100\mathrm{\mu s}$, we find that the resulting infidelity due to overlapping basis states at the end of the tracking sequence amounts to roughly $4\textrm{-}5\%$.  For earlier times, this error rapidly decreases toward $0$, indicating that even for small cat sizes, the approximation that the basis states are orthogonal is still quite accurate.

\subsection{Code-space Leakage}\label{leakage}
To a good approximation, for $\bar{n}\lesssim2$ the basis states of the cat code can be interpreted as superpositions of only the Fock states $\ket{0}_f\rightarrow\ket{7}_f$.  In turn, this restricted space can be described in a binary representation that requires just three physical qubits (Fig.~S\ref{fig:Circuit_b}a), with coefficients given by the Poisson distribution of a coherent state of $\alpha\lesssim\sqrt{2}$ and with $d_0$ the least significant bit in $\ket{d_2d_1d_0}$ (e.g. $\ket{110}\equiv\ket{6}$).  In this formulation, $d_0$ is the ``Parity Bit:" when $d_0=0$ the parity of the state is even and when $d_0=1$ the parity is odd.  Note that the even and odd logical basis states are still all mutually orthogonal.  

Although in principle such an encoding scheme can be fashioned in a cQED system with three transmons, the error processes would be completely different, dominated by single transmon energy decay and dephasing rather than the correlated shift of bits arising from the action of some effective lowering operator.  The utility of this representation, however, is that it emphasizes the possibility of excursions out of the code space; for increasing deviations $\epsilon_n$ of the coefficients $c_n$ from their ideal values as specified by the Poisson distribution, the overlap $(\bra{C_\alpha^+}+\sum_{n=0}^7\epsilon_n\bra{n})\ket{C_\alpha^+}\rightarrow 0$.  This effect is of course continuous.

One may note that the Kerr of the resonator immediately changes these coefficients at a rate $K_s$.  This effect, however, is deterministic, does not change the parity of the state, and in fact periodically brings the coefficients back to their original values (minus the effect of amplitude decay)~\cite{Kirchmair:2013gu}.  It therefore does not constitute a source of dephasing since it can be taken as just a continuous change of basis in time.  As long as we take this into account when decoding the encoded state back onto the ancilla at the end of our protocol, no information is lost (sec.~\ref{grape}).  There are, however, several non-deterministic effects that do constitute dephasing, all arising from undesired interactions of the resonator with the ancilla, with the readout resonator, and again with itself (a second effect of Kerr to be described in the subsequent section).  Some of these sources of loss are possible to partially recover from even in the current implementation of the experiment, while others are a central vehicle of non-fault-tolerance in this system.

\subsubsection{Undesired Couplings -- Self-Kerr \& Cross-Kerr}\label{undesired_couplings}
\begin{figure*}[!ht]
\centering
\includegraphics[width=5in]{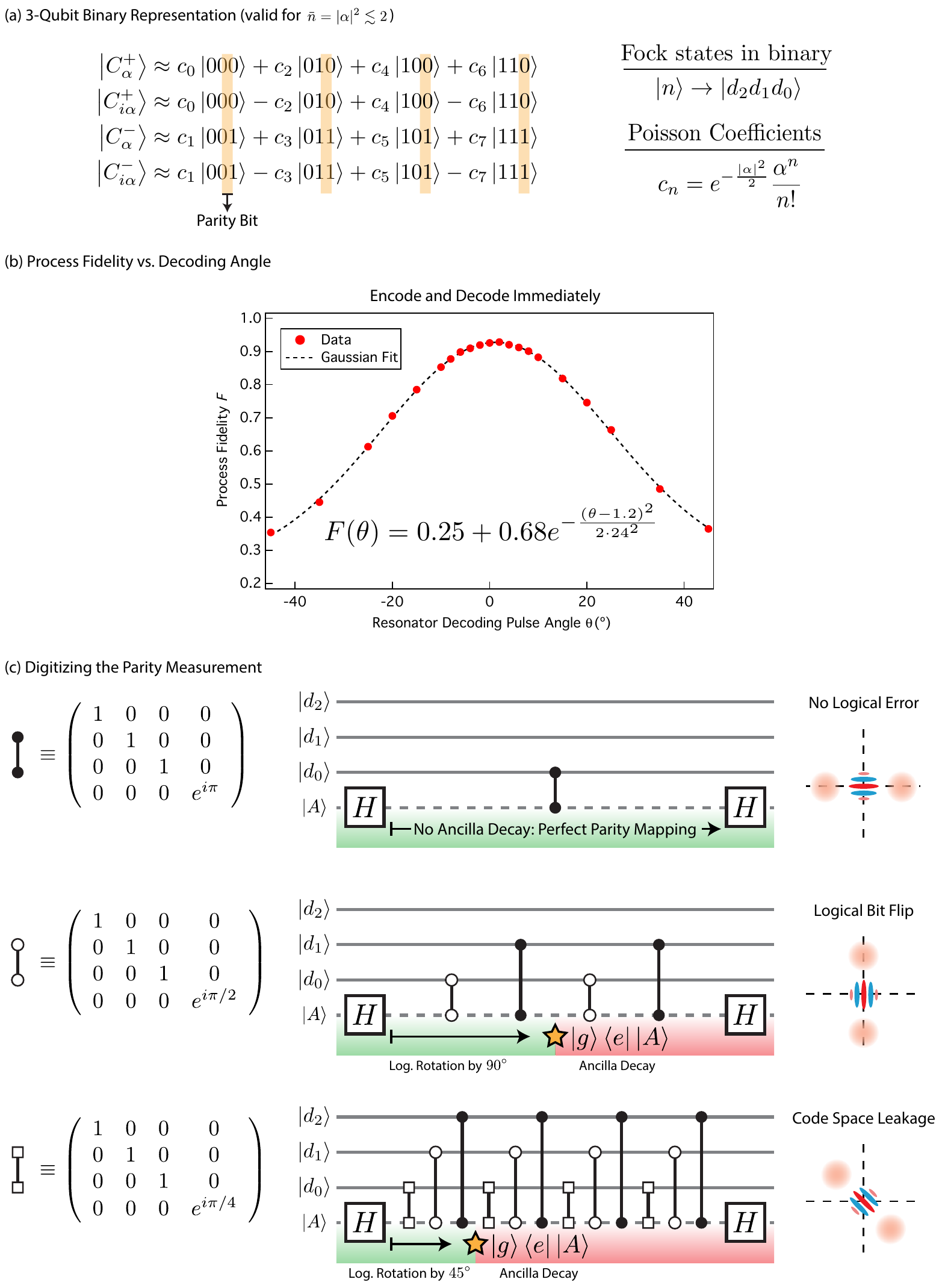}
\caption{\footnotesize
    \textbf{Leaking out of the code space.  (a)}  The cat code logical basis states $\ket{C_{\alpha}^\pm}$ and $\ket{C_{i\alpha}^\pm}$ can be expanded in the Fock basis, where each component is rewritten in binary.  For small cat sizes of $|\alpha|^2\lesssim2$, only three physical qubits are necessary to realize such a representation to high accuracy, as the Poisson coefficients $c_n$ for Fock states greater than $\ket{111}\equiv\ket{7}$ become vanishingly small.  Should these coefficients deviate from their specified values without our knowledge, the integrity of the quantum information may start to suffer.  \textbf{(b)}  Measured loss of fidelity as a function of an intentional phase offset $\theta$ of a decoding pulse that immediately follows qubit encoding.  With a standard deviation of $\sim24^\circ$, the broad Gaussian fit shows that for small deflections the fidelity suffers only quadratically.  Hence, the Kerr-induced deflection per photon jump is not a major source of dephasing for low jump numbers even with a $t_\mathrm{w}\approx20\mathrm{\mu s}$.  However, as ancilla decay during mapping can deflect the state by any angle, it causes a substantial degradation in process fidelity.  \textbf{(c)}  Code space leakage can be particularly acute if the ancilla $\ket{A}$ undergoes energy decay (or excitation) during the parity mapping.  Shown in the first panel is an ideal parity mapping using the binary representation.  The least significant bit, $d_0$, is the parity bit; the parity is even for $d_0=0$ and odd for $d_0=1$, regardless of $d_1$ and $d_2$.  The parity mapping is thus a simple cNOT gate, depicted here as a controlled phase (solid black circles, $\pi$ phase shift) between two Hadamard gates (H).  Such a circuit representation belies the fact that the mapping is finite in time, lasting $\pi/\chi_{sa}\approx250\mathrm{ns}$.  To obtain a better approximation of the true dynamics, we split the cPHASE into two pieces, where the two empty circles are now controlled phase gates with a $\pi/2$ phase shift.  A simple calculation in the Fock basis demonstrates that if the ancilla suddenly decays to $\ket{g}$ exactly halfway through the mapping, a logical bit flip occurs.  For arbitrary decay times, we witness code space leakage, where the cat state is aligned with neither the real nor the imaginary axis.  The third row shows an example of this for one more layer of granulation.}
\label{fig:Circuit_b}
\end{figure*}

The non-commutativity of the resonator's Kerr Hamiltonian and the annihilation operator $[\frac{K_s}{2}\hat{a}_s^{2\dag}\hat{a}^2_s,\hat{a}_s]\neq0$ results in the following relation~\cite{Leghtas:2012ff} (excluding an irrelevant global phase):
\begin{align}
\hat{a}e^{-i\frac{K}{2}t\hat{a}^{\dag2}\hat{a}^2}&=e^{-i\frac{K}{2}t\hat{a}^{\dag2}\hat{a}^2}e^{-iKt\hat{a}^\dag\hat{a}}\hat{a}.
\end{align}
This means that every time a photon jumps at a time $t_j$, the resonator state is rotated in phase space by an angle $\theta=K_st_j$.  Without an infinite cadence of measurement, however, there is always some finite uncertainty in $t_j$, $\delta t_j$, and consequently in $\theta$, $\delta\theta$.  A non-zero $\delta\theta$ results in an angular spread of cat code basis states in phase space.  On a shot-by-shot case, it means that we lose track of the phase of the resonator state within the angular window defined by $K_st_M$.  In other words, the state leaks out of the code space.  We study the effects of such leakage by encoding a qubit into our codeword and then immediately thereafter intentionally decoding back at the wrong phase (Fig.~S\ref{fig:Circuit_b}b).  The resulting Gaussian curve allows us to quantify the sensitivity of the cat code to uncertainties in the jump time.

If we were to measure the error syndrome very quickly, we would know $t_j$ to high accuracy and the resulting error due to uncertainty in jump time would be negligible given the second-order dependence of process fidelity on decoding angle.  With a $t_M\approx1\mathrm{\mu s}$, the average angle of deflection $\bar{\theta}\approx1^\circ$.  For this cadence, we can assume a uniform probability distribution that gives an uncertainty in angle of $\delta\theta\approx0.5^\circ$.  Using the result of the fit in Fig.~S\ref{fig:Circuit_b}b, we average over the Gaussian distribution within a window of $\pm0.8^\circ$ to find a probability of dephasing $p_K\approx0.02\%$, an expectedly minor contribution.  Weighting $p_K$ by the probability to have a jump within $t_M$, which is on the order of $1\%$, the predicted gain in such a scenario where this is the only source of error is consequently very high: $G_K=t_M/(p_K\cdot0.01\cdot\tau_{f01})\approx2000$, as in the table in the main text.  

Given the necessity of spacing out parity measurements in time by $t_\mathrm{w}$ in order to maximize lifetime gain (see sec.~\ref{optimal_rate}), however, the absolute time of the jump and thus the value of $\theta$ inherit some non-negligible uncertainty, resulting in code space leakage.  In other words, the coefficients $c_n$ now deviate from the Poisson distribution by $\epsilon_n$ that are unknown.  For a typical $t_\mathrm{w}\approx 20\mathrm{\mu s}$ (see Table 1 in the main text) and the value of the resonator's Kerr (Table~\ref{table_params}) the uncertainty in jump angle is $\sim10^\circ$, resulting in a $\sim3\%$ loss of fidelity.  This loss of course increases the more errors occur.  Assuming the probability of detecting a photon jump is again $\sim 20\%$ per step, the loss in process fidelity is $p_K\approx0.2\times0.03$.  The gain $G_K=t_M/(p_K\tau_{f01})\approx10$, as in the table in the main text.  The rate $K_s$, however, is on the order of several kHz, and so in principle we can completely recover from this minor dephasing by interleaving the parity measurements with the dissipative pumping scheme demonstrated in~\cite{Leghtas:2015uf}, which pumps and refocuses slightly dephased cat states back to the original logical basis (restoring their amplitude as well).

Likewise, code space leakage occurs whenever a coherent state is injected into the readout resonator to measure the state of the ancilla.  As coherent states have an uncertainty in their average photon number of $\sqrt{\bar{n}}$, the cross-Kerr interaction leads to a dephasing of the encoded state at a rate proportional to $\chi_{sr}$.  Quantitatively, per measurement we see an average deflection of $\bar{\phi}=\bar{n}\chi_{sr}\tau_\mathrm{meas}\approx20^\circ$ of the resonator state in phase space for a readout pulse duration $\tau_\mathrm{meas}$.  Given the value of $\chi_{sr}$ in Table~\ref{table_params}, we estimate that each readout pulse contains $\bar{n}\approx70$ photons.  The uncertainty in the deflection scales as the square root of $\bar{n}$: $\delta\bar{\phi}=\sqrt{\bar{n}}\chi_{sr}\tau_\mathrm{meas}\approx2^\circ$.  With $n$ measurements, the total uncertainty in the angle is $\sqrt{n}\delta\bar{\phi}$.  For example, after ten measurements this uncertainty is still much smaller than the standard deviation of the Gaussian in Fig.~S\ref{fig:Circuit_b}b.  Given the minimal effect on the process fidelity, this source of dephasing is excluded from the discussion in the main text.

We therefore treat the resonator anharmonicity and coupling to the readout resonator not as sources of non-fault tolerance, but rather as necessary technical trade-offs that can in principle be fully compensated.  In fact, by applying real-time feedback to mitigate some of these effects we are able to substantially enhance the fidelities our QEC system can offer; these features are detailed in sec.~\ref{record_error_time}.

\subsubsection{Forward Propagation}\label{forward_prop}
Infidelities due to ancilla dephasing outlined in sec.~\ref{readout_errors} and the forward propagation of errors to be discussed in this section have a common denominator: ancilla decoherence.  In the former, both phase flips and amplitude decay of the ancilla contribute to a decrease in parity measurement fidelity.  In the latter, one can see by looking at the system Hamiltonian (see sec.~\ref{hamiltonian}) that the frequency of the resonator depends on the state of the transmon.  Any random change in ancilla energy due to decay ($\sigma_-$) or excitation therefore rotates the resonator state at a rate $\chi_{sa}$, while phase flips ($\sigma_z$) have no effect.  Figure S\ref{fig:Circuit_b}c depicts how one can model a parity measurement in the digitized version of the cat code.  Employing a single ancilla, the parity measurement is nothing more than a cNOT gate between this ancilla and the parity bit $d_0$, which specifies the state's symmetry with respect to a $180^\circ$ rotation.  The cNOT is written here equivalently as a controlled phase gate between two Hadamard gates (H)~\cite{NielsenQI}.  The higher parity bits, $d_1$ and $d_2$, provide further information about the state's symmetry properties with respect to $90^\circ$ and $45^\circ$ degrees.  The first panel shows that with no ancilla energy decay, the parity mapping is perfect since it does nothing to coefficients $c_n$ at the end of the protocol.

The length of this mapping, $\pi/\chi_{sa}\approx 250\mathrm{ns}$, however, is a small but non-negligible fraction of the ancilla $T_1$.  One can model this finite gate time by splitting the phase gate into two ``controlled-$\pi/2$" gates and adding two phase gates to the next parity bit $d_1$.  With a perfect parity mapping one obtains the exact same results as in the first row.  If the ancilla decays exactly half-way through the sequence, however, the resonator state inherits a phase of $\pi/2$ in phase space; this is a logical bit flip in our basis.  One can continue splitting the gate into smaller and smaller pieces (e.g. third panel), where now the ancilla $T_1$ decay rotates the resonator state by an arbitrary angle that is known only if the time of ancilla decay is known.  Of course, as we depend on the entangling interaction between the ancilla and resonator throughout the parity mapping time, in this implementation we have no way of detecting when this decay occurs.  Equivalently, the photon number parity operator $\hat{P}=e^{i\pi\hat{a}^\dag_s\hat{a}_s}$ commutes with any rotation in phase space: $[\hat{P},e^{i\theta\hat{a}^\dag_s\hat{a}_s}]=0$.  The environment gains information that we do not.  Beyond the risk of $T_1$ decay during the parity mapping, the equally detrimental effect of $T_1$ decay of the ancilla during the readout pulse (before ancilla reset, see sec.~\ref{smart_tracking}) reduces the fidelity of all trajectories where one or more photon jumps occur.  In addition, unknown ancilla excitations from $\ket{g}$ to $\ket{e}$ (or higher states), which occur at a rate proportional to $\Gamma_\uparrow$, dephase the resonator state similarly.

We can comb through the probabilities of a $T_1$ event in each of the three main steps throughout the entire duration of the sequence and calculate a gain in a manner similar to calculations in previous sections.  During the parity mapping, the probability of ancilla decay is $p_{\downarrow a,1}\approx\pi/(\chi_{sa}\cdot2T_1)=0.004$; the probability of measuring ancilla $\ket{e}$ equals the probability of measuring a photon jump, and so the contribution to dephasing is $p_{\downarrow a,2}\approx(\bar{n}\kappa_s t_M)\tau_\mathrm{meas}/T_1=0.0001$, the probability of a photon jump times the probability of $T_1$ decay in a duration $\tau_\mathrm{meas}$; finally, during $t_M$ there is always the risk of ancilla excitation, with a probability $p_{\uparrow a,3}\approx t_M\Gamma_\uparrow=0.001$.  The total probability of forward propagation is $p_{fp}(t_M)\approx p_{\downarrow a,1}+p_{\downarrow a,2}+p_{\uparrow a,3}$.  For $\bar{n}=2$, we find that $p_{fp}(1\mathrm{\mu s})\approx0.5\%$.  Defining the gain $G_{fp}(t_M)=t_M/(p_{fp}\tau_{f01})$, we find $G_{fp}(1\mathrm{\mu s})\approx0.7$.  Performing the same calculation for $t_M\approx21\mathrm{\mu s}$, we find $G_{fp}(21\mathrm{\mu s})\approx2$.  We thus obtain the numbers in the final row of Table 1 in the main text, and arrive at the key constraint of our system: measuring more frequently enhances the likelihood of forward propagation of errors.  As seen in the preceding sections, by mitigating this decoherence channel we stand to gain substantially in all other aspects with faster syndrome measurements.


\section{Quantum Control Architecture}\label{quantum_control}
Experiments in quantum computation typically involve a continuous and carefully choreographed generation of pulses in order to implement a certain operation.  This process may often require modifying the pulse generation in real-time as a response to the real-time analysis of returned signals (real-time feedback).  Given that the time evolution of our quantum systems can be counted in nanoseconds, crucial to the success of the quantum experiment is the efficiency of collecting, interpreting, and reacting to the returned signals.  The architecture demonstrated here is comprised of four major components: Digital-to-Analog converters (DACs) that output pulses; Analog-to-Digital converters (ADCs) that sample input signals; digital inputs/outputs (DIG-IOs) that enable inter-card communication as well as the triggering of certain digital RF components; and finally a Field Programmable Gate Array (FPGA) that dictates the flow of the experiment in time, orchestrating the three previous components to steer the quantum system to some desired state in real-time (see Fig.~\ref{fig:setup} for a schematic).  This is our quantum controller.

Each hardware unit, or ``card," is an independent agent.  It combines the functionality of an instrument like a commercially available Arbitrary Waveform Generator (AWG), a data sampling card, and certain data analysis functions crucial for efficient feedback all on one piece of equipment.  Such a design dramatically enhances the possible levels of control, sophistication, and complexity of a quantum experiment.  Furthermore, all cards run in parallel with no inherent dependency on each other.  They may produce pulses that are sent to manipulate a particular aspect of the quantum system.  Incoming signals from the quantum system may also be routed as inputs to the cards in some pre-defined way.  Each card thus produces and analyzes different signals, and it can then distribute its findings among the other cards in real-time through a dedicated digital communication layer.

The common denominator in this scheme is the set of instructions loaded onto each card prior to the experiment, which coordinates how the cards work together.  Once the experiment starts there is no one master card that must dictate the flow; each card can decide what to do independently.  The result is a decentralized network of classical computation that provides a fast, efficient, and flexible platform to interface with the quantum system.  Thus, by properly coordinating the signals sent and received by this network of cards, the user ultimately coordinates the interactions between distinct entities of the quantum system, all accomplished on time scales of just a few hundred nanoseconds.

\subsection{Hardware Specifications}
We use three Innovative Integration X6-1000M boards housed in a VPXI-ePC chassis with a VPX-COMEX module that produces a 1GHz clock and sharp, synchronized triggers.  Each board contains two 1 GS/s ADCs, two 1 GS/s DAC channels, and digital inputs/outputs that are controlled by a Xilinx VIRTEX-6 FPGA loaded with in-house logic.  The three boards are synchronized to control the storage resonator, readout resonator, and ancilla transmon. The readout signals are routed to the ADCs on the readout resonator board, whereafter the FPGA demodulates and thresholds the signal to determine the state of the ancilla ($\ket{g}$, $\ket{e}$, and higher).  The feedback latency between the last sample in to the FPGA and the first sample out is approximately 200ns, providing us with a powerful tool to mitigate the effects of ancilla $T_1$ decay post-measurement (see section~\ref{feedback_applications}).  

\subsection{Quantum Error Correction Process}\label{exptflow}
Here we summarize our full protocol (Fig.~S\ref{fig:timings}).  Each run of the experiment cycles through the following steps:
\begin{enumerate}
	\item System and ancilla reset -- using feedback, we make sure that the ancilla is in ground state $\ket{g}$ and the resonator is in vacuum $\ket{0}_f$.
	\item Qubit initialization -- in each realization of the experiment we apply a single-qubit gate on the ancilla to encode one of the six cardinal points on the qubit Bloch sphere.  This over-complete set of states allows us to perform process tomography of the QEC system (see sec.~\ref{process_tomo}), and is equivalent to characterizing the action of a system on the qubit $\ket{\psi_\mathrm{init}}=c_0\ket{0}+c_1\ket{1}$.
	\item Encoding -- we transfer the qubit from the ancilla into a superposition of cat states in the resonator. At the end of this step the state of the resonator is $\ket{\psi_\mathrm{init}}\rightarrow c_0\ket{C_\alpha^+}+c_1\ket{C_{i\alpha^+}}$, while the ancilla, to the best of our ability given experimental realities (see sec. ~\ref{grape_implementation}), ends in $\ket{g}$, ideally completely disentangled from the resonator state.
	\item Parity monitoring -- we identify photon jumps, or errors in our codeword, by monitoring the parity of the logical state in the resonator. This is done using the adaptive parity monitoring scheme (Fig.~2 of the main text, and elaborated upon in~\ref{smart_tracking}).
	\item Decoding and correction -- the quantum bit of information is brought back onto the ancilla using the knowledge we gather while monitoring the error syndrome (see sec.~\ref{adaptive_decoding}).  A different decoding pulse is used for each point in time due to the changing amplitude and Kerr evolution of the cat states.  At the end of the decoding pulse, the resonator should ideally be completely in vacuum and with the measured error record the qubit is corrected following the cat code prescription (see sec.~\ref{cat_code}).
	\item Tomography -- we perform qubit state tomography on the ancilla to compare the final qubit $\ket{\psi_\mathrm{fin}}$ with the initial state $\ket{\psi_\mathrm{init}}$.  Using the results we fully characterize the QEC system process (see sec.~\ref{process_tomo}).
\end{enumerate}

Steps 1-2 prepare the system in its ground state to high accuracy; ancilla reset is more than $99.8\%$ effective and no residual thermal population is measured after resonator reset. Steps 3-5 are the error correction part of the experiment.  This part does not assume any knowledge about the quantum state it is designed to protect or the decoherence mechanisms.  In the final step we measure the ancilla that is ideally back in the initial state.  Any deviation leads to a decay of the process fidelity in time.  While the entire experiment is implemented as one big state machine, only the exact durations of the system and ancilla reset step are not predetermined.

\subsection{Feedback Details}\label{feedback_applications}
\subsubsection{System and Ancilla Reset}\label{reset}
 As detailed in Table~\ref{coherence}, the thermal populations of the ancilla and the resonator are $\sim4\%$ and $<2\%$, respectively.  This is enough to adversely affect not only our encoding pulses (see sec.~\ref{grape}), but also subsequent error syndrome detection.  The protocol starts with the quantum controller measuring the state of the ancilla.  If the result is the excited state $\ket{e}$, the controller applies a fast $\pi$ pulse (Gaussian envelope with $\sigma=2\mathrm{ns}$) to return the ancilla to $\ket{g}$ and measures again; if the pulse is not successful the loop is repeated, while if the pulse is successful the experiment continues.  With feedback latencies of just $\sim200$ns (last sample in, first sample out), a readout pulse duration of $\tau_\mathrm{meas}\approx400$ns, and latencies due to cables into and out of the experimental setup totaling $\sim100\mathrm{ns}$, we are able to reset the ancilla to $>99.8\%$ in $\ket{g}$.  This protocol was also demonstrated in~\cite{Riste:2012wc}.

Second, we use the now initialized ancilla to project the resonator state to the vacuum by applying long $\pi$ pulses on the ancilla (Gaussian envelope with $\sigma=600\mathrm{ns}>1/\chi_{sa}$) that address only the $\ket{0}_f$ Fock state~\cite{Schuster:2007ki}.  If the result of a subsequent ancilla measurement is $\ket{e}$, with high probability the resonator is in vacuum.  These pulses, however, have a lower fidelity ($\sim90-95\%$) owing to the ancilla $T_2$, and so we repeat this experiment until we measure $\ket{e}$ three times consecutively.  Once this occurs, we once again employ the protocol above to reset the ancilla to $\ket{g}$ and continue to the encoding step.  With a thermal population of $\sim2\%$ and a lifetime of $250\mathrm{\mu s}$, the average rate of excitation of the resonator from $\ket{0}_f\rightarrow\ket{1}_f$ is on the order of $10\mathrm{ms}$, and so we are unable to measure any residual population in $\ket{1}_f$ after this purification procedure.


\subsubsection{Adaptive Parity Monitoring State Machine}\label{smart_tracking}

In measuring photon number parity as the error syndrome, two protocols may be used that both employ a Ramsey-style pulse sequence to map opposite parities to opposite poles of the ancilla Bloch sphere; they differ only in the sign of the second $\pi/2$ pulse.  For example, when the resonator starts in an even parity cat state and the sign of the second $\pi/2$ pulse is positive, the ancilla ends up in $\ket{e}$ at the end of the protocol:
\begin{align}
\ket{\psi}=&\ket{g}\frac{1}{\sqrt{2}}(\ket{\alpha}+\ket{-\alpha})\\
\xrightarrow{+\pi/2}&\frac{1}{2}(\ket{g}+\ket{e})(\ket{\alpha}+\ket{-\alpha})\\
\xrightarrow{wait}&\frac{1}{2}\ket{g}(\ket{\alpha}+\ket{-\alpha})+\frac{1}{2}\ket{e}(\ket{\alpha e^{i\chi_{sa}t}}+\ket{-\alpha e^{i\chi_{sa}t}})\\
\xrightarrow{t=\pi/\chi_s}&\frac{1}{2}\ket{g}(\ket{\alpha}+\ket{-\alpha})+\frac{1}{2}\ket{e}(\ket{\alpha e^{i\chi_{sa}(\pi/\chi_{sa})}}+\ket{-\alpha e^{i\chi_{sa}(\pi/\chi_{sa})}})\\
=&\frac{1}{2}\ket{g}(\ket{\alpha}+\ket{-\alpha})+\frac{1}{2}\ket{e}(\ket{\alpha e^{i\pi}}+\ket{-\alpha e^{i\pi}})\\
=&\frac{1}{2}\ket{g}(\ket{\alpha}+\ket{-\alpha})+\frac{1}{2}\ket{e}(\ket{-\alpha}+\ket{\alpha})\\
\xrightarrow{+\pi/2}&(\ket{g}+\ket{e})(\ket{\alpha}+\ket{-\alpha})+(\ket{e}-\ket{g})(\ket{-\alpha}+\ket{\alpha})\\
=&\ket{e}\frac{1}{\sqrt{2}}(\ket{\alpha}+\ket{-\alpha})
\end{align}

Note, however, that if the parity is odd the ancilla ends up in $\ket{g}$; and likewise, if the parity is even but the sign of the second pulse is negative, the ancilla again ends up in $\ket{g}$.  Thus, in implementing our QEC system, simply repeating just one of the two protocols during the error syndrome monitoring does not suffice, since with either one the ancilla spends much more time in the excited state for one of the two parities.  This asymmetry provides a strong motivation for using real-time feedback.

Starting in even parity, and given the low probability ($\sim20\%$) to have an error between two consecutive syndrome measurements for the optimal measurement cadence, the controller plays the pulse sequence that maps even parity to ancilla $\ket{\psi_a}=\ket{g}$ and odd parity to ancilla $\ket{\psi_a}=\ket{e}$.  When an error occurs and the parity changes, the controller pulses the ancilla from $\ket{e}$ back to $\ket{g}$ and then continues monitoring errors by employing the opposite protocol, which instead maps odd parity to $\ket{g}$ and even parity to $\ket{e}$.  Therefore, throughout a single measurement trajectory, counting the number of errors amounts to just counting the number of times $\ket{\psi_a}=\ket{e}$ occurs.  

\begin{figure}[h!]
\centering
\includegraphics[page=1,width=\linewidth]{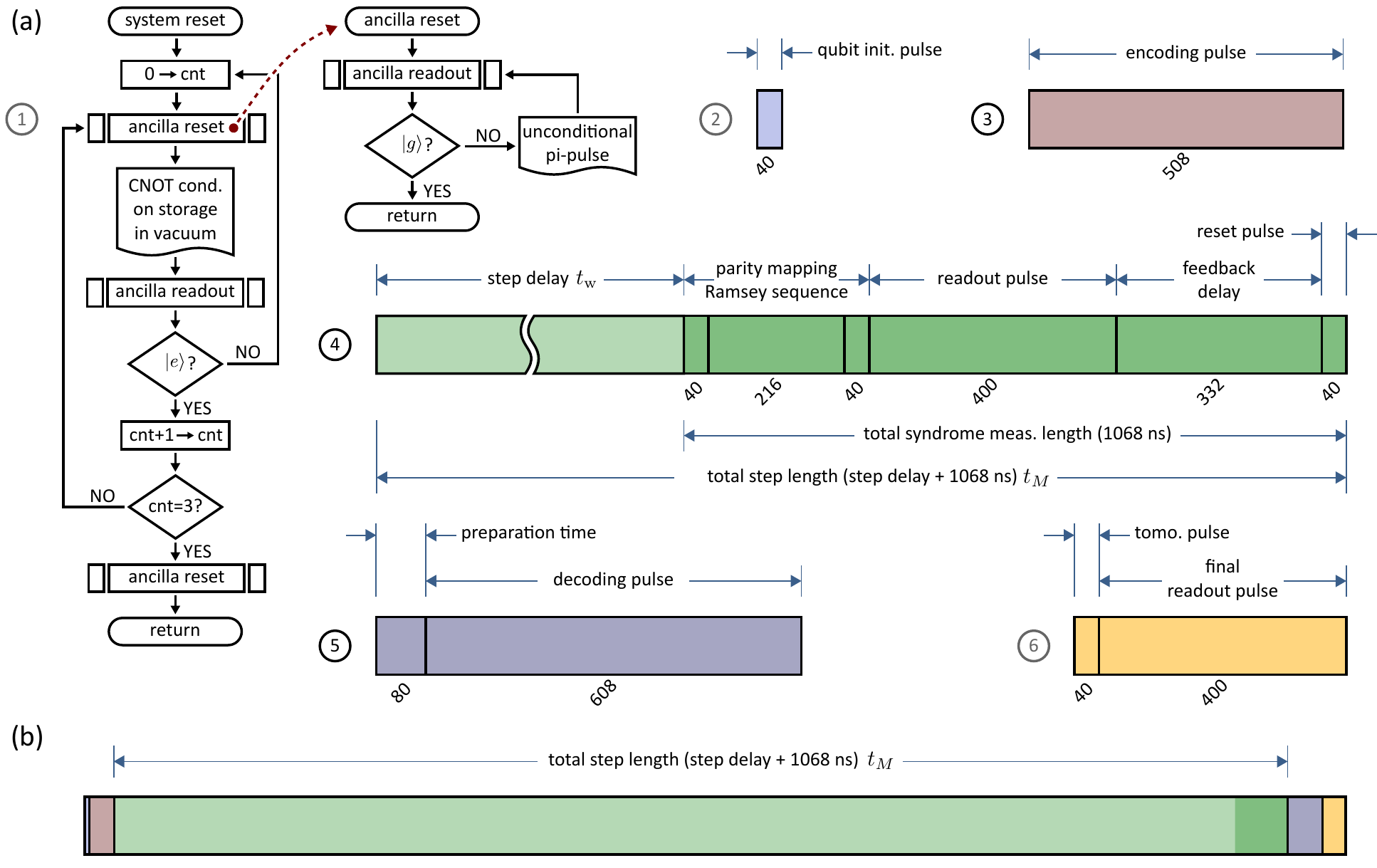}
\caption{\footnotesize
    \textbf{Experimental flow.  (a)} The six steps of the QEC protocol using a standard flow-chart convention.  1. System reset: we use a long selective $R_\pi^y$ pulse on the ancilla ($\sigma\sim600$ns) that addresses only the Fock state $\ket{0}_f$ of the resonator~\cite{Schuster:2007ki} to verify that it is indeed in the vacuum.  In order to boost our confidence, we require three consecutive verifications to trust the results (counter ``cnt" must be incremented from 0 to 3 for the process to continue).  We then perform the ancilla reset protocol by measuring it and applying a short pulse ($\sigma=2$ns) to return to to $\ket{g}$ if it is found to be in $\ket{e}$.  2. Ancilla initialization: we apply a short pulse ($\sigma=2$ns) to encode the ancilla into one of 6 cardinal points on the qubit Bloch sphere.  3. Encoding: an optimized control pulse of length 508ns transfers the quantum information from the ancilla to the resonator, leaving the ancilla in $\ket{g}$.  4. Parity monitoring: we repeat the adaptive parity monitoring protocol. The number of steps and their durations are optimized for each total duration (e.g. three steps for $54\mathrm{\mu s}$ as in Fig.~4 of the main text), which we specify at the beginning of the experiment.  Each monitoring step begins with a delay of some duration, followed by the right Ramsey-like sequence that maps the ancilla back to $\ket{g}$ if there was no photon jump during the delay.  We then readout the ancilla; if we find it in $\ket{e}$, we reset it as soon as possible. This happens $332$ns from the moment the readout pulse ends ($200\mathrm{ns}$ of FPGA calculation latency, plus experimental delays such as finite microwave cable lengths).  5. Decoding: after a short delay to finalize the estimation of current state in the resonator, the decoding pulse is chosen in real-time and is played with a best estimate of a corrected resonator phase to account for Kerr-induced deflection for non-zero error cases (see sec.~\ref{undesired_couplings}).  6. Qubit state tomography: measuring the ancilla after a pre-rotation to find the final qubit density matrix.  \textbf{(b)}  The whole protocol set to scale, shown to emphasize that we interrogate the system for only fraction of the entire sequence duration.}
	\label{fig:timings}
\end{figure}

The benefits of employing the adaptive protocol, depicted in the main text (Fig.~2), cannot be overstated.  Feedback latencies of just $\sim200\mathrm{ns}$ mean that the qubit spends just $\sim700\mathrm{ns}$ in $\ket{e}$ per error.  Without feedback, this time can be far greater, perhaps as much as $\sim50\mathrm{\mu s}$ per error in a $100\mathrm{\mu s}$-long experiment, effectively guaranteeing cat state dephasing with our ancilla coherence times (see sec.~\ref{forward_prop}).  As shown in Fig.~S\ref{fig:timings}, with the adaptive parity monitoring scheme and the ancilla reset described above, the full timeline of our measurement sequence is designed to have the ancilla in the ground state as much as possible.  We therefore regard the role of the quantum controller to be crucial to our goal of realizing a QEC scheme without the use of any post-selection or corrections for measurement inefficiencies.

\subsubsection{Recording Error Time}\label{record_error_time}

As described in section~\ref{undesired_couplings}, the resonator state $\ket{\psi(t)}$ is rotated by an angle $\theta=K_st_j$ in phase space every time a photon jumps at time $t_j$: $\ket{\psi(t)}\xrightarrow{\hat{a}_s}e^{i\theta\hat{a}_s^\dag\hat{a}_s}(\hat{a}_s\ket{\psi(t)})$.  When the difference in time between syndrome measurements $t_\mathrm{w}$ is non-zero, the uncertainty in jump time, which grows with increasing $t_\mathrm{w}$, leading to the aforementioned dephasing.  For $n$ total syndrome measurements spaced by $t_\mathrm{w}$ there is, however, still a known average angle of rotation for a jump that is measured at step $j$: $\bar{\theta}_j=K_s(j-1/2)t_\mathrm{w}$.  In other words, $K_s(j-1/2)t_\mathrm{w}$ is our best estimate of $t_j$ given the measurement cadence.  We can and must take this angle into account to prevent substantially greater excursions out of the logical subspace. 

In order to do so, our controller must record the step in the monitoring at which the photon jump occurs, or equivalently the time.  Then, in real-time it must apply a rotation to the coordinate system of the resonator's phase space by an angle $\bar{\theta}(k)=\sum_{j=1}^k\bar{\theta}_j$ for $k$ jumps so that the decoding pulse at the end of the sequence is applied correctly.  For $n$ monitoring steps there are $l=\frac{n!}{k!(n-k)!}$ different combinations of jump times for $k\leq n$ errors, and thus the controller must individually align all $l$ error trajectories that correspond to $k$ photon jumps on top of one another.  For example, when $n=3$ one can have $2^n=8$ different monitoring outcomes $(0\equiv \mathrm{``no~error"}$ and $1\equiv \mathrm{``error"})$: $000,100,010,001,110,101,011,111$; in this case, the feedback rotates $100$ by $\bar{\theta}_1=K_st_\mathrm{w}/2$, $010$ by $\bar{\theta}_2=3K_s t_\mathrm{w}/2$, and $001$ by $\bar{\theta}_3=5K_st_\mathrm{w}/2$ so all three can be decoded with a single pulse.  Employing this scheme increases the process fidelity of cases where jumps are detected by as much as $10-15\%$.

This aspect of the feedback highlights the complexity of the calculations that the controller does in real-time.  Furthermore, it can in principle handle an unlimited number of steps; as the number of combinations of jump times grows exponentially, it is a testament to the capability of the logic to efficiently perform and store the results of such calculations.  In the future, when measurement rates become much faster, this will be an indispensable feature.

\subsubsection{Adaptive Decoding}\label{adaptive_decoding}
Finally, based on the knowledge of how many errors occurred, the controller decides in real-time to apply one of two decoding pulses, depending on even or odd parity, to map the resonator state back onto the ancilla.  The necessity of these features stems from the fact that a unitary operation cannot map two initially orthogonal states onto a single final state (see sec.~\ref{grape}).  Although this feature is simple to implement, it is in some sense the most crucial; applying the wrong pulse does not disentangle the resonator and ancilla at the end of the decoding, leading to a complete incoherent qubit mixture when tracing over the resonator state.  

\section{Optimizing Cat Code Performance}\label{optimal_rate}

The presence of forward propagating errors in our system due to ancilla decoherence substantially alters the optimal strategy one normally seeks to employ in a QEC system.  Typically, the goal is to suppress the occurrence of errors within the codeword to second order with measurements performed at the maximum rate permitted by the parameters of the system.  With such a strategy, however, one necessarily entangles the logical states with the ancillary systems needed to extract the error syndromes for a substantial fraction of the QEC protocol's duration.  Since the rate of photon jumps in our system is much lower than $1/\tau_\mathrm{meas}$, the probability that two errors occur within $\tau_\mathrm{meas}$ is considerably lower than the probability of ancilla-induced decoherence.  Such a strategy thus results in a net-loss; indeed, by measuring as quickly as possible, we end up dephasing the qubit more quickly than had we encoded it in the resonator's Fock states $\ket{0}_f$, $\ket{1}_f$.  

We thus explore a different approach, one that instead slows down the syndrome measurement cadence to find an optimal balance between errors in the code and ancilla induced dephasing.  We take the point of view that experimentally our task is to preserve a quantum bit of information for a total time $T$.  The analytical model we present below then calculates the optimal measurement cadence and the predicted lifetime.  It takes into consideration the basic measured parameters in our system: resonator and ancilla coherence properties, thermal populations, and measurement fidelities.  The predictions we arrive at closely match the data we present in the main text.  Such results indicate that not only can we successfully optimize and employ a measurement strategy that preserves a qubit beyond the break-even point, but given only a basic set of assumptions about the sources of loss in the system, namely those outlined in sec.~\ref{losses}, we capture the dominant mechanisms that set the performance metrics of our QEC system.  Using this model, we can then predict the potential gains we expect to witness when certain key system parameters are enhanced.  In particular, by improving ancilla coherence times to levels of $\sim100\mathrm{\mu s}$ (well within the range of current transmon technology), the cat code promises to provide gains of over an order of magnitude.

At the moment of decoding, what we have available is a record of measurements. As a function of this classical information we then act on the system and decode the state to the best of our knowledge. There are two questions we can ask:
\begin{enumerate}
	\item [a.] For a given trajectory of photon jumps, what is the probability that the conclusion we obtained, based on the measurement record, is correct?
	\item [b.] What is the probability distribution of possible trajectories that may produce a given measurement record?   
\end{enumerate}

The first question relates to the optimization process. We wish to maximize the probability of correctly identifying the actual resonator error trajectory.  Trajectories with either many errors or consecutive errors result in a lower probability of success, as those events are rare and can be polluted by measurement errors.

The second question relates to our confidence in the output (see sec.~\ref{confidence}). There are different error trajectories that can produce the same measurement record. The best strategy is to simply choose the most probable one. Our confidence in the output is then the probability that this scenario occurred conditioned on the measured record. The output is thus not only the final state, but also a measure of confidence. We can either ignore this extra classical information and treat the whole process as a black box (red curve, Fig.~4a of the main text) or we can also use this information to post-select the data relative to some required confidence constraints (purple curve, Fig.~4a of the main text).

\subsection{The Optimized Configuration}\label{opt_config}
In this section we summarize our findings from an analytical model fully derived in section~\ref{optconfig_deriv}. For a fixed, desired time $T$ for which we would like to correct a qubit, we define a configuration to be the combination of the following parameters:

\begin{enumerate}
	\item The initial Cat size, $\nbar_0=\left|\alpha\right|^2$.
	\item The number of parity tracking steps $S$.
	\item The step durations $\{t_1,t_2,\dots,t_S\}$, where $\sum_{k=1}^St_k=T$.
\end{enumerate}

The process fidelity, $F_{process}$ decays exponentially from 1 to 1/4 (which is the process fidelity for a completely mixed final state). Here we derive and optimize the scaled version which decays from 1 to 0 at exactly the same rate, which we denote as the \emph{fidelity}:
\begin{equation}
F \equiv \frac{F_{process}-1/4}{3/4}.
\end{equation}
This is the probability we successfully corrected the state. We can write the total fidelity as a product of four terms:
\begin{equation}
F(T,\nbar_0,S,\left\{t_k\right\}_{t=1}^S)=F_{\Gamma_\uparrow}(T)\cdot F_{ED}(T,\nbar_0)\cdot F_T(T,\nbar_0,S,\left\{t_k\right\}_{t=1}^S)\cdot F_{KD}(T,\nbar_0,S,\left\{t_k\right\}_{t=1}^S).
\end{equation}

\begin{itemize}
	\item [$F_{\Gamma_\uparrow}$] Whenever the ancilla is excited to $\ket{e}$ we lose the encoded information (see sec.~\ref{forward_prop}). This affects our protocol and also the $\ket{0},\ket{1}$ Fock states encoding equally the same. It depends on $T$ alone and equals $e^{-T\cdot\Gamma_\uparrow}$.
	\item [$F_{ED}$] The fidelities of the encoding and decoding pulses depend on the initial and final cat sizes which are $\nbar_0$ and $\nbar_0 e^{-\kappa_s\,T}$. The non-orthogonality is simulated and taken into account numerically (see sec.~\ref{orthogonality}).
	\item [$F_T$] The loss of fidelity due to the monitoring itself depends on the ancilla's figures of merit throughout the time-scale of the parity mapping and projective measurement. It also depends on the cat size and $\kappa_s$ through the probability to miss photon jumps during a single step. 
	\item [$F_{KD} $] The uncertainty in the angle due to the Kerr deflection decreases the fidelity of the decoding pulse (see sec.~\ref{undesired_couplings}). We calculate the Kerr deflection distribution from the number of expected photon jumps and the step lengths together with the measured fidelity of the decoding pulse as a function of the angle (Fig.~S\ref{fig:Circuit_b}b).
\end{itemize}

If we ignore $F_{KD}$ we can show that the optimal fidelity can be written in the following form:
\begin{equation}
F^{OPT}(T,\nbar_0)=e^{-T \Gamma_\uparrow}\cdot F_{ED}(T,\nbar_0)\cdot e^{-\nbar_0[1-e^{-\kappa_s\,T}]/G},
\label{eq:opt}
\end{equation}
where $G$, the system gain, is a function of the other system parameters ($\chi_{sa}, T_1,\dots$), a constant of the system. When $\kappa_s T\ll 1$ we can approximate the optimized fidelity as:
\begin{equation}
F^{OPT}(T,\nbar_0)=e^{-T \Gamma_\uparrow}\cdot F_{ED}(0,\nbar_0)\cdot e^{-{\kappa_s\,T}\cdot \frac{\nbar_0}{G}},
\end{equation}
which shows that the decay rate of the quantum error corrected information is $G/\nbar_0$ slower compared to storage cavity decay rate $\kappa_s$. The process fidelity decay rate of the $\ket{0}_f,\ket{1}_f$ Fock state encoding decays $3/2$ slower than $\kappa_s$. The break-even condition is therefore:
\begin{equation}
\frac{2G}{3\nbar_0}>1.
\end{equation}

As we increase $\nbar_0$, the gain in lifetime decreases.  On the other hand, in order to have sufficient orthogonality between the logical basis states, $\nbar_0$ has to be high enough. Hence, an optimal $\nbar_0$ exists. Since eq. \ref{eq:opt} ignores $F_{KD}$, it expresses an upper bound for the fidelity, and thus $G$ has to be even larger in order to get an actual gain in lifetime. Better ancilla coherence times will increase the optimal measurement cadence and make $F_{KD}$ approach unity (see sec.~\ref{losses}). In this limit, eq. \ref{eq:opt} is exact and defines the limits of our scheme.

\begin{figure}[h!]
\centering
\includegraphics[width=6in]{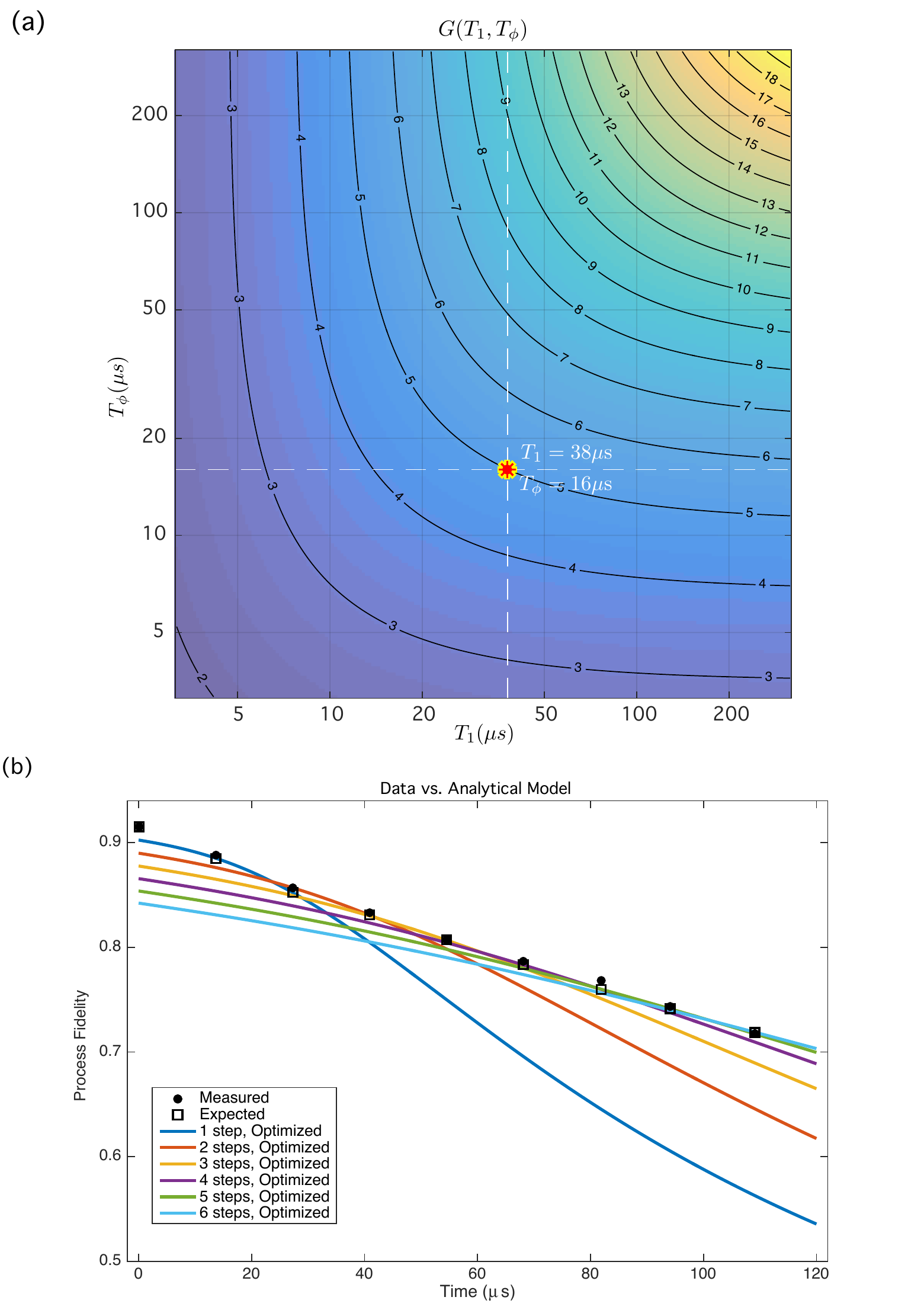}
\caption{\footnotesize
    \textbf{Simulation results.  (a)}  The gain $G$ as a function of $T_1$ and $T_\phi$ of the ancilla.  With the parameters of our ancilla we get $G=4.96$, and the predicted gain over the Fock state encoding is $1.65$.  This value is higher than the measured $10\%$ improvement since this plot does not include the effects of dephasing due to Kerr or the degradation of information due to overlapping logical basis states.  A key point is that with ancilla coherence on the order of $100\mathrm{\mu s}$, we already expect to see gains of an order of magnitude.  \textbf{(b)}  The optimal predicted process fidelity for our system given ancilla and resonator coherence times, and resonator Kerr (squares).  Comparing this simulation with the data (circles) shown in Fig.~4a of the main text, we find that our model faithfully predicts the measurement results at each point.  We also display the expected process fidelity we would have obtained had we fixed the number of steps, in colored curves.  Commensurate with the top axis Fig.~4a of the main text, we chose optimal configuration for each of the total time durations.}
	\label{fig:analyticalopt}
\end{figure}

Figure S\ref{fig:analyticalopt}a shows how $G$ depends on $T_1$ and $T_\phi$ of the ancilla. With our ancilla's coherence times, $G$ is about 5. With $\nbar_0=2$, the ratio $2G/3\nbar_0$ equals 1.65. The actual gain is lower, since we need to take into account the effect of the Kerr deflection and the degradation of the decoding pulse due to loss of orthogonality. The decay due to $\Gamma_\uparrow$ affects both the cat code and the $\ket{0}_f,\ket{1}_f$ Fock state encoding and also lowers the gain in measured lifetimes. 

We can also optimize the configuration when fixing the number of steps and taking into account both $F_{KD}$ and $F_{ED}$. Figure S\ref{fig:analyticalopt}b displays the measured fidelity and our model's prediction for the same configurations. It also displays the optimal expectations when forcing different numbers of steps. Our model predicts the measured data accurately and lets us find the optimal configuration as a function of the total duration $T$.  We end up with a total predicted gain of 10\%, in line with the results in Fig.~4 of the main text.

\subsection{Deriving the Optimized Configuration}\label{optconfig_deriv}
Here we derive the optimal fidelity $F$ assuming $F_{KD}=1$. We follow a pessimistic approach. Any error that may happen will be regarded as a total loss of information. An error in the parity measurement could, in principle, be corrected by repeating it several times and taking a majority vote. In practice, although our measurement fidelities are very high, the errors that these extra measurements introduce are larger than the those we wish to correct for, so it is better to blindly trust any result. This approach simplifies the analysis greatly, and we can calculate the probability to successfully keep the information for each step independently and take the product to get the final total fidelity.

A successful step starts with the ancilla in $\ket{g}$, followed by either no jumps or a single photon jump during the delay, an accurate parity measurement and the ancilla back in $\ket{g}$ at the end. There are several failure mechanisms:

\begin{enumerate}
	\item While waiting, the ancilla may have been excited to $\ket{e}$
	\item Two or more photon jumps occurred during the step delay
	\item The parity measurement returned the wrong answer
	\item A successful parity measurement brought the ancilla to $\ket{e}$, but the following reset pulse failed to return it back to $\ket{g}$
\end{enumerate}

The probabilities to have zero, one, or more jumps are a function of the cat size at the beginning of the step and the step length. We express the step success probability as:
\begin{equation}
F_k = e^{-t_k \Gamma_\uparrow}\left[P_k(0)\cdot f_0+P_k(1)\cdot f_1\right],
\end{equation}
where $P_k(n)$ is the probability to have $n$ photon jumps in the $k$'th step, $f_0$ ($f_1$) is the conditional success probability when no (a single) photon jump occurred. The final success probability is then:
\begin{equation}
F/F_{ED} = \prod_{k=1}^Se^{-t_k \Gamma_\uparrow}\left[P_k(0)\cdot f_0+P_k(1)\cdot f_1\right]= \underbrace{e^{-T \Gamma_\uparrow}}_{F_{\Gamma_\uparrow}}\cdot\underbrace{\prod_{k=1}^S\left[P_k(0)\cdot f_0+P_k(1)\cdot f_1\right]}_{F_T}.
\end{equation}
From this point onward we focus only on $F_T$.

The success of the parity measurement depends primarily on the ancilla's $T_2$. On top of that, there is the readout fidelity (which is different for the ground and the excited states). When a single photon jump occurs, the ancilla ends up in $\ket{e}$. It may decay back to $\ket{g}$ before the reset pulse, which means that the reset pulse inadvertently returns it back to $\ket{e}$. This is a critical period of time where we are vulnerable to $T_1$ decay of the ancilla:
\begin{eqnarray}
f_0&\approx&e^{-\frac{\pi}{\chi_{sa} T_2}}\cdot M_{gg}\\
f_1&\approx&e^{-\frac{\pi}{\chi_{sa} T_2}}\cdot M_{ee}\cdot e^{-\frac{\tau_{\mathrm{meas}}+T_{_{FB}}}{T_1}},
\end{eqnarray} 
where $M_{gg}$ ($M_{ee}$) is the probability to measure correctly $\ket{g}$ ($\ket{e}$), $\tau_{\mathrm{meas}} = 400$ns is the readout pulse length and $T_{_{FB}}=332$ns is the feedback latency that includes delays due to the experimental setup (cables, etc.). Ancilla $T_1$ decay causes code failure no matter when it happens, which is why we take into account the whole duration until the ancilla is back in $\ket{g}$.

At the beginning of the $k^{th}$ step, the averaged photon number in the cavity is given by ($\nbar_0$ is the cat size at the beginning of the $k=1$ step):
\begin{equation}
\nbar_{k-1}=\nbar_0\cdot \exp\left(-\kappa_s \sum_{i=1}^{k-1}t_i\right).
\end{equation}

During the step delay, the number of photon jumps has a Poissonian distribution with the following mean value (the decay during the step itself is taken into account):
\begin{equation}
\lambda_k=\nbar_{k-1}\cdot\left[1-e^{-\kappa_s t_i}\right].
\end{equation} 

We expect no photon jumps with probability $e^{-\lambda_k}$ and a single jump with probability $\lambda_k e^{-\lambda_k}$. Hence, the step fidelity is simply given by:

\begin{eqnarray}
F_T^{(k)} &=& f_0\, e^{-\lambda_k} + f_1\, \lambda_k e^{-\lambda_k}=\left(f_0+f_1\,\lambda_k\right)e^{-\lambda_k}\\
\nonumber&\downarrow&\\
F_T\left(T,\nbar_0,S,\left\{t_k\right\}_{t=1}^S
\right)&=&\prod_{k=1}^S\left[\left(f_0+f_1\,\lambda_k\right)e^{-\lambda_k}\right]
\label{eq:F_tilde_total}
\end{eqnarray}

We can now optimize the step length for a fixed number of steps. For $S=1$, a single step, the solution is forced to be $t_1=T$. For two steps we will need to optimize the following expression:

\begin{eqnarray}
\max_{t_1} F_T\left(T,\nbar_0,S,\left\{t_1,t_2=T-t_1\right\}
\right) &=&\left[\left(f_0+f_1\,\lambda_1\right)e^{-\lambda_1}\right]\cdot\left[\left(f_0+f_1\,\lambda_2\right)e^{-\lambda_2}\right]\\
&=&\left(f_0+f_1\,\lambda_1\right)\cdot\left(f_0+f_1\,\lambda_2\right)e^{-(\lambda_1+\lambda_2)} \\
&=&\left(f_0+f_1\,\lambda_1\right)\cdot\left(f_0+f_1\,\lambda_2\right)e^{-(1-e^{-\kappa_s T})}
\end{eqnarray}
The expected number of photon jumps during the two steps sums up to the expected number of jumps during the whole duration. This is independent of how we partition the whole duration into two steps. We can simply maximize the multiplication of  $\left(f_0+f_1\,\lambda_1\right)$ and $\left(f_0+f_1\,\lambda_2\right)$. Since the sum of these terms is constant, the maximum is achieved when they are equal, meaning $\lambda_1=\lambda_2$. In other words, we need to maintain a constant rate of photon jumps between the steps. This will be true for any number of steps; for $S$ steps the mean number of photon jumps per step is given by: 
\begin{equation}
\lambda_k=\frac{\nbar_0\left[1-e^{-\kappa_s T}\right]}{S}.
\end{equation}

Substituting this expression into eq. \ref{eq:F_tilde_total} and using the optimal step lengths, we obtain:
\begin{eqnarray}
F_T\left(T,\nbar_0,S\right) &=&\left[\left(f_0+f_1\,\frac{\nbar_0\left[1-e^{-\kappa_s T}\right]}{S}\right)\cdot e^{-{\nbar_0\left[1-e^{-\kappa_s T}\right]}/{S}}\right]^S\\
&=&\underbrace{\left(f_0+f_1\,\frac{\nbar_0\left[1-e^{-\kappa_s T}\right]}{S}\right)^S}_{\equiv F'_T}\cdot e^{-\nbar_0\left[1-e^{-\kappa_s T}\right]}
\end{eqnarray}
As an exercise, we can see how this expression behaves for $S$ much greater than the number of expected photon jumps during the whole duration, $S\gg\nbar_0[1-e^{-\kappa T}]$:
\begin{eqnarray}
\nonumber F_T(\nbar_0,T,S)&=&\lim_{S\rightarrow\infty}f_0^S   \left(1+\frac{f_1\nbar_0[1-e^{-\kappa_s T}]}{f_0\,S}\right)^S\,e^{-\nbar_0[1-e^{-\kappa_s T}]} \\
\nonumber &\approx&f_0^S\,e^{\nbar_0[1-e^{-\kappa_s T}]\frac{f_1}{f_0}} \,e^{-\nbar_0[1-e^{-\kappa_s T}]} \\
\nonumber&=&f_0^S\,{e^{-(1-f_1/f_0)\nbar_0[1-e^{-\kappa_s T}]} } \\
\nonumber&\approx&f_0^S \left(\frac{f_1}{f_0}\right)^{\nbar_0[1-e^{-\kappa_s T}]} \\
&=&f_0^{S-\nbar_0[1-e^{-\kappa_s T}]}\,f_1^{\nbar_0[1-e^{-\kappa_s T}]}
\end{eqnarray} 
What this limit means is that when we measure frequently enough, the fidelity will fall off by a factor $f_0$ for any steps where no jumps happened and by $f_1$ when a single jump occurs. As the step size is so short, errors due to double photon jumps are negligible and therefore excluded.

We continue with $F'_T$:
\begin{eqnarray}
n_j&\equiv&\nbar_0\left[1-e^{-\kappa_s T}\right]\cdot\frac{f_1}{f_0}\\
\nonumber&\downarrow&\\
F'_T\left(T,\nbar_0,S\right)&=&f_0^S\left(1+\frac{n_j}{S}\right)^S,
\end{eqnarray}
where $n_j$ is the number of expected photon jumps during the whole duration up to a scale factor of order 1. We will treat the number of steps as a continuous variable and find its optimum. In practice, we will use the closest integer:

\begin{eqnarray}
\nonumber \frac{d}{dS}F'_T&=&F'_T\cdot\frac{d}{dS}\left[S\cdot \log(f_0)+S\cdot \log\left(1+\frac{n_j}{S}\right)\right] \\
\nonumber &=&F'_T \cdot \left[ \log(f_0)+\log\left(1+\frac{n_j}{S}\right)+\frac{S}{\left(1+\frac{n_j}{S}\right)}\cdot\frac{-n_j}{S^2}\right] \\
&=&F'_T \cdot \left[ \log(f_0)+\log\left(1+\frac{n_j}{S}\right)-\frac{n_j}{S+n_j}\right]\mathop{=}^{want}0
\end{eqnarray}

We need to solve:
\begin{eqnarray}
\log(f_0)+\log\left(1+\frac{n_j}{S}\right)-\frac{n_j}{S+n_j}&=&0\\
\nonumber &\downarrow& \\
\log(f_0)+\log\left(1+\frac1r\right)-\frac{1}{1+r}&=&0
\label{eq:u}
\end{eqnarray}
where $r\equiv S/n_j$ (the number of steps per photon jump up to small correction). Although we cannot solve for $r$ explicitly, it is a function of $f_0$, the success probability of the parity measurement conditioned on no photon jumps ($r(f_0)$). The optimal number of steps is then:
\begin{equation}
S^{OPT}=r(f_0)\cdot n_j=\frac{r(f_0)\cdot f_1}{f_0}\nbar_0\left[1-e^{-\kappa_s T}\right],
\end{equation}
leading to the optimal average number of photon jumps per step:
\begin{equation}
\lambda_k=\frac{\nbar_0\left[1-e^{-\kappa_s T}\right]}{S^{OPT}}=\frac{f_0}{r(f_0)\cdot f_1}.
\end{equation}

In the limit $f_0\rightarrow1$, $r(f_0)$ approaches infinity. We can expand eq. \ref{eq:u} in $1/r$:

\begin{eqnarray}
\nonumber \log(f_0)&=&\frac{1}{1+r}-\log\left(1+\frac{1}{r}\right)\\
\nonumber \log(1-[1-f_0])&=&\frac{1}{r}\cdot\frac{1}{1+\frac1r}-\left[\frac1r-\frac1{2r^2}+\cdots\right]\\
\nonumber  -[1-f_0]-\frac12[1-f_0]^2-\cdots&=& \frac1r-\frac1{r^2}+\cdots -\left[\frac1r-\frac1{2r^2}+\cdots\right]\\
1-f_0&\approx&\frac{1}{2r^2}
\end{eqnarray}
For $f_0,f_1\sim1$, the optimized average number of jumps per step is simply $1/r$.  Hence, $1/2r^2$ is the Poissonian probability to have two jumps. The last approximation states that in this limit we need to match the parity measurement infidelity with the double-jump probability. As long as $f_0$ is above 90\%, this approximation will be correct within 50\% of the right value.

We can now substitute the optimal steps number and get the maximal success probability as a function of the total duration and initial cat size:
\begin{eqnarray}
\nonumber {F'_T}{}^{OPT}(T,\nbar_0)\equiv F'_T (T,\nbar_0,S^{OPT})&=&\left[f_0\left(1+\frac{n_j}{S^{OPT}}\right)\right]^{S^{OPT}} \\
\nonumber \log\left({F'_T}{}^{OPT}\right)&=&S^{OPT}\underbrace{\left[\log(f_0)+\log\left(1+\frac1r\right)\right]}_{=\frac1{1+r}}\\
&=&\frac{r(f_0)}{1+r(f_0)}\frac{f_1}{f_0}\cdot \nbar_0[1-e^{-\kappa_s T}]\\
\nonumber&\downarrow&\\
F_T{}^{OPT}(T,\nbar_0)= {F'_T}{}^{OPT}(T,\nbar_0)\cdot e^{-\nbar_0[1-e^{-\kappa_s T}]}&=&  e^{+\nbar_0[1-e^{-\kappa_s T}]\frac{r(f_0)}{1+r(f_0)}\frac{f_1}{f_0}}\cdot e^{-\nbar_0[1-e^{-\kappa_s T}]}.
\end{eqnarray}
The second term is the uncorrected cat state decay, which is $\nbar_0$ faster than $\kappa_s$. The first term counteracts this decay, and so we identify it as the action of the QEC. 

We define the unit-less parameter $G$ as follows:
\begin{eqnarray}
G&\equiv& \frac{1}{1-\frac{f_1}{f_0}\cdot\frac{r(f_0)}{1+r(f_0)}}\\
\nonumber&\downarrow&\\
F_T{}^{OPT}&=& e^{-\frac{\nbar_0[1-e^{-\kappa_s T}]}{G}}\mathop{\approx}_{\kappa_s T\ll1}e^{-\kappa_s T\cdot\frac{\nbar_0 }{G}}.
\end{eqnarray}
In other words, we slow down the decay of the cat state by a factor of $G$.  This number is a constant of the system, a function of the various infidelities. It approaches infinity as $f_0,f_1$ get closer to 1. 



\section{QEC System Process Tomography}\label{process_tomo}
In this section we focus on the details of the process fidelity results presented in the main text.  In addition to expanding upon the calculations used to characterize the process of the QEC system, we address the consequences of post-selection and what it allows us to deduce about the confidence of a given measurement trajectory.
\subsection{Quantifying the Process Fidelity}\label{proc_fid_calc}
\subsubsection{The Process Matrix}

We seek to characterize the entire process of the QEC system on a qubit.  The density matrix of the final state, $\rho_\mathrm{fin}=\ket{\psi_\mathrm{fin}}\bra{\psi_\mathrm{fin}}$, is the output of the entire QEC system process (see sec.~\ref{exptflow}) $\mathcal{E}(\rho_\mathrm{init})$: $\rho_\mathrm{fin}=\mathcal{E}(\rho_\mathrm{init})$, where $\rho_\mathrm{init}=\ket{\psi_\mathrm{init}}\bra{\psi_\mathrm{init}}$.  Ideally, $\mathcal{E}(\rho_\mathrm{init})=\rho_\mathrm{init}$, where the process is simply given by the identity operator $\hat{I}$, corresponding to perfect error correction.  In reality, however, due to decoherence in conjunction with experimental imperfections, $\mathcal{E}(\rho_\mathrm{init})$ is a combination of non-unitary and unitary operations on the encoded state.

In order to characterize the full process $\mathcal{E}(\rho_\mathrm{init})$, we find $\rho_\mathrm{fin}$ by performing state tomography of the ancilla following the correction step to measure the components of the final qubit Bloch vector $\vec{r}=\{r_x,r_y,r_z\}$: $\rho_\mathrm{fin}=(\hat{I}+r_x\hat{\sigma}_x+r_y\hat{\sigma}_y+r_z\hat{\sigma}_z)/2$, where $\hat{\sigma}_x,\hat{\sigma}_y,\hat{\sigma}_z$ are the single qubit Pauli operators.  The results allow us to represent $\mathcal{E}(\rho_\mathrm{init})$ in the chi ($X$) matrix representation using the operator-sum notation~\cite{NielsenQI}: $\mathcal{E}(\rho_\mathrm{init})=\sum_{jk}\tilde{E}_j\rho_\mathrm{init}\tilde{E}_k^\dag X_{jk}$, where for a single qubit $\tilde{E}_0=\hat{I},\tilde{E}_1=\hat{\sigma}_x,\tilde{E}_2=-i\hat{\sigma}_y,\tilde{E}_3=\hat{\sigma}_z$ and the coefficients $X_{jk}$ comprise the process matrix $X$.  This is a complex $4\times4$ matrix of trace $\mathrm{Tr}(X)=1$ that completely describes the action of our QEC system on an arbitrary input state.  We define the fidelity $F$ to be the overlap of the measured chi matrix, $X^M$, with $X_0$, the ideal identity process: $F=\mathrm{Tr}(X^MX_0)$.  In principle, only four cardinal points are needed to determine $X^M$, the two at the poles of the Bloch sphere ($+\vec{z}$, $-\vec{z}$) and those along $\hat{\sigma}_x$ ($+\vec{x}$) and $\hat{\sigma}_y$ ($+\vec{y}$).  Following the derivation presented in~\cite{NielsenQI}, we can also find a simple formula for the $(0,0)$ entry of $X^M$, $X^M_{00}$, which is equivalent to the expression above for the fidelity to the identity process.  It requires the results of qubit state tomography, $\vec{r}^{~\hat{n}}=\{r_x^{\hat{n}},r_y^{\hat{n}},r_z^{\hat{n}}\}$, for the four cardinal points ($\hat{n}=+\vec{x},+\vec{y},+\vec{z},-\vec{z}$):
\begin{align}\label{form:entanglement_fid}
X_{00}& = \frac{1}{4}(1+(r_x^{+\vec{x}}-\frac{r_x^{+\vec{z}}+r_x^{-\vec{z}}}{2})+(r_y^{+\vec{y}}-\frac{r_y^{+\vec{z}}+r_y^{-\vec{z}}}{2})+\frac{r_z^{+\vec{z}}-r_z^{-\vec{z}}}{2})
\end{align}
We perform these calculations with both $+\vec{x}$, $+\vec{y}$ and $-\vec{x}$,$-\vec{y}$, however, to verify that there are no unexpected asymmetries in the cat code.  Indeed we find identical process fidelities regardless of whether or not we rotate the coordinate system by $\pi$.

\subsubsection{Maximizing the Fidelity in Software}
Given that the optimal control encoding and decoding pulses do not realize the intended unitary perfectly at each time step, there could be some overall unintended rotation of the final qubit.  Following the approach in~\cite{Schindler:2011ch}, we allow for one and the same rotation to be applied in software to all six cardinal points simultaneously that maximizes the overlap $\mathrm{Tr}(X^MX_0)$.  This is a simple change of reference frame that in no way compensates for measurement infidelity or an artificial enhancement of performance.  It is equivalent to applying a fixed pre-determined unitary operation on the decoded qubit to adjust its orientation that most closely matches that of the initial state.  In practice, rotations do not exceed several degrees around each axis and have only a small effect on the reported results.

\subsubsection{Depolarization Errors}\label{depolarization}
\begin{figure*}[!ht]
\centering
\includegraphics[width=7.2in]{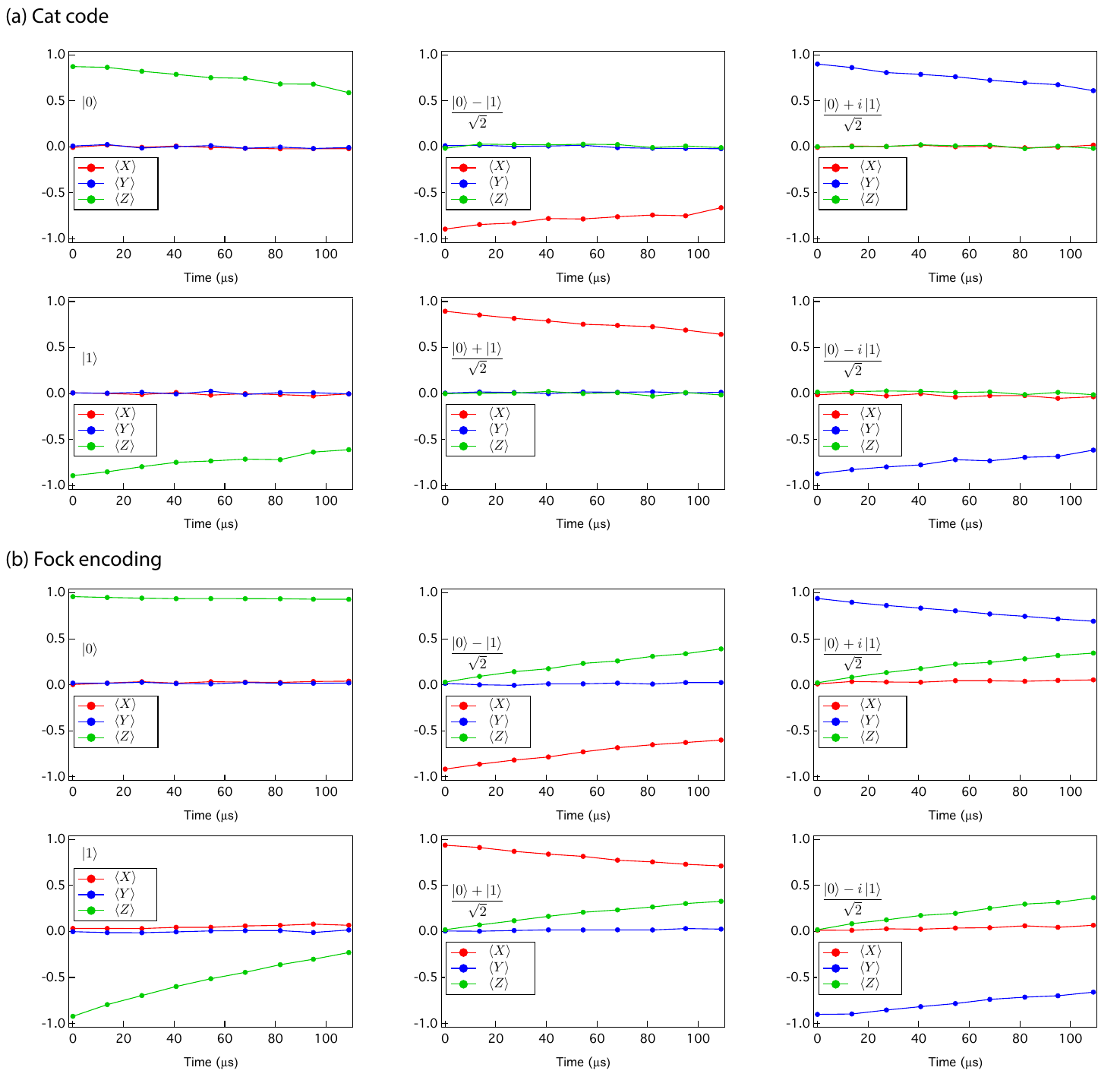}
\caption{\footnotesize
    \textbf{Qubit depolarization.  (a)}  This data demonstrates that depolarization, in which every Bloch vector shrinks uniformly toward a fully mixed state at the origin, is the dominant error channel in our system.  The six plots show the decay in time of the average $X$, $Y$, and $Z$ components of the qubit Bloch vector ($\braket{X}$, $\braket{Y}$, $\braket{Z}$ respectively) for each cardinal point after using the cat code QEC system.  These six points are initialized by applying the identity ($I$), $\pi$ pulse about the $Y$ axis ($R^y_\pi$), or $\pm\pi/2$ rotations about the $Y$ or $X$ axes ($R^y_{\pi/2}$,$R^y_{-\pi/2}$,$R^x_{\pi/2}$,$R^x_{-\pi/2}$) on the ancilla prior to the error monitoring (see sec.~\ref{exptflow}).  This data is used to calculate the process matrix $X^M$ of the corrected qubit in Fig.~4a of the main text and produce the images in Fig.~4b.  In each of the six cases, only the non-zero coordinate of the Bloch vector at zero time decays while the other two remain at $0$ throughout the entire tracking duration.  We find the decay rate of cat states along $\ket{\pm X^+_L}$ to be slightly more robust as these states are symmetric about both axes in the resonator's phase space, while $\ket{C_{\alpha}^+}$, $\ket{C_{i\alpha}^+}$ and $\ket{\pm Y^+_L}$ are symmetric about only one.  Thus, rotations in phase space are somewhat less detrimental for $\ket{\pm X^+_L}$.  \textbf{(b)}  In contrast, the Fock state encoding shows decay curves typical of amplitude damping, or $T_1$-type decoherence, in which all coordinates preferentially decay to the energetically favorable resonator ground state $\ket{0}_f$.  One can see in each plot that the value of the $\braket{Z}$ coordinate monotonically increases towards (or stays at) $+1$ regardless of the initial state, while every other coordinate decays to $0$.}
\label{fig:depolarization}
\end{figure*}

As depicted in Fig.~4b of the main text and Fig.~S\ref{fig:depolarization}a, while monitoring the error syndrome of the cat code for longer times we find that all of the cardinal points on the qubit Bloch sphere shrink uniformly towards the fully mixed state $\rho_\mathrm{fin}=\hat{I}/2$.  This feature is characteristic of a depolarization error channel~\cite{NielsenQI}:
\begin{align}
\mathcal{E}(\rho)&=\frac{p\hat{I}}{2}+(1-p)\rho\\
&=(1-\frac{3p}{4})\rho+\frac{p}{4}(\hat{\sigma}_x\rho\hat{\sigma}_x+\hat{\sigma}_y\rho\hat{\sigma}_y+\hat{\sigma}_z\rho\hat{\sigma}_z)\\
&=\sum_{j}\tilde{E}_j\rho\tilde{E}_j^\dag X_{jj}\\
&\rightarrow X=
\begin{pmatrix}
1-\frac{3p}{4}&0&0&0\\
0&\frac{p}{4}&0&0\\
0&0&\frac{p}{4}&0\\
0&0&0&\frac{p}{4}\\
\end{pmatrix}
\end{align}

This simple formula shows that the signature of depolarization errors is a diagonal process matrix $X$ in which the value of $X_{00}$ decreases with $p$, while the remaining diagonal components increase with $p$, for $p$ a probability between $0$ and $1$.  As the data in Fig.~S\ref{fig:depolarization}a and Fig.~4a of the main text demonstrates, for the cat code of initial size $\bar{n}_0=2$ we have a $X^M_{00}(t)=1-3p(t)/4=1/4+3e^{-t/\tau_{qec}}/4$, where $\tau_{qec}\approx 320\mathrm{\mu s}$.  Equivalently, each term in eq.~\ref{form:entanglement_fid} decays with the time constant $\tau_{qec}$ as $X_{00}$ asymptotes to $0.25$.

The dominant source of depolarization stems from the incorrect application of the decoding pulse at the end of the QEC sequence due to either the forward propagation of errors or an incorrect knowledge of the number of errors that has occurred.  If the decoding pulse is applied at either the wrong angle (quantified in sec.~\ref{leakage}) or attempts to decode the wrong parity (see sec.~\ref{grape_implementation}), at the end of the decoding sequence we are left with a completely mixed qubit after tracing over the resonator state.  Although at first glance the basis states along the logical $Z$ axis, 2-cats, should be immune to double-errors within the waiting time between syndrome measurements, the resulting Kerr deflection after two photon jumps is sufficient to appreciably rotate the cat states out of the logical space.  Thus qubit $T_1$ decay, errant syndrome mapping, as well as double jumps substantially degrade the efficacy of the decoding pulse in faithfully transferring the quantum information from the resonator back to the ancilla.

In contrast, both the Fock state and the transmon encodings are susceptible to generalized amplitude damping, which is given by the following process~\cite{NielsenQI}:
\begin{align}
\mathcal{E}(\rho) &= E_0\rho E_0^\dag + E_1\rho E_1^\dag + E_2\rho E_2^\dag + E_3\rho E_3^\dag\\
&\rightarrow E_0=\sqrt{1-n_{th}}
\begin{pmatrix}
1&0\\
0&\sqrt{1-f(t)}
\end{pmatrix}\\
&\rightarrow E_1=\sqrt{1-n_{th}}
\begin{pmatrix}
0&\sqrt{f(t)}\\
0&0,
\end{pmatrix}\\
&\rightarrow E_2=\sqrt{n_{th}}
\begin{pmatrix}
\sqrt{1-f(t)}&0\\
0&1
\end{pmatrix}\\
&\rightarrow E_3=\sqrt{n_{th}}
\begin{pmatrix}
0&0\\
\sqrt{f(t)}&0,
\end{pmatrix}
\end{align}
where $f(t)$ is a function of time of the form $f(t)=1-e^{t/t_0}$; for the transmon $t_0=T_1$ and $n_{th}=n^a_{th}$; for the Fock state $t_0=\tau_s$ and $n_{th}=n^s_{th}$ (see sec.~\ref{fidelities}).  As seen in Fig.~S\ref{fig:depolarization}b, all Bloch sphere vectors preferentially decay toward the ground state of the codeword.  The off-diagonal elements in the density matrices of these encodings decay with a time constant that in addition to the amplitude damping also includes the pure dephasing in the system; these combined rates are $T_2$ for the transmon and $T_2^s$ for the Fock state components.  Thus, we find that $r_x^{+\vec{x}}$ and $r_y^{+\vec{y}}$ in eq. ~\ref{form:entanglement_fid} decay at $T_2^s$ ($T_2$) for the Fock state (transmon) encodings, while $(r_z^{+\vec{z}}-r_z^{-\vec{z}})/2$ decays at $\tau_s$ ($T_1$).  The process fidelity as a function of time $X^M_{00}(t)$ therefore decays at two different rates, resulting in a double-exponential behavior.  The single decay times reported in the main text (Fig.~4a) for each encoding strategy are simply the harmonic mean of the coherence times given in Table~\ref{coherence}, $\tau=3/(1/T_1+2/T_2)$.  Visually one can only discern this trend in the decay of the transmon fidelity, as the time constant of the Fock state encoding is too high to see a double-exponential for the comparatively short time scales in the plot.


\subsection{Error Record vs. Error History}

\subsubsection{Confidence Level Estimation}\label{confidence}
With the calculations below, we expand upon why the process fidelity of the $2$-error case after two syndrome measurements, shown in Fig.~3 of the main text, differs substantially from that of the $0$ and $1$ error cases.  We invoke Bayes' rule and the measured fidelity of successfully mapping parity in the presence of $\bar{n}=3$ photons in the cavity: $97.7\%$.  Although very high, this number still leads to the somewhat surprising differences in the confidence of certain measurement results over others.  It becomes clear that measurements of ``error" confirmed by subsequent measurements of ``no error" have a $\sim10\%$ higher fidelity than those without such a confirmation.  Moreover, if two consecutive ``error" measurements are recorded, the probability drops substantially by $\sim20-30\%$.  With these findings, many of the features in the data fall into place.  One can see the effects of these conditional probabilities by looking at the 1-error cases in the Wigner tomography in Fig.~3 of the main text, where the parity and fringe contrast of the $01$ case appear to be less negative and sharp than that of $10$( $0\equiv \mathrm{``no~error"}$ and $1\equiv \mathrm{``error"}$).  Along the same lines, the case $11$ has by far the lowest fidelity, as confirmed by the process tomography in Fig.~3e.  The single-shot records that come with each repetition of the monitoring sequence thus provide us with crucial information beyond simply how to bin each result.  Indeed, if one can tolerate certain levels of post-selection, by removing the trajectories with the lowest confidence we enhance the qubit lifetime by nearly a factor of two over the Fock state encoding.

We start by assuming that we have encoded a qubit in cat states of size $\bar{n}=3$ and that the ancilla is in $\ket{g}$ prior to the parity mapping.  This is the system state after Fig.~3a in the main text.  After a round of error monitoring where we use the parity protocol that maps even parity to $\ket{g}$ and odd parity to $\ket{e}$ (indicated by the superscript ``$-$"; see sec.~\ref{smart_tracking}), the probabilities to measure ancilla $\ket{g}$, $\ket{e}$ are given by:
\begin{align}
p(g)=&p^-(g|0\varepsilon)p(0\varepsilon)+p^-(g|1\varepsilon)p(1\varepsilon)\\
p(e)=&p^-(e|0\varepsilon)p(0\varepsilon)+p^-(e|1\varepsilon)p(1\varepsilon),
\end{align}
where $p(0\varepsilon)=e^{-(\bar{n}e^{-t/\tau_s})t_\mathrm{w}/\tau_s}$ is the probability that the resonator state had $0$ parity jumps, $p(1\varepsilon)=1-p(0\varepsilon)$, and $p^-(g|0\varepsilon)$ and $p^-(e|1\varepsilon)$ are respectively the probabilities to measure $\ket{g}$ when the resonator state had $0$ parity jumps and $\ket{e}$ when the resonator had $1$ parity jump. Likewise, when we use the parity protocol that maps odd parity to $\ket{g}$ and even parity to $\ket{e}$ (indicated by the superscript ``$+$"), the probabilities to measure ancilla $\ket{g}$, $\ket{e}$ are given by:
\begin{align}
p(g)=&p^+(g|0\varepsilon)p(0\varepsilon)+p^+(g|1\varepsilon)p(1\varepsilon)\\
p(e)=&p^+(e|0\varepsilon)p(0\varepsilon)+p^+(e|1\varepsilon)p(1\varepsilon),
\end{align}
where $p^+(g|0\varepsilon)$ and $p^+(e|1\varepsilon)$ are respectively the probabilities to measure $\ket{g}$ when the resonator had $0$ parity jumps and $\ket{e}$ when the resonator had $1$ parity jump.  In our system, for average photon number $\bar{n}=3$ in the resonator, $p^-(g|0\varepsilon)=p^+(g|0\varepsilon)=0.983$ and $p^+(e|1\varepsilon)=p^-(e|1\varepsilon)=0.971$.

We seek to predict the statistics for monitoring errors for two steps and an initial cat size of average photon number $\bar{n}=3$, as presented in Fig.~3e of the main text.  Following the flow of Fig.~3a-d, assuming at time $t=0$ we start with an even parity state in the resonator and perform the first round of error monitoring that maps even parity to $\ket{g}$, after $13.8\mathrm{\mu s}$ $p(g)=0.841$ and $p(e)=0.159$.  Using Bayes' rule, we can now calculate the conditional probabilities for the resonator to be in a certain parity state given the measurement outcome:
\begin{align}
p^-(0\varepsilon|g)=&\frac{p^-(g|0\varepsilon)p(0\varepsilon)}{p(g)}=0.995\\
p^-(1\varepsilon|e)=&\frac{p^-(e|1\varepsilon)p(1\varepsilon)}{p(e)}=0.910
\end{align}
The key point here is the difference in the confidence as to the true occurrence of an error when the ancilla ends up in $\ket{e}$.  The small measurement infidelities together with the relatively low probability to have an error in the first place leads to a considerable difference of $\sim8\%$ between the two conditional probabilities $p^-(0\varepsilon|g)$ and $p^-(1\varepsilon|e)$.  This difference leads to a higher likelihood for the parity meter to suggest the occurrence of another error in the encoded state in the subsequent measurement, and thus leads to a substantially lower confidence in any trajectory that indicates consecutive errors, as we show next.

If we measure $\ket{g}$, we continue using the same protocol, but now to obtain $p^{\pm}(g)$ and $p^{\pm}(e)$ (probabilities to measure $\ket{g}$ and $\ket{e}$ for the two different parity mapping protocols) we no longer have the luxury of knowing that we start in an even state and thus must use the conditional probabilities obtained above for the following:
\begin{align}
p^-(g)=&[p^-(g|0\varepsilon)p(0\varepsilon)+p^-(g|1\varepsilon)p(1\varepsilon)]p^-(0\varepsilon|g)+[p^-(g|1\varepsilon)p(0\varepsilon)+p^-(g|0\varepsilon)p(1\varepsilon)][1-p^-(0\varepsilon|g)]=0.834,
\end{align}
and $p^-(e)=1-p^-(g)=0.166$.  Similarly for the case where we instead measure $\ket{e}$ and the protocol is flipped in the second round:
\begin{align}
p^+(g)=&[p^+(g|0\varepsilon)p(0\varepsilon)+p^+(g|1\varepsilon)p(1\varepsilon)]p^-(1\varepsilon|e)+[p^+(g|1\varepsilon)p(0\varepsilon)+p^+(g|0\varepsilon)p(1\varepsilon)][1-p^-(1\varepsilon|e)]=0.777,
\end{align}
and $p^+(e)=1-p^+(g)=0.223$.


We now have the probabilities to obtain the following measurement records, which closely match those presented in Fig.~3e of the main text:
\begin{align}
p_{0\varepsilon}=&p(g)p^-(g)=0.841\times0.834\\
&=0.701\\
p_{1\varepsilon}=&p(g)p^-(e)+p(e)p^+(g)=0.841\times0.166+0.159\times0.777\\
&=0.263\\
p_{2\varepsilon}=&p(e)p^+(e)=0.159\times0.223\\
&=0.036
\end{align}
Beyond telling us that we understand the statistics of our system, this calculation also provides crucial information as to the confidence of certain trajectories over others.  First, one may immediately note the slight asymmetry between measuring $\ket{g}$ and then $\ket{e}$ ($0.841\times0.166=0.140$) vs. the reverse order ($0.159\times0.777=0.124$).  Indeed, with the following conditional probabilities for all possible error histories ($gg,eg,ge,ee$) we see the huge benefit of a ``confirmation" $g$ measurement on the probability that the measured trajectory faithfully reflects the error trajectory of the encoded state:
\begin{align}
p^-(0\varepsilon|gg)=&\frac{p^-(g|0\varepsilon)p(0\varepsilon)p^-(0\varepsilon|g)}{p^-(g)}=0.993\\
p^+(1\varepsilon|eg)=&\frac{p^+(g|0\varepsilon)p(0\varepsilon)p^-(1\varepsilon|e)}{p^+(g)}=0.978\\
p^-(1\varepsilon|ge)=&\frac{p^-(e|1\varepsilon)p(1\varepsilon)p^-(0\varepsilon|g)}{p^-(e)}=0.869\\
p^+(2\varepsilon|ee)=&\frac{p^+(e|1\varepsilon)p(1\varepsilon)p^-(1\varepsilon|e)}{p^+(e)}=0.592
\end{align}

We can extend these calculations to a general simulation that handles an arbitrary number of correction steps to show that this simple approach captures many of the features we see in our data.  A detailed look at tracking for $\sim68\mathrm{\mu s}$ in Fig.~S\ref{fig:bayesian}a, for example, details the individual process fidelities we expect to measure for every possible measurement record.   A particularly noteworthy conclusion is that trajectories such as $1010$ have a higher expected fidelity than $0001$; although the former suggests more errors in the codeword, each measurement result $1$ is confirmed by a subsequent $0$, whereas this is not the case in the latter.  In summary, with each error syndrome measurement, all measurement infidelities are pushed onto records that report higher and higher error numbers, and with time, these records become more and more common.  As seen in Fig.~S\ref{fig:bayesian}b, the occurrence of high-confidence trajectories falls in time, albeit slowly.  In this sense, although the post-selection substantially improves the quality of the final qubit, it nonetheless results in an exponential decay of acceptable trajectories, an expected trade-off.

\subsubsection{The Increasing Fidelity of the Two-Error Case}
\begin{figure}[h!]
\centering
\includegraphics[width=\linewidth]{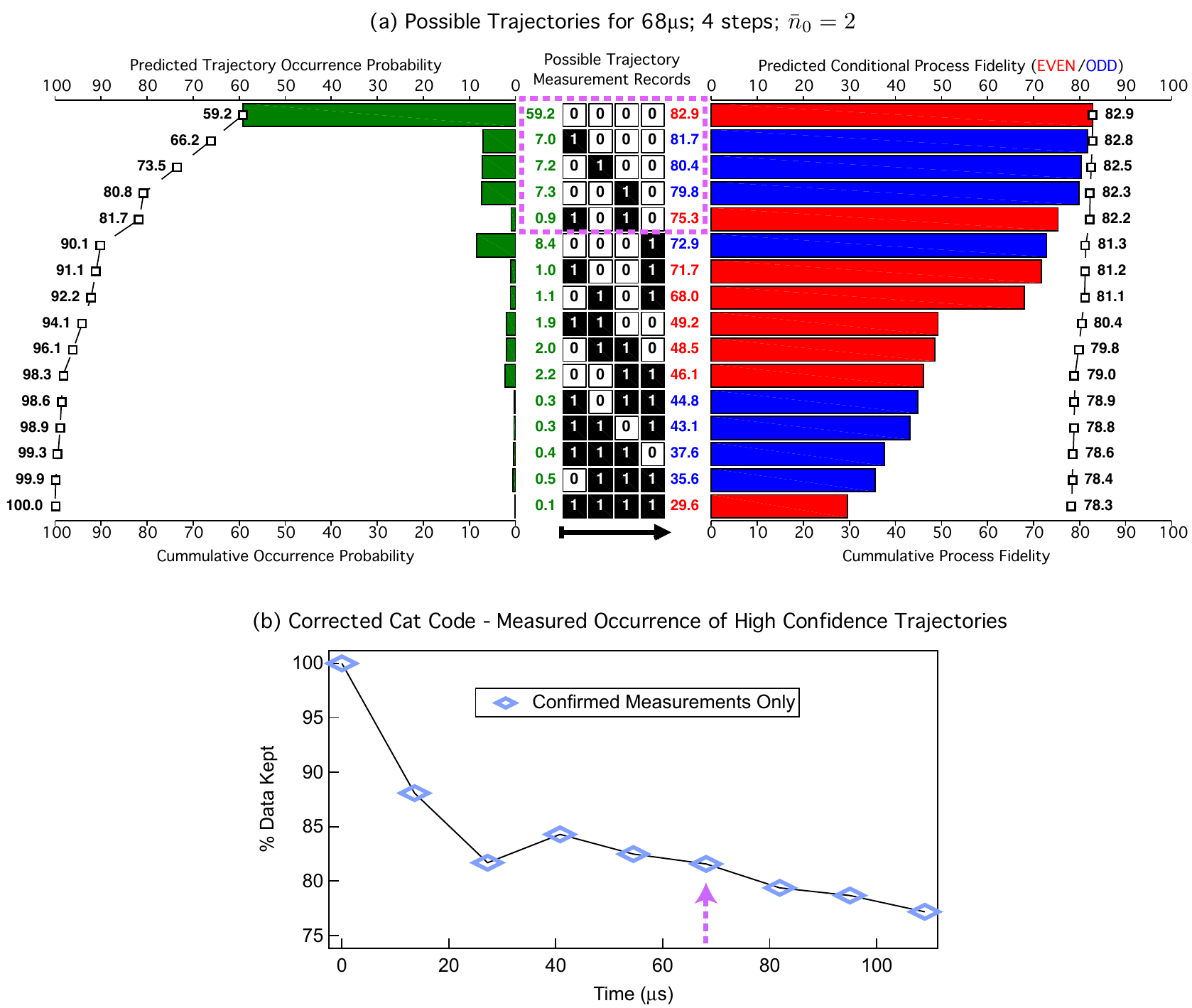}
\caption{\footnotesize
    \textbf{Assessing measurement record confidence.  (a)}  Predicted statistics and confidence for the corrected cat code after four syndrome measurements over $68\mathrm{\mu s}$ of monitoring; $\bar{n}_0=2$.  After four rounds of error correction there are sixteen possible result records: $0000, 1000,\dots,1111$.  The left plot shows the predicted probability to measure each of these records individually (green bars, top axis), and the cumulative probability (bottom axis).  In the right plot, we also show the predicted process fidelity conditioned on measuring each record (red bars are even parity, blue bars are odd parity).  This conditioned fidelity corresponds to our confidence in the output.  In the column separating the plots, the numbers in green correspond to the actual values of each individual green bar (left plot), and the numbers in red and blue correspond to the values of the red and blue bars (right plot).  The axis of the right plot, cumulative process fidelity, is the cumulative sum of the predicted conditional probabilities weighted by the trajectory occurrence probability.  It is interesting to compare the records $1010$ and $0001$. The first suggests two photon jumps (during the first and third steps) and the second suggests a single photon jump during the last step.  The conditional process fidelity for $1010$ is actually higher. This is because measuring ancilla $\ket{g}$, which indicates no change in parity, has a higher probability of being correct than measuring $\ket{e}$, which does indicate a change of parity.  Thus, every error ($1$) in $1010$ is "verified" by the subsequent measurement of no error ($0$), while an error in the last step of $0001$ has a higher likelihood to be a faulty measurement.  Outlined in purple is the set of data we accept as ``high-confidence" trajectories (Fig.~4, main text), wherein every $1$ is confirmed by a subsequent $0$.  \textbf{(b)}  This data shows the percentage of trajectories we accept in the post-selected data of Fig.~4 in the main text.  The purple arrow corresponds to the predictions for $68\mathrm{\mu s}$ in (a).  Although the percentage of high confidence trajectories falls off exponentially, it does so very slowly, and after $\sim100\mathrm{\mu s}$ we still keep $\sim80\%$ of the data.}
	\label{fig:bayesian}
\end{figure}

\begin{figure*}[!ht]
\centering
\includegraphics[width=7.2in]{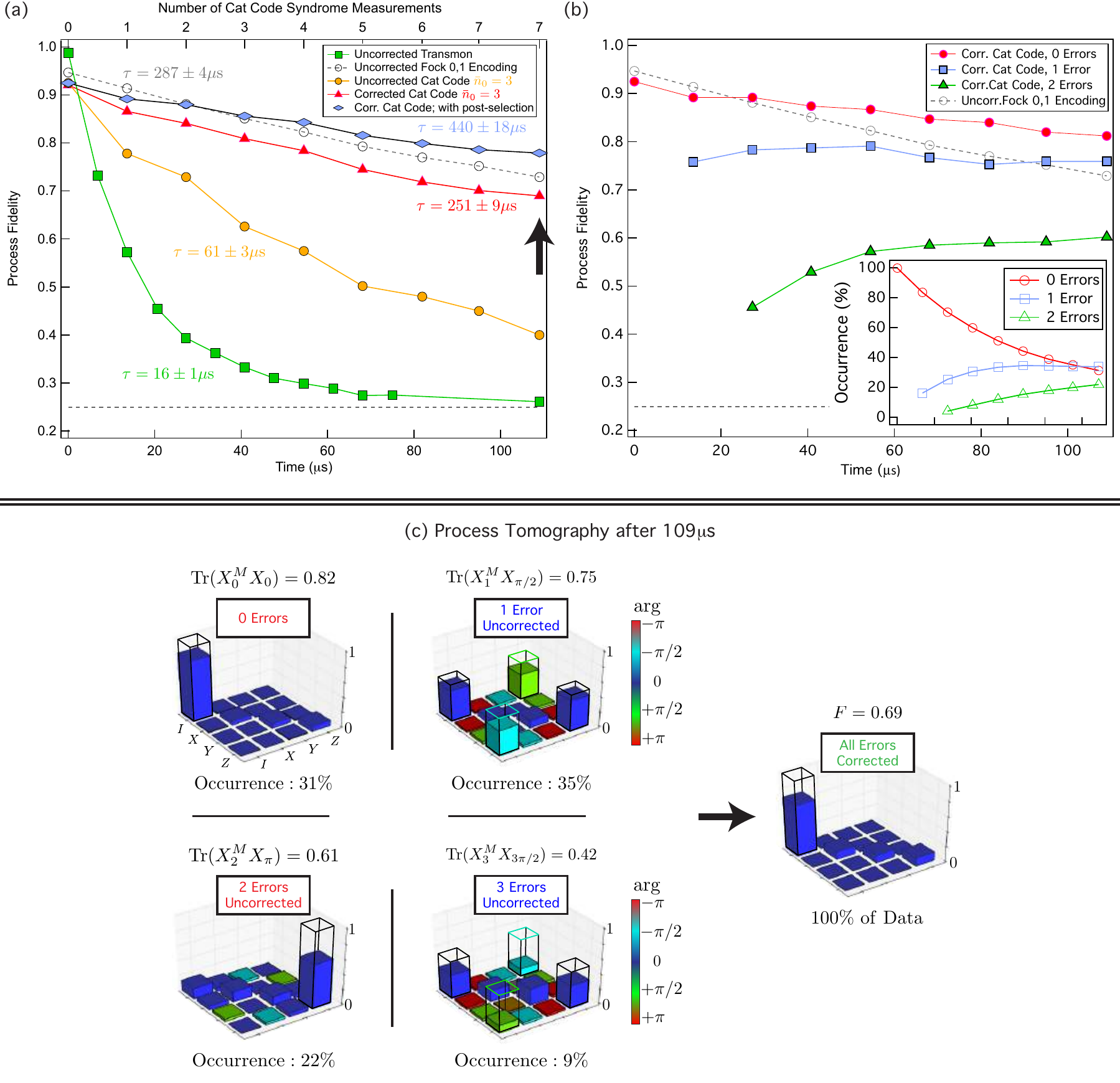}
\caption{\footnotesize
    \textbf{QEC program process tomography for $\bar{n}_0=3$.  (a)}  An identical plot to that shown in Fig.~4a of the main text, except the initial cat size here is $\bar{n}_0=3$.  Every time constant involving cat code is lower than for $\bar{n}_0=2$ due to the increased error rate in the codeword for higher photon numbers, $\gamma=\bar{n}\kappa_s\approx3\kappa_s$ versus $\gamma\approx2\kappa_s$.  As in the main text, on the top axis we plot the number of syndrome measurements used for each point in the corrected cat code; note that for this larger encoding we typically use more measurements for a given time step than for cat states of $\bar{n}_0=2$.  \textbf{(b)}  This plot shows the process fidelities conditioned on individual error trajectories for $j=0$, $1$, and $2$ errors (red circles, blue squares, and green triangles respectively).  Unsurprisingly, the $0$ error case has the highest fidelity, followed by the $1$ and $2$ error cases.  The (initially) surprising feature here is that the process fidelity of error cases increases with time, in particular for the $2$ error case.  This is precisely a consequence of error syndrome measurement fidelity, wherein with more $2$ error trajectories we have more result records that have the aforementioned confirmation measurements.  In other words, with more statistics (inset) we have greater knowledge that measured $2$ error cases in fact correspond to actual $2$ error cases in the encoded state.  Non-monotonic variations in the data points throughout the entire curve are attributed to variations in the efficacy of the decoding pulses at different points in time.  \textbf{(c)}  Measured process matrices $X^M_j$ for $j=0$, $1$, $2$, and $3$ errors after $109\mathrm{\mu s}$ for an initial encoding size of $\bar{n}_0=3$; ideal processes are given by $X_j$ and are wire-outlined.  As compared with the data in Fig.~3e of the main text, note that the $0$ error case has the greatest drop in fidelity, the $1$ error case goes down slightly, and the $2$ error case increases substantially.  Note the $3$ error case also exhibits clear signatures of the correct $X_3$ form.  The substantial drop in fidelity from $0.84$ of Fig.~3e in the main text to $0.69$ here is primarily due to the drop of $0$ error cases with time.}
\label{fig:chibyjump}
\end{figure*}

In this section we present process fidelity decay curves akin to those of Fig.~4a, here shown for encodings of initial size $\bar{n}_0=3$ (Fig.~S\ref{fig:chibyjump}a).  The time constant of the QEC curve is lower since by using as $\bar{n}=3$ rather than $\bar{n}=2$ (main text), the rate of errors the QEC system must handle is higher and the parity measurement fidelity for $\bar{n}=3$ is $0.4\%$ lower than for $\bar{n}=3$ (see sec.~\ref{fidelities}).  Taking this same data and plotting the fidelity decay conditioned on the numbers of errors detected, as shown in Fig.~S\ref{fig:chibyjump}b demonstrates the contrasting trends for the $0$-error case versus the $2$-error case.  As expected, the $0$-error case decays with a time constant of $\sim 630\mathrm{\mu s}$, consistent with double jumps and dephasing due to qubit excitation as the dominant sources of error.  This is the longest time constant in our error correction system, and demonstrates the high fidelity we can achieve if we use the cat code only as an error indicator, whether that error be due to photon jumps or ancilla dephasing.  The fidelity of the $2$-error case, however, increases with time as the occurrence of two-error trajectories that contain higher confidence ``confirmation" measurements within them increases as well.  For example, with four monitoring steps, one has the trajectory $1010$, which has a substantially higher fidelity than anything containing a $11$ (Fig.~S\ref{fig:bayesian}).  Thus the paradoxical rise in fidelity for the two jump case with time simply amounts to an increase in the knowledge that two $\ket{e}$ results actually correspond to two errors in the encoded state.  In terms of measured process matrices $X^M$, note that those shown in Fig.~S\ref{fig:chibyjump}c after $109\mathrm{\mu s}$ have the same form as those in Fig.~3e of the main text, except the fidelity of the two error case in the former is notably higher.  The lower final process fidelity is due to the diminishing number of $0$-error cases with time.

Crucially, these results do not suggest that the cat code is ill-equipped to handle multiple errors throughout a monitoring sequence, but rather highlight the trade-off we make between mitigating the effects of ancilla back-action at the expense of statistics.  Indeed, should we choose to post-select to remove any cases where we measure either anything with $11$ or any trajectory that ends in a $1$, we see a remarkable enhancement in the lifetime, and still keep a majority of the data (inset Fig.~S\ref{fig:chibyjump}b; Fig.~4b of the main text).  

\newpage

\begin{table}[h]
\begin{tabular}{cc}

\hline\hline

 Term & Measured  \\
      & (Prediction) \\

 \hline

 $\omega_a/2\pi$ & 6.2815 GHz \\
 $\omega_s/2\pi$ & 8.3056 GHz \\
 $\omega_r/2\pi$ & 9.3149 GHz \\

 \hline

 $K_a/2\pi$ & 297 MHz \\
 $K_s/2\pi$ & 4.5 kHz \\
 $K_r/2\pi$ & (0.5 kHz)\\ \hline

 $\chi_{sa}/2\pi$ & 1.97 MHz \\
 $\chi_{ra}/2\pi$ & 1 MHz \\
 $\chi_{sr}/2\pi$ & (2 kHz)\\

 \hline

    \caption{\textbf{Hamiltonian parameters} }%

\label{table_params}
\end{tabular}
\end{table}


\begin{table}[h]
\begin{tabular}{cccc}
\hline\hline
 & Ancilla & Storage & Readout \\ \hline
$T_1$ & $35\mu$s & - & -\\
$T_2$& $12\mu$s & - & -\\
$\tau_s$ & - & $250\mu$s & $100$ns\\
$T^s_2$ & - & $330\mu$s & $ - $\\
\hline
ground state (\%)& $96\%$ & $>98\%$ & $>99.3\%$ \\
\hline
  \caption{\textbf{Coherence and thermal properties}}
\label{coherence}
\end{tabular}
\end{table}

\clearpage
\footnotesize
\bibliographystyle{naturemag}
\bibliography{arxiv_QEC_final}

\begin{thebibliography}{10}
\expandafter\ifx\csname url\endcsname\relax
  \def\url#1{\texttt{#1}}\fi
\expandafter\ifx\csname urlprefix\endcsname\relax\def\urlprefix{URL }\fi
\providecommand{\bibinfo}[2]{#2}
\providecommand{\eprint}[2][]{\url{#2}}

\bibitem{Shor:1995vn}
\bibinfo{author}{Shor, P.~W.}
\newblock \bibinfo{title}{Scheme for reducing decoherence in quantum computer
  memory}.
\newblock \emph{\bibinfo{journal}{Phys. Rev. A}} \textbf{\bibinfo{volume}{52}},
  \bibinfo{pages}{R2493--R2496} (\bibinfo{year}{1995}).

\bibitem{NielsenQI}
\bibinfo{author}{Nielsen, M.} \& \bibinfo{author}{Chuang, I.}
\newblock \emph{\bibinfo{title}{{Quantum computation and quantum information}}}
  (\bibinfo{publisher}{Cambridge University Press}, \bibinfo{year}{2010}).

\bibitem{Devoret:2013jz}
\bibinfo{author}{Devoret, M.~H.} \& \bibinfo{author}{Schoelkopf, R.~J.}
\newblock \bibinfo{title}{{Superconducting Circuits for Quantum Information: An
  Outlook}}.
\newblock \emph{\bibinfo{journal}{Science}} \textbf{\bibinfo{volume}{339}},
  \bibinfo{pages}{1169--1174} (\bibinfo{year}{2013}).

\bibitem{Cory:1998kw}
\bibinfo{author}{Cory, D.} \emph{et~al.}
\newblock \bibinfo{title}{{Experimental Quantum Error Correction}}.
\newblock \emph{\bibinfo{journal}{Phys. Rev. Lett.}}
  \textbf{\bibinfo{volume}{81}}, \bibinfo{pages}{2152--2155}
  (\bibinfo{year}{1998}).

\bibitem{Knill:2001tr}
\bibinfo{author}{Knill, E.}, \bibinfo{author}{Laflamme, R.},
  \bibinfo{author}{Martinez, R.} \& \bibinfo{author}{Negrevergne, C.}
\newblock \bibinfo{title}{Benchmarking quantum computers: The five-qubit error
  correcting code}.
\newblock \emph{\bibinfo{journal}{Phys. Rev. Lett.}}
  \textbf{\bibinfo{volume}{86}}, \bibinfo{pages}{5811--5814}
  (\bibinfo{year}{2001}).

\bibitem{Moussa:2011jj}
\bibinfo{author}{Moussa, O.}, \bibinfo{author}{Baugh, J.},
  \bibinfo{author}{Ryan, C.~A.} \& \bibinfo{author}{Laflamme, R.}
\newblock \bibinfo{title}{{Demonstration of Sufficient Control for Two Rounds
  of Quantum Error Correction in a Solid State Ensemble Quantum Information
  Processor}}.
\newblock \emph{\bibinfo{journal}{Phys. Rev. Lett.}}
  \textbf{\bibinfo{volume}{107}}, \bibinfo{pages}{160501}
  (\bibinfo{year}{2011}).

\bibitem{Leung:1999vz}
\bibinfo{author}{Leung, D.} \emph{et~al.}
\newblock \bibinfo{title}{Experimental realization of a two-bit phase damping
  quantum code}.
\newblock \emph{\bibinfo{journal}{Phys. Rev. A}} \textbf{\bibinfo{volume}{60}},
  \bibinfo{pages}{1924--1943} (\bibinfo{year}{1999}).

\bibitem{Chiaverini:2004hy}
\bibinfo{author}{Chiaverini, J.} \emph{et~al.}
\newblock \bibinfo{title}{{Realization of quantum error correction}}.
\newblock \emph{\bibinfo{journal}{Nature}} \textbf{\bibinfo{volume}{432}},
  \bibinfo{pages}{602--605} (\bibinfo{year}{2004}).

\bibitem{Schindler:2011ch}
\bibinfo{author}{Schindler, P.} \emph{et~al.}
\newblock \bibinfo{title}{{Experimental Repetitive Quantum Error Correction}}.
\newblock \emph{\bibinfo{journal}{Science}} \textbf{\bibinfo{volume}{332}},
  \bibinfo{pages}{1059--1061} (\bibinfo{year}{2011}).

\bibitem{Nigg:2014eb}
\bibinfo{author}{Nigg, D.} \emph{et~al.}
\newblock \bibinfo{title}{{Quantum computations on a topologically encoded
  qubit}}.
\newblock \emph{\bibinfo{journal}{Science}} \textbf{\bibinfo{volume}{345}},
  \bibinfo{pages}{302--305} (\bibinfo{year}{2014}).

\bibitem{Waldherr:2014kt}
\bibinfo{author}{Waldherr, G.} \emph{et~al.}
\newblock \bibinfo{title}{{Quantum error correction in a solid-state hybrid
  spin register}}.
\newblock \emph{\bibinfo{journal}{Nature}} \textbf{\bibinfo{volume}{506}},
  \bibinfo{pages}{204--207} (\bibinfo{year}{2014}).

\bibitem{Taminiau:2014up}
\bibinfo{author}{Taminiau, T.~H.}, \bibinfo{author}{Cramer, J.},
  \bibinfo{author}{van~der Sar, T.}, \bibinfo{author}{Dobrovitski, V.~V.} \&
  \bibinfo{author}{Hanson, R.}
\newblock \bibinfo{title}{{Universal control and error correction in
  multi-qubit spin registers in diamond}}.
\newblock \emph{\bibinfo{journal}{Nat. Nano.}} \textbf{\bibinfo{volume}{9}},
  \bibinfo{pages}{171--176} (\bibinfo{year}{2014}).

\bibitem{Cramer:2015uk}
\bibinfo{author}{Cramer, J.} \emph{et~al.}
\newblock \bibinfo{title}{{Repeated quantum error correction on a continuously
  encoded qubit by real-time feedback}}.
\newblock \emph{\bibinfo{journal}{arXiv:quant-ph/1508.01388}}
  (\bibinfo{year}{2015}).

\bibitem{Pittman:2005du}
\bibinfo{author}{Pittman, T.~B.}, \bibinfo{author}{Jacobs, B.~C.} \&
  \bibinfo{author}{Franson, J.~D.}
\newblock \bibinfo{title}{{Demonstration of quantum error correction using
  linear optics}}.
\newblock \emph{\bibinfo{journal}{Phys. Rev. A}} \textbf{\bibinfo{volume}{71}},
  \bibinfo{pages}{052332} (\bibinfo{year}{2005}).

\bibitem{Aoki:2009ew}
\bibinfo{author}{Aoki, T.} \emph{et~al.}
\newblock \bibinfo{title}{{Quantum error correction beyond qubits}}.
\newblock \emph{\bibinfo{journal}{Nat. Phys.}} \textbf{\bibinfo{volume}{5}},
  \bibinfo{pages}{541--546} (\bibinfo{year}{2009}).

\bibitem{Reed:2012hu}
\bibinfo{author}{Reed, M.~D.} \emph{et~al.}
\newblock \bibinfo{title}{{Realization of three-qubit quantum error correction
  with superconducting circuits}}.
\newblock \emph{\bibinfo{journal}{Nature}} \textbf{\bibinfo{volume}{482}},
  \bibinfo{pages}{382--385} (\bibinfo{year}{2012}).

\bibitem{Kelly:2015tg}
\bibinfo{author}{Kelly, J.} \emph{et~al.}
\newblock \bibinfo{title}{{State preservation by repetitive error detection in
  a superconducting quantum circuit}}.
\newblock \emph{\bibinfo{journal}{Nature}} \textbf{\bibinfo{volume}{519}},
  \bibinfo{pages}{66--69} (\bibinfo{year}{2015}).

\bibitem{Corcoles:2015bg}
\bibinfo{author}{C{\'o}rcoles, A.~D.} \emph{et~al.}
\newblock \bibinfo{title}{{Demonstration of a quantum error detection code
  using a square lattice of four superconducting qubits}}.
\newblock \emph{\bibinfo{journal}{Nat. Comm.}} \textbf{\bibinfo{volume}{6}},
  \bibinfo{pages}{6979} (\bibinfo{year}{2015}).

\bibitem{Riste:2015uh}
\bibinfo{author}{Rist{\`e}, D.} \emph{et~al.}
\newblock \bibinfo{title}{{Detecting bit-flip errors in a logical qubit using
  stabilizer measurements}}.
\newblock \emph{\bibinfo{journal}{Nat. Comm.}} \textbf{\bibinfo{volume}{6}},
  \bibinfo{pages}{6983} (\bibinfo{year}{2015}).

\bibitem{book:haroche06}
\bibinfo{author}{Haroche, S.} \& \bibinfo{author}{Raimond, J.-M.}
\newblock \emph{\bibinfo{title}{{Exploring the Quantum: Atoms, Cavities, and
  Photons}}} (\bibinfo{publisher}{Oxford University Press},
  \bibinfo{year}{2006}).

\bibitem{Leghtas:2013ff}
\bibinfo{author}{Leghtas, Z.} \emph{et~al.}
\newblock \bibinfo{title}{{Hardware-Efficient Autonomous Quantum Memory
  Protection}}.
\newblock \emph{\bibinfo{journal}{Phys. Rev. Letters}}
  \textbf{\bibinfo{volume}{111}}, \bibinfo{pages}{120501}
  (\bibinfo{year}{2013}).

\bibitem{Mirrahimi:2014js}
\bibinfo{author}{Mirrahimi, M.} \emph{et~al.}
\newblock \bibinfo{title}{{Dynamically protected cat-qubits: a new paradigm for
  universal quantum computation}}.
\newblock \emph{\bibinfo{journal}{New J. Phys.}} \textbf{\bibinfo{volume}{16}},
  \bibinfo{pages}{045014} (\bibinfo{year}{2014}).

\bibitem{Vlastakis:2013ju}
\bibinfo{author}{Vlastakis, B.} \emph{et~al.}
\newblock \bibinfo{title}{{Deterministically Encoding Quantum Information Using
  100-Photon Schr{\"o}dinger Cat States}}.
\newblock \emph{\bibinfo{journal}{Science}} \textbf{\bibinfo{volume}{342}},
  \bibinfo{pages}{607--610} (\bibinfo{year}{2013}).

\bibitem{Sun:2013ha}
\bibinfo{author}{Sun, L.} \emph{et~al.}
\newblock \bibinfo{title}{{Tracking Photon Jumps with Repeated Quantum
  Non-Demolition Parity Measurements}}.
\newblock \emph{\bibinfo{journal}{Nature}} \textbf{\bibinfo{volume}{511}},
  \bibinfo{pages}{444--448} (\bibinfo{year}{2013}).

\bibitem{Steane:1996wk}
\bibinfo{author}{Steane, A.~M.}
\newblock \bibinfo{title}{{Error correcting codes in quantum theory}}.
\newblock \emph{\bibinfo{journal}{Phys. Rev. Lett.}}
  \textbf{\bibinfo{volume}{77}}, \bibinfo{pages}{793--797}
  (\bibinfo{year}{1996}).

\bibitem{Fowler:2012fi}
\bibinfo{author}{Fowler, A.~G.}, \bibinfo{author}{Mariantoni, M.},
  \bibinfo{author}{Martinis, J.~M.} \& \bibinfo{author}{Cleland, A.~N.}
\newblock \bibinfo{title}{{Surface codes: Towards practical large-scale quantum
  computation}}.
\newblock \emph{\bibinfo{journal}{Phys. Rev. A}} \textbf{\bibinfo{volume}{86}},
  \bibinfo{pages}{032324} (\bibinfo{year}{2012}).

\bibitem{Gambetta:2015ue}
\bibinfo{author}{Gambetta, J.~M.}, \bibinfo{author}{Chow, J.~M.} \&
  \bibinfo{author}{Steffen, M.}
\newblock \bibinfo{title}{{Building logical qubits in a superconducting quantum
  computing system}}.
\newblock \emph{\bibinfo{journal}{arXiv:quant-ph/1510.04375}}
  (\bibinfo{year}{2015}).

\bibitem{Chow:2012ug}
\bibinfo{author}{Chow, J.~M.} \emph{et~al.}
\newblock \bibinfo{title}{Universal quantum gate set approaching fault-tolerant
  thresholds with superconducting qubits}.
\newblock \emph{\bibinfo{journal}{Phys. Rev. Lett.}}
  \textbf{\bibinfo{volume}{109}}, \bibinfo{pages}{060501}
  (\bibinfo{year}{2012}).

\bibitem{Barends:2014fu}
\bibinfo{author}{Barends, R.} \emph{et~al.}
\newblock \bibinfo{title}{{Superconducting quantum circuits at the surface code
  threshold for fault tolerance}}.
\newblock \emph{\bibinfo{journal}{Nature}} \textbf{\bibinfo{volume}{508}},
  \bibinfo{pages}{500--503} (\bibinfo{year}{2014}).

\bibitem{Lloyd:2003kv}
\bibinfo{author}{Lloyd, S.} \& \bibinfo{author}{Braunstein, S.~L.}
\newblock \bibinfo{title}{Quantum computation over continuous variables}.
\newblock \emph{\bibinfo{journal}{Phys. Rev. Lett.}}
  \textbf{\bibinfo{volume}{82}}, \bibinfo{pages}{1784--1787}
  (\bibinfo{year}{1999}).

\bibitem{Gottesman:2001jb}
\bibinfo{author}{Gottesman, D.}, \bibinfo{author}{Kitaev, A.} \&
  \bibinfo{author}{Preskill, J.}
\newblock \bibinfo{title}{{Encoding a qubit in an oscillator}}.
\newblock \emph{\bibinfo{journal}{Phys. Rev. A}} \textbf{\bibinfo{volume}{64}},
  \bibinfo{pages}{012310} (\bibinfo{year}{2001}).

\bibitem{Menicucci:2006ir}
\bibinfo{author}{Menicucci, N.} \emph{et~al.}
\newblock \bibinfo{title}{{Universal Quantum Computation with
  Continuous-Variable Cluster States}}.
\newblock \emph{\bibinfo{journal}{Phys. Rev. Lett.}}
  \textbf{\bibinfo{volume}{97}}, \bibinfo{pages}{110501}
  (\bibinfo{year}{2006}).

\bibitem{LundRalph:2008}
\bibinfo{author}{Lund, A.~P.}, \bibinfo{author}{Ralph, T.~C.} \&
  \bibinfo{author}{Haselgrove, H.~L.}
\newblock \bibinfo{title}{Fault-tolerant linear optical quantum computing with
  small-amplitude coherent states}.
\newblock \emph{\bibinfo{journal}{Phys. Rev. Lett.}}
  \textbf{\bibinfo{volume}{100}}, \bibinfo{pages}{030503}
  (\bibinfo{year}{2008}).

\bibitem{Leghtas:2015uf}
\bibinfo{author}{Leghtas, Z.} \emph{et~al.}
\newblock \bibinfo{title}{Confining the state of light to a quantum manifold by
  engineered two-photon loss}.
\newblock \emph{\bibinfo{journal}{Science}} \textbf{\bibinfo{volume}{347}},
  \bibinfo{pages}{853--857} (\bibinfo{year}{2015}).

\bibitem{Wallraff:2004dy}
\bibinfo{author}{Wallraff, A.} \emph{et~al.}
\newblock \bibinfo{title}{{Strong coupling of a single photon to a
  superconducting qubit using circuit quantum electrodynamics}}.
\newblock \emph{\bibinfo{journal}{Nature}} \textbf{\bibinfo{volume}{431}},
  \bibinfo{pages}{162--167} (\bibinfo{year}{2004}).

\bibitem{Paik:2011hd}
\bibinfo{author}{Paik, H.} \emph{et~al.}
\newblock \bibinfo{title}{{Observation of High Coherence in Josephson Junction
  Qubits Measured in a Three-Dimensional Circuit QED Architecture}}.
\newblock \emph{\bibinfo{journal}{Phys. Rev. Lett.}}
  \textbf{\bibinfo{volume}{107}}, \bibinfo{pages}{240501}
  (\bibinfo{year}{2011}).

\bibitem{Vijay:2011gl}
\bibinfo{author}{Vijay, R.}, \bibinfo{author}{Slichter, D.~H.} \&
  \bibinfo{author}{Siddiqi, I.}
\newblock \bibinfo{title}{Observation of quantum jumps in a superconducting
  artificial atom}.
\newblock \emph{\bibinfo{journal}{Phys. Rev. Lett.}}
  \textbf{\bibinfo{volume}{106}}, \bibinfo{pages}{110502}
  (\bibinfo{year}{2011}).

\bibitem{Hatridge:2013ke}
\bibinfo{author}{Hatridge, M.} \emph{et~al.}
\newblock \bibinfo{title}{{Quantum Back-Action of an Individual
  Variable-Strength Measurement}}.
\newblock \emph{\bibinfo{journal}{Science}} \textbf{\bibinfo{volume}{339}},
  \bibinfo{pages}{178--181} (\bibinfo{year}{2013}).

\bibitem{Bergeal:2010iu}
\bibinfo{author}{Bergeal, N.} \emph{et~al.}
\newblock \bibinfo{title}{{Phase-preserving amplification near the quantum
  limit with a Josephson ring modulator}}.
\newblock \emph{\bibinfo{journal}{Nature}} \textbf{\bibinfo{volume}{465}},
  \bibinfo{pages}{64--68} (\bibinfo{year}{2010}).

\bibitem{Sayrin:2011jx}
\bibinfo{author}{Sayrin, C.} \emph{et~al.}
\newblock \bibinfo{title}{{Real-time quantum feedback prepares and stabilizes
  photon number states}}.
\newblock \emph{\bibinfo{journal}{Nature}} \textbf{\bibinfo{volume}{477}},
  \bibinfo{pages}{73--77} (\bibinfo{year}{2011}).

\bibitem{Vijay:2012bv}
\bibinfo{author}{Vijay, R.} \emph{et~al.}
\newblock \bibinfo{title}{{Stabilizing Rabi oscillations in a superconducting
  qubit using quantum feedback}}.
\newblock \emph{\bibinfo{journal}{Nature}} \textbf{\bibinfo{volume}{490}},
  \bibinfo{pages}{77--80} (\bibinfo{year}{2012}).

\bibitem{Riste:2013if}
\bibinfo{author}{Rist{\`e}, D.} \emph{et~al.}
\newblock \bibinfo{title}{{Deterministic entanglement of superconducting qubits
  by parity measurement and feedback}}.
\newblock \emph{\bibinfo{journal}{Nature}} \textbf{\bibinfo{volume}{502}},
  \bibinfo{pages}{350--354} (\bibinfo{year}{2013}).

\bibitem{Shulman:2014ua}
\bibinfo{author}{Shulman, M.~D.} \emph{et~al.}
\newblock \bibinfo{title}{{Suppressing qubit dephasing using real-time
  Hamiltonian estimation}}.
\newblock \emph{\bibinfo{journal}{Nat. Comm.}} \textbf{\bibinfo{volume}{5}},
  \bibinfo{pages}{5156} (\bibinfo{year}{2014}).

\bibitem{Liu:2015}
\bibinfo{author}{Liu, Y.} \emph{et~al.}
\newblock \bibinfo{title}{{Comparing and combining measurement-based and
  driven-dissipative entanglement stabilization}}.
\newblock \emph{\bibinfo{journal}{arXiv:quant-ph/1509.00860}}
  (\bibinfo{year}{2015}).

\bibitem{Lutterbach:1997cn}
\bibinfo{author}{Lutterbach, L.~G.} \& \bibinfo{author}{Davidovich, L.}
\newblock \bibinfo{title}{{Method for direct measurement of the Wigner function
  in cavity QED and ion traps}}.
\newblock \emph{\bibinfo{journal}{Phys. Rev. Lett.}}
  \textbf{\bibinfo{volume}{78}}, \bibinfo{pages}{2547--2550}
  (\bibinfo{year}{1997}).

\bibitem{Cahill:1969vq}
\bibinfo{author}{Cahill, K.~E.} \& \bibinfo{author}{Glauber, R.~J.}
\newblock \bibinfo{title}{Density operators and quasiprobability
  distributions}.
\newblock \emph{\bibinfo{journal}{Phys. Rev.}} \textbf{\bibinfo{volume}{177}},
  \bibinfo{pages}{1882--1902} (\bibinfo{year}{1969}).

\bibitem{Reagor:2015}
\bibinfo{author}{Reagor, M.} \emph{et~al.}
\newblock \bibinfo{title}{{A quantum memory with near-millisecond coherence in
  circuit QED}}.
\newblock \emph{\bibinfo{journal}{arXiv:quant-ph/1508.05882}}
  (\bibinfo{year}{2015}).

\bibitem{Gottesman:1998gt}
\bibinfo{author}{Gottesman, D.}
\newblock \bibinfo{title}{{Theory of fault-tolerant quantum computation}}.
\newblock \emph{\bibinfo{journal}{Phys. Rev. A}} \textbf{\bibinfo{volume}{57}},
  \bibinfo{pages}{127--137} (\bibinfo{year}{1998}).

\bibitem{Bertet:2002df}
\bibinfo{author}{Bertet, P.} \emph{et~al.}
\newblock \bibinfo{title}{{Direct Measurement of the Wigner Function of a
  One-Photon Fock State in a Cavity}}.
\newblock \emph{\bibinfo{journal}{Phys. Rev. Lett.}}
  \textbf{\bibinfo{volume}{89}}, \bibinfo{pages}{200402}
  (\bibinfo{year}{2002}).

\bibitem{Haroche:2007uc}
\bibinfo{author}{Haroche, S.}, \bibinfo{author}{Brune, M.} \&
  \bibinfo{author}{Raimond, J.~M.}
\newblock \bibinfo{title}{{Measuring the photon number parity in a cavity: from
  light quantum jumps to the tomography of non-classical field states}}.
\newblock \emph{\bibinfo{journal}{J. Mod. Opt.}} \textbf{\bibinfo{volume}{54}},
  \bibinfo{pages}{2101--2114} (\bibinfo{year}{2007}).

\bibitem{Vlastakis:2015tw}
\bibinfo{author}{Vlastakis, B.} \emph{et~al.}
\newblock \bibinfo{title}{{Characterizing entanglement of an artificial atom
  and a cavity cat state with Bell's inequality}}.
\newblock \emph{\bibinfo{journal}{Nat. Comm.}} \textbf{\bibinfo{volume}{6}},
  \bibinfo{pages}{8970} (\bibinfo{year}{2015}).

\bibitem{Vijay:2009ko}
\bibinfo{author}{Vijay, R.}, \bibinfo{author}{Devoret, M.~H.} \&
  \bibinfo{author}{Siddiqi, I.}
\newblock \bibinfo{title}{{The Josephson bifurcation amplifier}}.
\newblock \emph{\bibinfo{journal}{Rev. Sci. Inst.}}
  \textbf{\bibinfo{volume}{80}}, \bibinfo{pages}{111101}
  (\bibinfo{year}{2009}).

\bibitem{Schuster:2007ki}
\bibinfo{author}{Schuster, D.~I.} \emph{et~al.}
\newblock \bibinfo{title}{{Resolving photon number states in a superconducting
  circuit}}.
\newblock \emph{\bibinfo{journal}{Nature}} \textbf{\bibinfo{volume}{445}},
  \bibinfo{pages}{515--518} (\bibinfo{year}{2007}).

\bibitem{Blais:2004tl}
\bibinfo{author}{Blais, A.}, \bibinfo{author}{Huang, R.-S.},
  \bibinfo{author}{Wallraff, A.}, \bibinfo{author}{Girvin, S.~M.} \&
  \bibinfo{author}{Schoelkopf, R.~J.}
\newblock \bibinfo{title}{Cavity quantum electrodynamics for superconducting
  electrical circuits: An architecture for quantum computation}.
\newblock \emph{\bibinfo{journal}{Phys. Rev. A}} \textbf{\bibinfo{volume}{69}},
  \bibinfo{pages}{062320} (\bibinfo{year}{2004}).

\bibitem{Nigg:2012jja}
\bibinfo{author}{Nigg, S.~E.} \emph{et~al.}
\newblock \bibinfo{title}{Black-box superconducting circuit quantization}.
\newblock \emph{\bibinfo{journal}{Phys. Rev. Lett.}}
  \textbf{\bibinfo{volume}{108}}, \bibinfo{pages}{240502}
  (\bibinfo{year}{2012}).

\bibitem{Sears:2012cm}
\bibinfo{author}{Sears, A.~P.} \emph{et~al.}
\newblock \bibinfo{title}{{Photon shot noise dephasing in the strong-dispersive
  limit of circuit QED}}.
\newblock \emph{\bibinfo{journal}{Phys. Rev. B}} \textbf{\bibinfo{volume}{86}},
  \bibinfo{pages}{180504} (\bibinfo{year}{2012}).

\bibitem{Knill:2001is}
\bibinfo{author}{Knill, E.}, \bibinfo{author}{Laflamme, R.} \&
  \bibinfo{author}{Milburn, G.~J.}
\newblock \bibinfo{title}{{A scheme for efficient quantum computation with
  linear optics}}.
\newblock \emph{\bibinfo{journal}{Nature}} \textbf{\bibinfo{volume}{409}},
  \bibinfo{pages}{46--52} (\bibinfo{year}{2001}).

\bibitem{Lloyd:1998ub}
\bibinfo{author}{Lloyd, S.} \& \bibinfo{author}{Slotine, J.}
\newblock \bibinfo{title}{Analog quantum error correction}.
\newblock \emph{\bibinfo{journal}{Phys. Rev. Lett.}}
  \textbf{\bibinfo{volume}{80}}, \bibinfo{pages}{4088--4091}
  (\bibinfo{year}{1998}).

\bibitem{Braunstein:1998uo}
\bibinfo{author}{Braunstein, S.~L.} \& \bibinfo{author}{Kimble, H.~J.}
\newblock \bibinfo{title}{Teleportation of continuous quantum variables}.
\newblock \emph{\bibinfo{journal}{Phys. Rev. Lett.}}
  \textbf{\bibinfo{volume}{80}}, \bibinfo{pages}{869--872}
  (\bibinfo{year}{1998}).

\bibitem{Ralph:1999ex}
\bibinfo{author}{Ralph, T.~C.}
\newblock \bibinfo{title}{{Continuous variable quantum cryptography}}.
\newblock \emph{\bibinfo{journal}{Phys. Rev. A}} \textbf{\bibinfo{volume}{61}},
  \bibinfo{pages}{010303} (\bibinfo{year}{1999}).

\bibitem{Braunstein:2000cw}
\bibinfo{author}{Braunstein, S.} \& \bibinfo{author}{Kimble, H.}
\newblock \bibinfo{title}{{Dense coding for continuous variables}}.
\newblock \emph{\bibinfo{journal}{Phys. Rev. A}} \textbf{\bibinfo{volume}{61}},
  \bibinfo{pages}{042302} (\bibinfo{year}{2000}).

\bibitem{Braunstein:1998cd}
\bibinfo{author}{Braunstein, S.~L.}
\newblock \bibinfo{title}{{Quantum error correction for communication with
  linear optics}}.
\newblock \emph{\bibinfo{journal}{Nature}} \textbf{\bibinfo{volume}{394}},
  \bibinfo{pages}{47--49} (\bibinfo{year}{1998}).

\bibitem{Ralph:2011ct}
\bibinfo{author}{Ralph, T.~C.}
\newblock \bibinfo{title}{{Quantum error correction of continuous-variable
  states against Gaussian noise}}.
\newblock \emph{\bibinfo{journal}{Phys. Rev. A}} \textbf{\bibinfo{volume}{84}},
  \bibinfo{pages}{022339} (\bibinfo{year}{2011}).

\bibitem{Michael:2016}
\bibinfo{author}{Michael, M.} \emph{et~al.}
\newblock \bibinfo{title}{{New class of quantum error-correcting codes for a
  bosonic mode}}.
\newblock \emph{\bibinfo{journal}{arXiv:quant-ph/1602.00008}}
  (\bibinfo{year}{2016}).

\bibitem{Deleglise:2008gt}
\bibinfo{author}{Del{\'e}glise, S.} \emph{et~al.}
\newblock \bibinfo{title}{{Reconstruction of non-classical cavity field states
  with snapshots of their decoherence}}.
\newblock \emph{\bibinfo{journal}{Nature}} \textbf{\bibinfo{volume}{455}},
  \bibinfo{pages}{510--514} (\bibinfo{year}{2008}).

\bibitem{Hofheinz:2009ba}
\bibinfo{author}{Hofheinz, M.} \emph{et~al.}
\newblock \bibinfo{title}{{Synthesizing arbitrary quantum states in a
  superconducting resonator}}.
\newblock \emph{\bibinfo{journal}{Nature}} \textbf{\bibinfo{volume}{459}},
  \bibinfo{pages}{546--549} (\bibinfo{year}{2009}).

\bibitem{Jensen:2011ba}
\bibinfo{author}{Jensen, K.} \emph{et~al.}
\newblock \bibinfo{title}{{Quantum memory for entangled continuous-variable
  states}}.
\newblock \emph{\bibinfo{journal}{Nat. Phys.}} \textbf{\bibinfo{volume}{7}},
  \bibinfo{pages}{13--16} (\bibinfo{year}{2011}).

\bibitem{Heeres:2015}
\bibinfo{author}{Heeres, R.~W.} \emph{et~al.}
\newblock \bibinfo{title}{Cavity state manipulation using photon-number
  selective phase gates}.
\newblock \emph{\bibinfo{journal}{Phys. Rev. Lett.}}
  \textbf{\bibinfo{volume}{115}}, \bibinfo{pages}{137002}
  (\bibinfo{year}{2015}).

\bibitem{Brecht:2015_1}
\bibinfo{author}{Brecht, T.} \emph{et~al.}
\newblock \bibinfo{title}{{Multilayer microwave integrated quantum circuits for
  scalable quantum computing}}.
\newblock \emph{\bibinfo{journal}{arXiv:quant-ph/1519.01127}}
  (\bibinfo{year}{2015}).

\bibitem{Brecht:2015_2}
\bibinfo{author}{Brecht, T.} \emph{et~al.}
\newblock \bibinfo{title}{{Demonstration of superconducting micromachined
  cavities}}.
\newblock \emph{\bibinfo{journal}{arXiv:cond-mat.supr-con/1509.01119}}
  (\bibinfo{year}{2015}).

\bibitem{Minev:2015}
\bibinfo{author}{Minev, Z.~K.} \emph{et~al.}
\newblock \bibinfo{title}{{2.5D circuit quantum electrodynamics/1519.01619}}.
\newblock \emph{\bibinfo{journal}{arXiv:cond-mat.supr-con/}}
  (\bibinfo{year}{2015}).

\bibitem{Khaneja:2005vd}
\bibinfo{author}{Khaneja, N.}, \bibinfo{author}{Reiss, T.},
  \bibinfo{author}{Kehlet, C.}, \bibinfo{author}{Schulte-Herbr{\"u}ggen, T.} \&
  \bibinfo{author}{Glaser, S.~J.}
\newblock \bibinfo{title}{Optimal control of coupled spin dynamics: design of
  nmr pulse sequences by gradient ascent algorithms}.
\newblock \emph{\bibinfo{journal}{J. Mag. Res.}}
  \textbf{\bibinfo{volume}{172}}, \bibinfo{pages}{296 -- 305}
  (\bibinfo{year}{2005}).

\bibitem{deFouquieres:2011wm}
\bibinfo{author}{de~Fouquieres, P.}, \bibinfo{author}{Schirmer, S.~G.},
  \bibinfo{author}{Glaser, S.~J.} \& \bibinfo{author}{Kuprov, I.}
\newblock \bibinfo{title}{Second order gradient ascent pulse engineering}.
\newblock \emph{\bibinfo{journal}{J. Mag. Res.}}
  \textbf{\bibinfo{volume}{212}}, \bibinfo{pages}{412 -- 417}
  (\bibinfo{year}{2011}).

\bibitem{Johansson20131234}
\bibinfo{author}{Johansson, J.~R.}, \bibinfo{author}{Nation, P.~D.} \&
  \bibinfo{author}{Nori, F.}
\newblock \bibinfo{title}{{QuTiP 2: A Python framework for the dynamics of open
  quantum systems}}.
\newblock \emph{\bibinfo{journal}{Comp. Phys. Comm.}}
  \textbf{\bibinfo{volume}{184}}, \bibinfo{pages}{1234--1240}
  (\bibinfo{year}{2013}).

\bibitem{Johansson20121760}
\bibinfo{author}{Johansson, J.~R.}, \bibinfo{author}{Nation, P.~D.} \&
  \bibinfo{author}{Nori, F.}
\newblock \bibinfo{title}{Qutip: An open-source python framework for the
  dynamics of open quantum systems}.
\newblock \emph{\bibinfo{journal}{Comp. Phys. Comm.}}
  \textbf{\bibinfo{volume}{184}}, \bibinfo{pages}{1760 -- 1772}
  (\bibinfo{year}{2012}).

\bibitem{Kirchmair:2013gu}
\bibinfo{author}{Kirchmair, G.} \emph{et~al.}
\newblock \bibinfo{title}{{Observation of quantum state collapse and revival
  due to the single-photon Kerr effect}}.
\newblock \emph{\bibinfo{journal}{Nature}} \textbf{\bibinfo{volume}{495}},
  \bibinfo{pages}{205--209} (\bibinfo{year}{2013}).

\bibitem{Leghtas:2012ff}
\bibinfo{author}{Leghtas, Z.} \emph{et~al.}
\newblock \bibinfo{title}{{Hardware-efficient autonomous quantum error
  correction}}.
\newblock \emph{\bibinfo{journal}{Phys. Rev. Lett.}} \bibinfo{pages}{120501}
  (\bibinfo{year}{2012}).

\bibitem{Riste:2012wc}
\bibinfo{author}{Rist\`e, D.}, \bibinfo{author}{Bultink, C.~C.},
  \bibinfo{author}{Lehnert, K.~W.} \& \bibinfo{author}{DiCarlo, L.}
\newblock \bibinfo{title}{Feedback control of a solid-state qubit using
  high-fidelity projective measurement}.
\newblock \emph{\bibinfo{journal}{Phys. Rev. Lett.}}
  \textbf{\bibinfo{volume}{109}}, \bibinfo{pages}{240502}
  (\bibinfo{year}{2012}).

\end{thebibliography}

\end{document}